\documentclass[longauth]{aa} 

\usepackage{graphicx}
\usepackage{txfonts}
\usepackage[utf8]{inputenc}
\usepackage{array}
\usepackage{float}
\usepackage{multirow}
\usepackage[]{natbib}
\usepackage{amsmath}
\usepackage{epstopdf}
\usepackage[colorlinks=true,linkcolor=blue,citecolor=blue]{hyperref}
\begin{document} 

\title{Orbital and spectral characterization of the benchmark T-type brown dwarf HD\,19467B\thanks{Based on observations collected at the European Organisation for Astronomical Research in the Southern Hemisphere under ESO programmes 1100.C-0481, 0100.C-0234, 096.C-0602, 072.C-0488, 183.C-0972, 084.D-0965, 188.C-0265, 192.C-0852, and 0100.D-0444.}$^{,}$\thanks{The reduced images shown in Fig.~\ref{fig:images} are available in electronic form at the CDS via anonymous ftp to \url{cdsarc.u-strasbg.fr} (130.79.128.5) or via \url{http://cdsweb.u-strasbg.fr/cgi-bin/qcat?J/A+A/xxx/Axx}.}}

  \author{A.-L. Maire\inst{1,2,\thanks{F.R.S.-FNRS Postdoctoral Researcher.}}, K. Molaverdikhani\inst{2,\thanks{International Max Planck Research School for Astronomy and Cosmic Physics, Heidelberg, Germany.}}, S. Desidera\inst{3}, T. Trifonov\inst{2}, P. Molli\`ere\inst{2}, V. D'Orazi\inst{3}, N. Frankel\inst{2}, J.-L. Baudino\inst{4,5}, S. Messina\inst{6}, A. M\"uller\inst{2}, B. Charnay\inst{5}, A.~C. Cheetham\inst{2}, P. Delorme\inst{7}, R. Ligi\inst{8}, M. Bonnefoy\inst{7}, W. Brandner\inst{2}, D. Mesa\inst{3}, F. Cantalloube\inst{2}, R. Galicher\inst{5}, T. Henning\inst{2}, B.~A. Biller\inst{9,2}, J. Hagelberg\inst{10,7}, A.-M. Lagrange\inst{7}, B. Lavie\inst{10}, E. Rickman\inst{10}, D. S\'egransan\inst{10}, S. Udry\inst{10}, G. Chauvin\inst{7,11}, R. Gratton\inst{3}, M. Langlois\inst{12,13}, A. Vigan\inst{13}, M.~R. Meyer\inst{14}, J.-L. Beuzit\inst{13}, T. Bhowmik\inst{5}, A. Boccaletti\inst{5}, C. Lazzoni\inst{3,15}, C. Perrot\inst{16,17,5}, T. Schmidt\inst{18,5}, A. Zurlo\inst{19,20,13}, L. Gluck\inst{7}, J. Pragt\inst{21}, J. Ramos\inst{2}, R. Roelfsema\inst{21}, A. Roux\inst{7}, and J.-F. Sauvage\inst{22,14}
         }

   \institute{STAR Institute, Universit\'e de Li\`ege, All\'ee du Six Ao\^ut 19c, B-4000 Li\`ege, Belgium \\
          \email{almaire@uliege.be}
          \and
          Max-Planck-Institut f\"ur Astronomie, K\"onigstuhl 17, D-69117 Heidelberg, Germany
          \and
          INAF - Osservatorio Astronomico di Padova, Vicolo dell'Osservatorio 5, I-35122 Padova, Italy   
         \and
          Department of Physics, University of Oxford, Oxford, UK
          \and
          LESIA, Observatoire de Paris, Universit\'e PSL, CNRS, Sorbonne Universit\'e, Univ. Paris Diderot, Sorbonne Paris Cit\'e, 5 place Jules Janssen, F-92195 Meudon, France
          \and
          INAF Catania Astrophysical Observatory, Via S. Sofia 78, 95123 Catania, Italy
          \and
          CNRS, IPAG, Univ. Grenoble Alpes, F-38000 Grenoble, France
          \and
         INAF - Osservatorio Astronomico di Brera, Via E. Bianchi 46, 23807, Merate (Lc), Italy
         \and
         Institute for Astronomy, The University of Edinburgh, Royal Observatory, Blackford Hill View, Edinburgh, EH9 3HJ, UK
         \and
         Geneva Observatory, University of Geneva, Chemin des Maillettes 51, 1290 Versoix, Switzerland
		\and
         Unidad Mixta Internacional Franco-Chilena de Astronom\'ia CNRS/INSU UMI 3386 and Departamento de Astronom\'ia, Universidad de Chile, Casilla 36-D, Santiago, Chile
         \and
         CRAL, UMR 5574, CNRS, Universit\'e Lyon 1, ENS Lyon, 9 av. Charles Andr\'e, 69561 Saint-Genis-Laval Cedex, France
         \and
         Aix-Marseille Universit\'e, CNRS, CNES,  LAM, Marseille, France
         \and
		Department of Astronomy, University of Michigan, 1085 S. University Ave, Ann Arbor, MI 48109-1107, USA
         \and
         Dipartimento di Fisica e Astronomia ``G. Galilei'', Universit\`a di Padova, Via Marzolo, 8, 35121 Padova, Italy
         \and
         Instituto de F\'isica y Astronom\'ia, Facultad de Ciencias, Universidad de Valpara\'iso, Av. Gran Breta\~{n}a 1111, Playa Ancha, Valpara\'iso, Chile
		\and
		N\'ucleo Milenio Formaci\'on Planetaria - NPF, Universidad de Valpara\'iso, Av. Gran Breta\~{n}a 1111, Playa Ancha, Valpara\'iso, Chile
		\and
          Hamburger Sternwarte, Gojenbergsweg 112, 21029 Hamburg, Germany
         \and
         N\'ucleo de Astronom\'ia, Facultad de Ingenier\'ia y Ciencias, Universidad Diego Portales, Av. Ejercito 441, Santiago, Chile
         \and
         Escuela de Ingenier\'ia Industrial, Facultad de Ingenier\'ia y Ciencias, Universidad Diego Portales, Av. Ejercito 441, Santiago, Chile
         \and
         NOVA Optical Infrared Instrumentation Group, Oude Hoogeveensedijk 4, 7991 PD Dwingeloo, The Netherlands
         \and
         DOTA, ONERA, Université Paris Saclay, F-91123, Palaiseau France
            }

   \date{Received 20 March 2020 / Accepted 16 May 2020}

 
  \abstract
   {Detecting and characterizing substellar companions for which the luminosity, mass, and age can be determined independently is of utter importance to test and calibrate the evolutionary models due to uncertainties in {their formation mechanisms}. HD\,19467 is a bright and nearby star hosting a cool brown dwarf companion detected with radial velocities and imaging, making it a valuable object for such studies.}
   {We aim to further characterize the orbital, spectral, and physical properties of the HD\,19467 system.}
   {We present new high-contrast imaging data with the SPHERE and NaCo instruments. We also analyze archival data from {the instruments} HARPS, NaCo, HIRES, UVES, and ASAS. Furthermore, we use proper motion data of the star from \textsc{Hipparcos} and \textit{Gaia}.}
  {We refined the properties of the host star and derived an age of 8.0$^{+2.0}_{-1.0}$~Gyr based on isochrones, gyrochronology, and chemical and {kinematic arguments}. This age estimate is slightly younger than previous age estimates of $\sim$9--11~Gyr based on isochrones. No orbital curvature is seen in the current imaging, radial velocity, and astrometric data. From a joint fit of the data, we refined the orbital parameters for HD\,19467B, including: a {period of 398$^{+95}_{-93}$~yr, an inclination of 129.8$^{+8.1}_{-5.1}$~deg, an eccentricity of 0.56$\pm$0.09, a longitude of {the} ascending node of 134.8$\pm$4.5~deg, and an argument of the periastron of 64.2$^{+5.5}_{-6.3}$~deg. We assess a dynamical mass of 74$^{+12}_{-9}$~$M_J$}. The fit with atmospheric models of the spectrophotometric data of the companion indicates an atmosphere without clouds or with very thin clouds, an effective temperature of 1042$^{+77}_{-71}$~K, and a high surface gravity of 5.34$^{+0.08}_{-0.09}$~dex. The comparison to model predictions of the bolometric luminosity and dynamical mass of HD\,19467B, assuming our system age estimate, indicates a better agreement with the Burrows et al. models; whereas, the other evolutionary models used tend to underestimate its cooling rate.}
   {}


\authorrunning{A.-L. Maire et al.}
\titlerunning{Orbital and spectral characterization of the benchmark brown dwarf HD\,19467B}

   \maketitle

\section{Introduction}

The mass of most substellar companions found around stars with high-contrast imaging techniques is inferred from the comparison of their measured luminosity and estimated age to evolutionary models \citep[e.g.,][]{Burrows1997, Baraffe2003, Marley2007, Baraffe2015}. However, uncertainties in the age estimates and in the initial conditions during the formation of these objects produce large uncertainties in the mass estimates, especially at the boundary of the planet and brown dwarf regimes. {To test} and calibrate the evolutionary models, the detection and the characterization of benchmark low-mass companions, for which the luminosity, mass, and age can be derived from independent methods, is of paramount importance.

HD\,19467 is a G3 main-sequence star {located at 32.03$\pm$0.11~pc\footnote{{The uncertainty includes} an additional uncertainty of 0.1~mas to account for potential parallax systematics, \url{https://www.cosmos.esa.int/web/gaia/dr2}.} \citep[][]{GaiaCollaboration2016b, GaiaCollaboration2018}. \citet{Crepp2014} infer an age of 4.6--10~Gyr from isochrones and gyrochronology, and a subsolar metallicity} of [Fe/H]\,=\,$-$0.15$\pm$0.04~dex. \citet{Mason2001} {did not find} evidence for stellar binarity from speckle interferometry. \citet{Crepp2014} report the discovery of a cool brown dwarf companion from a radial velocity (RV) trend measured with the Keck High Resolution Echelle Spectrometer (HIRES) and subsequently confirmed with near-infrared (NIR) high-contrast imaging with the Keck Near-InfraRed Camera (NIRC2). HD\,19467B has an angular separation to the star of $\sim$1.65$''$, corresponding to a projected separation of $\sim$53~au, a {flux ratio with respect to the star $\Delta K_s$\,=\,12.57$\pm$0.09~mag, with blue {$J$-$H$ and $J$-$K_s$} colors, as well as} a minimum dynamical mass of 51.9$^{+3.6}_{-4.3}$~$M_J$ inferred from the RV acceleration and the projected separation \citep[assuming a distance of 30.86$\pm$0.60~pc from \textsc{Hipparcos},][]{vanLeeuwen2007}. It is part of a growing group of imaged brown dwarfs of spectral type T with an {RV signature, which includes,} GJ\,758B \citep{Thalmann2009, Bowler2018}, HD\,4113C \citep{Cheetham2018b}, GJ\,229B \citep{Nakajima1995, Brandt2019c}, and HD\,13724B \citep{Rickman2020}{. Such objects} are valuable benchmarks for atmospheric and evolutionary models of cool substellar objects.

Subsequent observations of HD\,19467B with the integral field spectrometer (IFS) Project 1640 (P1640) {include} a low-resolution NIR spectrum ($R$\,=\,30){, covering} the $J$ and $H$ {bands, and indicate} a spectral type of T5.5$\pm$1.0 \citep{Crepp2015}. By fitting the IFS spectrum with BT-Settl models \citep{Allard2012} with solar metallicity, \citet{Crepp2015} also {estimated} an effective temperature of $T_{\rm{eff}}$\,=\,978$^{+20}_{-43}$~K. Nevertheless, they {deemed} their surface gravity constraints (log\,$g$\,=\,4.21--5.31~dex) to be unreliable by fitting spectra of template T dwarfs, which were degraded and trimmed, to the P1640 resolution and bandwidth. More recently, \citet{JensenClem2016} report a nondetection of the companion in polarized light in the $H$ band using Gemini Planet Imager (GPI) data. The authors {assess} a degree of linear polarization below 2.4\% at 99.73\% confidence{. This result} does not bring any further constraints on the atmospheric properties, in particular, the cloud structure {because} the expectations for such an object are below 1\%. \citet{Wood2019} report a stellar radius measurement of 1.295$\pm$0.048~$R_\sun$ using the Center for High Angular Resolution Astronomy (CHARA) interferometer. The authors also {find an isochronal age of the system of 10.06$^{+1.16}_{-0.82}$~Gyr and that} evolutionary models underpredict the bolometric luminosity of the companion ($-$5.19$^{+0.06}_{-0.07}$~dex) by $\sim$0.5~dex, assuming the isochronal age of the system and the minimum dynamical mass derived in \citet{Crepp2014}. Recently, \citet{Bowler2020} {present an orbital analysis of the companion using new and archival Keck/NIRC2 imaging data} spanning $\sim$6.5~yr. They {derived} (median values and 68.3\% credible intervals) a semi-major axis of 56$^{+15}_{-25}$~au, an eccentricity of 0.39$^{+0.26}_{-0.18}$, an inclination of 125.0$^{+9.4}_{-14.0}$~deg, a longitude of {the} node of 113$^{+16}_{-41}$~deg, and an argument of {the} periastron of 66$^{+32}_{-44}$~deg (the last two parameters {are} restricted to the interval [0,180) deg because of ambiguities due to the use of imaging data only). \citet{Mesa2020} present a long slit spectrum at a resolution of $\sim$350 over the $YJH$ bands, which {was} obtained with the Spectro-Polarimetric High-contrast Exoplanet REsearch (SPHERE) instrument. They {derived} a spectral type of T6$\pm$1 and an effective temperature of 1000$\pm$100~K by fitting BT-Settl spectra, in agreement with \citet{Crepp2015}. They also {derived} a surface gravity of 5.0$\pm$0.5~dex, in the high range of the values in \citet{Crepp2015}.

\begin{figure*}[t]
\sidecaption
\includegraphics[width=.45\textwidth, angle=90]{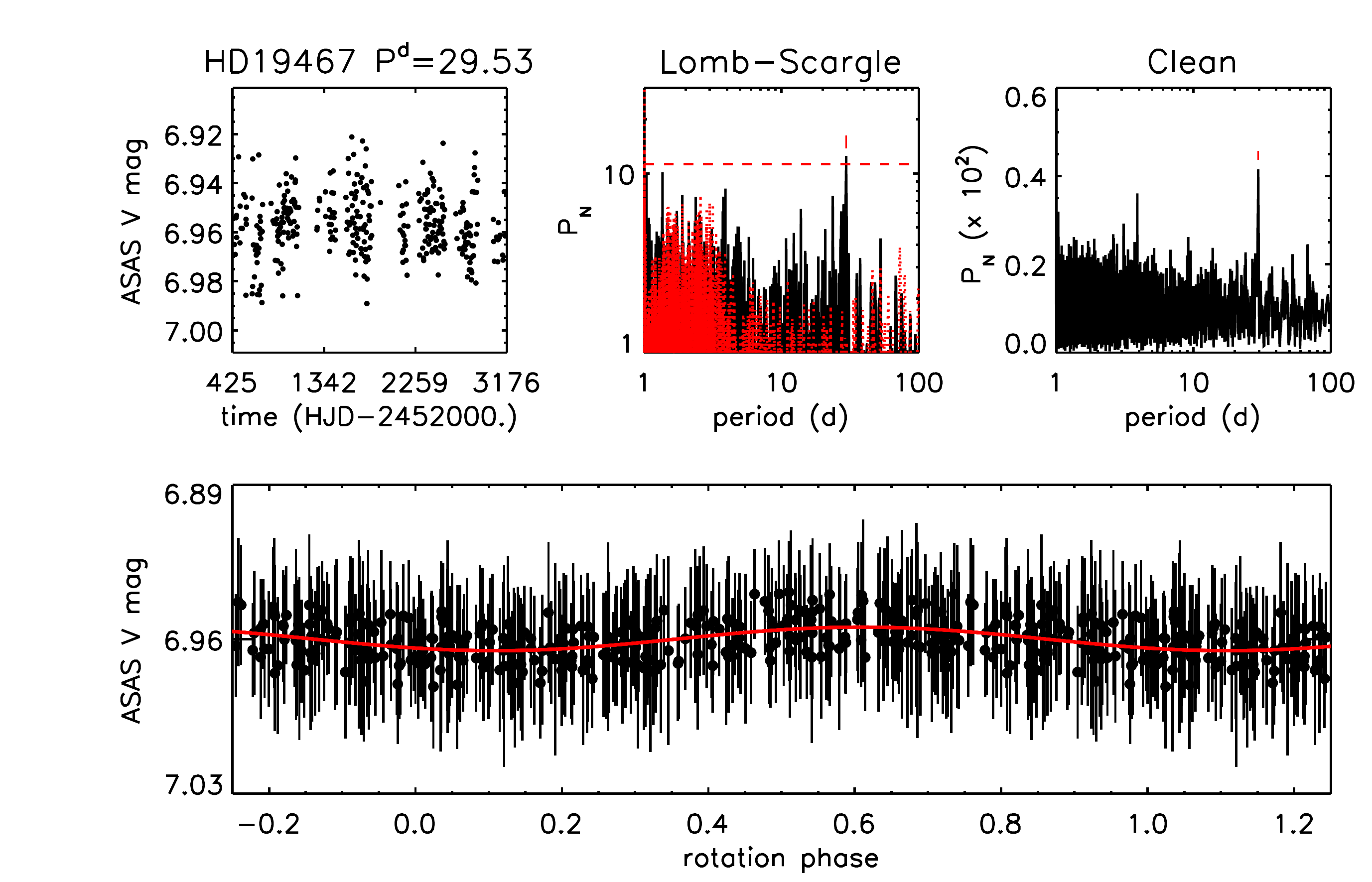}
\caption{Photometric analysis of HD\,19467 based on ASAS data. \textit{Top row from left to right}: $V$-band magnitude vs. Heliocentric Julian Day, Lomb-Scargle periodogram, and \textit{CLEAN} periodogram. For the Lomb-Scargle periodogram, we show the spectral window function (in red), the 99\% confidence level (horizontal dashed red line), and the peak corresponding to the rotation period (red vertical mark). \textit{Bottom panel}: Light curve phased with the rotation period. The solid red curve represents the sinusoidal fit.}
\label{fig:photometry_asas}
\end{figure*}

We present here NIR follow-up observations of HD\,19467B {with the} SPHERE instrument \citep{Beuzit2019} at the Very Large Telescope (VLT), {taken} as part of the SpHere INfrared survey for Exoplanets \citep[SHINE,][]{Chauvin2017b}. {We also present complementary} observations in the thermal IR {with the} Nasmyth Adaptive Optics System and Near-Infrared Imager and Spectrograph \citep[NaCo,][]{Rousset2003, Lenzen2003}. In addition, we {analyze} archival RV data from the High Accuracy Radial velocity Planet Searcher \citep[HARPS,][]{Mayor2003} and HIRES \citep{Vogt1994}, archival imaging data from NaCo, archival spectroscopic data from HARPS and the Ultraviolet and Visual Echelle Spectrograph \citep[UVES,][]{Dekker2000}, as well as archival photometric data from the All Sky Automated Survey \citep[ASAS,][]{Pojmanski1997}. {Furthermore, we use} proper motion measurements of the star from \textsc{Hipparcos} and \textit{Gaia}. We present an updated analysis of the properties of the host star in Sect.~\ref{sec:star_ppties}. We describe the new high-contrast imaging observations and the archival RV data that we {use} to further characterize HD\,19467B in Sect.~\ref{sec:data}. We {fit the SPHERE, NIRC2, HARPS, HIRES, and \textsc{Hipparcos}-\textit{Gaia} data simultaneously} and derive orbital parameters and a dynamical mass for HD\,19467B in Sect.~\ref{sec:orbit}. Section~\ref{sec:sed} discusses the spectral properties of the companion using our new photometric data and literature measurements. Finally, we compare the dynamical and spectral properties of HD\,19467B to model predictions in Sect.~\ref{sec:compa_models}.

\section{Properties of the host star}
\label{sec:star_ppties}

\citet{Crepp2014} {note} that HD\,19467 is a field star not associated to any moving group. They also {find} that given that it is located slightly above the median \textsc{Hipparcos}-based {main sequence} \citep{Wright2005} and has subsolar metallicity, it should {be older than the Sun (4.6~Gyr)}.

Considering the relevance of the stellar age for the goals of our study, we present here a comprehensive reassesment of the age of the system and other stellar properties. Our approach is based on the inclusion of a variety of indicators, as in \citet{Desidera2015}.

\subsection{Abundance analysis}

We retrieved and analyzed archival data from HARPS and UVES to perform a spectroscopic determination of stellar parameters ($T_{\rm{eff}}$, log\,$g$, and microturbulence velocity $\xi$) and elemental abundances for light, iron-peak, and $\alpha$ elements. The signal-to-noise ratio on the HARPS spectrum is $\sim$400. The analysis of the HARPS spectrum was carried out in the standard way, as described in our previous work \citep[{see, e.g.},][]{DOrazi2017}, by using the Kurucz set of model atmospheres (\citealt{Castelli2003}) and the code MOOG by \citet[][2017 version]{Sneden1973}. Briefly, effective temperature and microturbulence come from removing spurious trends between iron abundances from Fe{\sc i} lines and excitation potential and reduced equivalent width of the spectral lines, respectively. Surface gravity has been obtained via ionization equilibrium of Fe{\sc i} and Fe{\sc ii}. We refer the reader to \citet{DOrazi2017} for details on linelist, atomic parameters and error estimate computations.

We {derived} $T_{\rm{eff}}$=5770$\pm$80~K, log\,$g$=4.32$\pm$0.06~dex, and a microturbulent velocity $\xi$=1.00$\pm$0.15~km\,s$^{-1}$. The $T_{\rm{eff}}$ is slightly {higher and the log\,$g$ is slightly lower than those} derived in \citet{Crepp2014}: {$T_{\rm{eff}}$=5680$\pm$40~K and log\,$g$=4.40$\pm$0.06~dex}. We also {derived} [Fe/H]=-0.11$\pm$0.01~dex, [C/H]=-0.09$\pm$0.01~dex, [O/H]=-0.02$\pm$0.01~dex (non local thermal equilibrium corrections applied), [Na/H]=-0.03$\pm$0.01~dex, [Mg/H]=0.05$\pm$0.08~dex, [Al/H]=0.05$\pm$0.03~dex, [Si/H]=-0.04$\pm$0.04~dex, [S/H]=-0.09$\pm$0.07~dex, [Ca/H]=-0.04$\pm$0.05~dex, [Ti/H]I=0.03$\pm$0.05~dex, [Ti/H]II=0.09$\pm$0.02~dex, [Cr/H]I=-0.09$\pm$0.03~dex, [Cr/H]II=-0.07$\pm$0.06~dex, and [Ni/H]=-0.09$\pm$0.05~dex. The [$\alpha$/Fe] ratios are very weakly enhanced (at the level of $\sim$0.1~dex). This abundance pattern is not compatible with a thick disk membership and suggests membership to the thin disk population or to the small population intermediate between {the} thin disk and {the} thick disk proposed by \citet{Fuhrmann2019} and references therein.

The UVES spectrum was used to derive the oxygen abundance from the 7700\AA~triplet. The C/O ratio in number is 0.52, which is similar to the value for the Sun \citep[$\sim$0.54,][]{Asplund2009, Caffau2011}. This also argues against a thick disk {membership}.

Abundance ratios involving neutron capture elements can be used as age indicators.
From the measured [Y/Mg]=-0.15$\pm$0.07~dex, we infer an age of 8.5~Gyr following \citet{Nissen2016} and 7.7~Gyr using \citet{Spina2018}.

\subsection{Isochrone fitting}

As mentioned above, the star is slightly evolved above the main sequence, making it suitable for age determination using isochrones.
\citet{Crepp2014} {estimated} an age of 9$\pm$1~Gyr, while \citet{Wood2019} {derived} $10.06^{+1.16}_{-0.82}$~Gyr exploiting also the interferometric measurement of the stellar radius.
We obtained an independent determination using the models by \citet{Bressan2012} exploiting the online tool for Bayesian determination
of stellar parameters PARAM\footnote{\url{http://stev.oapd.inaf.it/cgi-bin/param_1.3}} \citep{daSilva2006}.
Using as input our spectroscopic effective temperature and metallicity, the $V$ band magnitude from \textsc{Hipparcos} (7.00~mag) and the \textit{Gaia} DR2 parallax, we {obtain} a stellar age of 9.3$\pm$1.6~Gyr and a stellar mass of 0.953$\pm$0.022~$M_{\odot}$.
All these estimates then converge on a very old age for the system. The RV time series and the adaptive optics observations presented here allow us to rule out that the position in the color-magnitude diagram is altered by binarity, the contribution of HD\,19467B to the integrated flux being negligible.

\begin{figure*}[t]
\centering
\includegraphics[width=.4\textwidth]{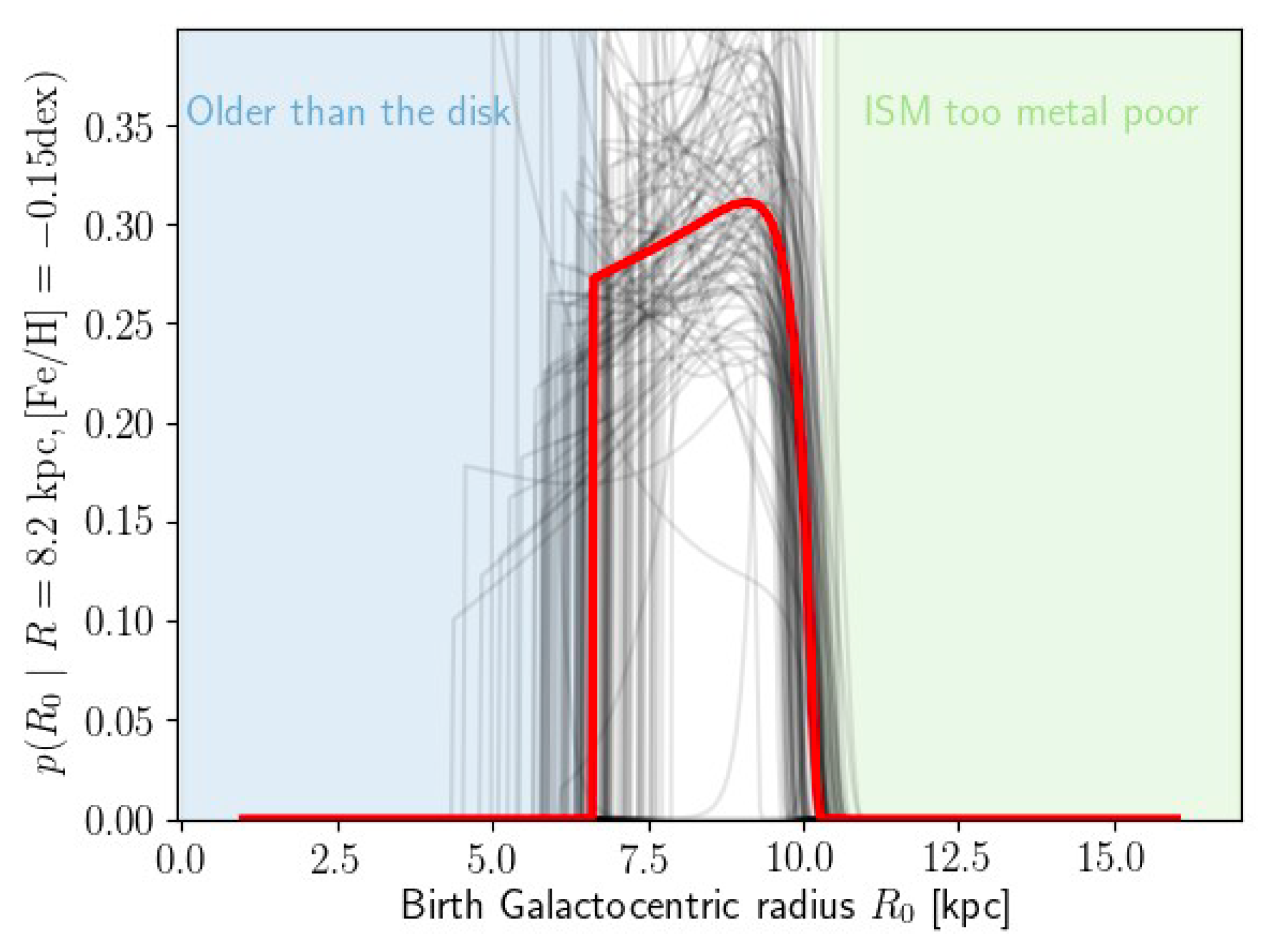}
\includegraphics[width=.4\textwidth]{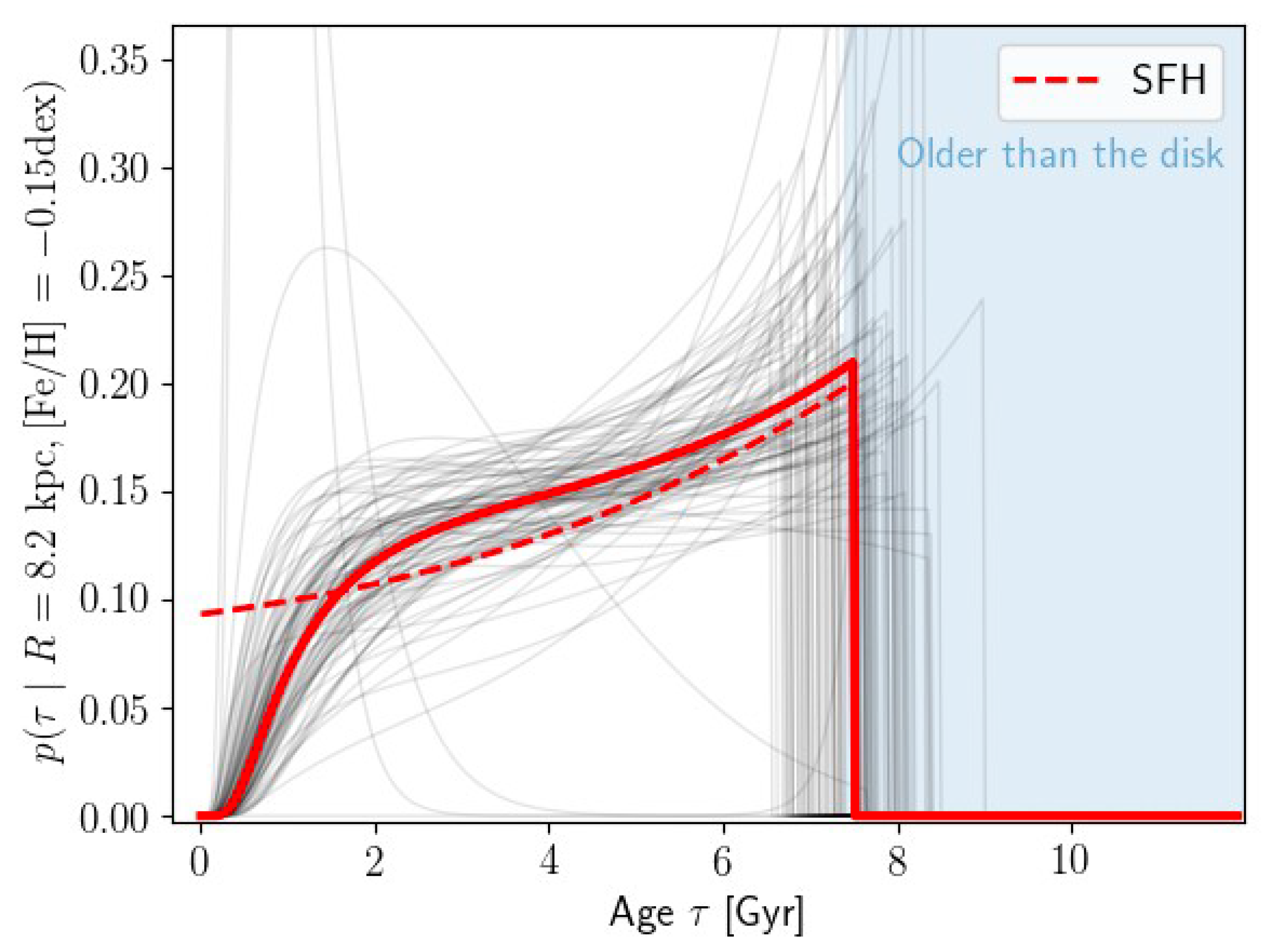}
\caption{PDF of the birth Galactocentric radius (\textit{left}) and of the age (\textit{right}) of HD\,19467 derived using an evolution model of the Milky Way (see text). In both panels, the red thick curves show the best-fit and the gray curves {are} drawn using the uncertainties in the model parameters. In the right panel, the red dashed curve represents the age PDF assuming only that HD\,19467 belongs to the Milky Way disk.}
\label{fig:MW_evolmodel}
\end{figure*}

\subsection{Photometric analysis}

{To better constrain} the stellar rotation period and age through gyrochronology, we analyzed photometric data from ASAS {using the approach in \citet{Messina2010}}. Figure~\ref{fig:photometry_asas} shows the results. Both the Lomb-Scargle periodogram \citep{Lomb1976, Scargle1982} and the \textit{CLEAN} periodogram \citep{Roberts1987} show a peak at $P_{\star}$\,=\,29.53\,$\pm$\,0.16~d. The uncertainty {was} estimated following the approach of \citet{Lamm2004}. Our rotation period estimate is longer by $\sim$1.8$\sigma$ than the indirect estimate of 24.9$\pm$2.5~d in \citet{Crepp2014} based on the measured chromospheric activity indicator log\,$R^{\prime}_{HK}$ and $B-V$ color \citep{Wright2004}. As a result, our gyrochronological age estimate of 5.6$\pm$0.8~Gyr points toward older ages than the age estimate of 3.1--5.3~Gyr in \citet{Crepp2014} based on the same model relations in \citet{Mamajek2008}. Our older age estimate is consistent with the slow $v$\,sin\,$i_{\star}$\,=\,1.6\,$\pm$\,0.5~km\,s$^{-1}$ of the star \citep{Crepp2014}. We also {derived} an age of 5.8$\pm$0.6~Gyr using the gyrochronological calibration in \citet{Delorme2011}. The uncertainty is dominated by the calibration errors calculated from the dispersion of periods around the Hyades and Praesepe calibration sample.

Combining the rotation period, the stellar radius \citep[1.295$\pm$0.048~$R_{\sun}$,][]{Wood2019}, and the projected rotational velocity, we infer using a Monte Carlo approach and Gaussian distributions for the three input parameters an inclination of the stellar rotation axis with respect to the line of sight {of} 46$^{+20}_{-15}$~deg or 137$^{+18}_{-17}$~deg with the uncertainties given at 68\%. The {wide ranges} are due to the large uncertainty on the projected rotational velocity. {The two sets of values are due to the degeneracy of the sin function \citep[see Eq.~(3) in][]{Bonnefoy2018}. As explained in \citet{Bowler2017} and \citet{Bonnefoy2018}, a lower limit on the relative orientation of a stellar spin axis and of the orbital angular momentum or true obliquity of a companion can be derived from the absolute difference between the posterior distribution of the orbital inclination of the companion (derived in Sect.~\ref{sec:orbit} for HD\,19467B) and of the posterior distribution of the inclination of the rotation axis of the star. For the latter, we only {consider} the range [90,180) deg. The resulting distribution extends down to zero. From the upper bound of the 68\% interval, we {infer} that the configuration of the star-brown dwarf system is compatible with a spin-orbit alignment or misalignment within 30$^{\circ}$ at 68\%.}

\subsection{Milky Way evolution model}

We also used a Milky Way evolution model approach \citep[][]{Frankel2018, Frankel2019} to constrain the stellar age given its present-day distance to the Milky Way center and its slight subsolar metallicity. Such an analysis is applicable for stars with low enhancements in $\alpha$ elements similar to HD\,19467. The model assumes a radial-dependent star formation history for the Milky Way, a relation for the stellar metallicity at their formation epoch as a function of their distance to the Milky Way center and their formation epoch, and the migration distance from their birth place as a function of time after birth.

The left panel of Fig.~\ref{fig:MW_evolmodel} shows the probability density function (PDF) of the birth distance of HD\,19467 at given present-day distance and metallicity (see red thick curve). The constraints come from two main aspects. First, given that HD\,19467 is metal poor, it cannot come from less than 6.8~kpc (or it would need to be older than the Milky Way thin disk). Then, it cannot either come from {farther} than 10~kpc, because beyond this distance the interstellar medium is metal poorer.

The right panel of Fig.~\ref{fig:MW_evolmodel} gives the PDF of the age of HD\,19467 at given present-day radius and metallicity. The red dashed curve represents the star formation history of the Milky Way disk{. It} would be the age PDF of HD\,19467 if we only assume that the star belongs to the Milky Way disk. If, additionally, information on the stellar metallicity and present-radius distance is available, the age PDF can be slightly narrowed down (see red thick curve) to a smoothed version of the local star formation history. However, {because} radial migration of stars belonging to the Milky Way disk is significant, the solar neighborhood is populated by a large {portion} of stars that come from very different birth distances with several star formation and metal enrichment histories. This implies that a Galactic evolution approach cannot put tight constraints on the age of HD\,19467. Moreover, the age of the Milky Way disk is not well constrained{. This} results in large uncertainties in the age PDF of HD\,19467 in addition to other uncertainties in the model parameters (shown as gray curves). The most robust information that we can derive from our approach is an estimate of the oldest age of HD\,19467. Our analysis suggests a value of 7.5$\pm$0.9~Gyr.

\subsection{Summary}

We summarize here the results of the various {methods of determining} the stellar age (see also Table~\ref{tab:stellarages}){. Using} ASAS photometric data of HD\,19467, we {derived} a direct gyrochronological age of 5.6$\pm$0.8~Gyr, which is older than the indirect estimate of 3.1--5.3~Gyr from activity indicators and colors in \citet{Crepp2014}. This still {disagrees} with isochronal age estimates \citep[9.2--11.2~Gyr,][]{Wood2019} and our isochronal age estimate (7.7--10.9~Gyr). Our abundance analysis from HARPS and UVES spectra indicates a C/O ratio similar to the Sun, which argues against a $\sim$10-Gyr age. Another argument against such an extremely old age comes from the chemical abundances and kinematics of the star, which suggest that it belongs to the thin disk population{. Literature} studies constrain the age of the oldest members {of this population} to 8~Gyr \citep[e.g.,][]{Fuhrmann2017}. The abundance of neutron capture elements also indicates an age similar to the thin-disk limit, 7.7--8.5~Gyr. The mild enhancement of $\alpha$ elements is also compatible with the intermediate population between {the} thin disk and {the} thick disk, while kinematic parameters would be unusual for a member of this population and more typical of a thin disk star. \citet{Fuhrmann2019} {derived} an age of about 10~Gyr for this intermediate population, similar to the age derived through the isochrone method.

Gyrochronology is expected to be reliable for old stars from theory (because they are less affected by the initial conditions), although in practice precise rotation period measurements are more difficult because stellar spots are usually smaller. This induces a larger scatter in the model relations at old ages. However, \citet{Amard2020} {show} that metal-poor stars spin down less effectively at ages older than $\sim$1~Gyr, making them appear younger than they are actually. For a star with an [Fe/H] similar to HD\,19467 (-0.11~dex) and a rotation period of 29.5~d, the age estimated from gyrochronology would shift from $\sim$5.8 to $\sim$6.5--6.8~Gyr according to the assumed wind-braking model \citep[see Fig.~2 in][]{Amard2020}. In addition, \citet{vanSaders2016} {show} that Sun-like stars older than 4--5~Gyr can experience weakened magnetic braking, which would also bias gyrochronological age estimates toward younger values. Assuming that the rotational evolution of HD\,19467 is not affected by weakened magnetic braking, our gyrochronological analysis would imply a nominal age of at least $\sim$6.5~Gyr from our measured rotation period. 

\begin{table}[t]
\caption{Summary of age estimates for HD\,19467.}
\label{tab:stellarages}
\begin{center}
\begin{tabular}{l c}
\hline\hline
Method & Age \\
 & (Gyr) \\
\hline
Isochrones & 9.3$\pm$1.6 \\
Kinematics & $\leq$8 \\
C/O ratio & $<$10 \\
Y/Mg ratio & 7.7--8.5 \\
Gyrochronology & 6.5$\pm$0.8 \\
Milky Way evolution model & <7.5$\pm$0.9 \\
\hline
Adopted value & 8.0$^{+2.0}_{-1.0}$ \\
\hline
\end{tabular}
\end{center}
\end{table}

\begin{table*}[t]
\caption{Observing log.}
\label{tab:obs}
\begin{center}
\begin{tabular}{l c c c c c c c c}
\hline\hline
UT date & $\epsilon$ ($''$) & $\tau_0$ (ms) & AM start/end & Instrument & Bands & DIT\,(s)\,$\times$\,Nfr & FoV rot. ($^{\circ}$) & SR \\
\hline
2015/12/20 & 0.6--0.9 & 2--3 & 1.05--1.04 & NaCo & $M^{\prime}$ & 0.08$\times$45\,000 & 90 & -- \\
2017/10/06 & 0.5--0.8 & 4--6 & 1.02--1.05 & NaCo & $L^{\prime}$ & 0.10$\times$32\,680 & 71.3 & -- \\
2017/11/04 & 0.3--0.9 & 1--25 & 1.06--1.02 & SPHERE & $K12$ & 64$\times$66 & 63.3 & 0.60--0.87 \\
2018/10/18 & 0.3--0.6 & 5--11 & 1.02--1.04 & SPHERE & $YJ$+$H23$ & 64(96)$\times$50(33) & 50.2 & 0.82--0.87 \\
\hline
\end{tabular}
\end{center}
\tablefoot{The columns provide the observing date, the seeing and coherence time measured by the differential image motion monitor (DIMM) at 0.5~$\mu$m, the airmass at the beginning and the end of the sequence, the observing mode, the spectral bands, the DIT (detector integration time) multiplied by the number of frames in the sequence, the field of view rotation, and the Strehl ratio measured by the adaptive optics system (at 1.6~$\mu$m, SPHERE data only). For the DIT$\times$Nfr column, the numbers {in parentheses} are for the IFS data.}
\end{table*}

We then consider the gyrochronological age as a lower limit. The age of the oldest thin disk stars (8~Gyr) is the most probable value for HD\,19467. The age of 8~Gyr seems very consistent with large scale surveys of Milky Way disk stars \citep{Pinsonneault2019}. The {conflict} with the isochrone results can be considered as marginal. We {adopt an age of 8.0$^{+2.0}_{-1.0}$~Gyr for HD\,19467}. The lower limit of about 7~Gyr is set by the isochrones and abundance of neutron-capture elements. The upper limit of 10~Gyr corresponds to the case of the star being a member of the intermediate population between {the} thin disk and {the} thick disk{. Asteroseismological} measurements of the star \citep{Ulrich1986} should provide independent clues on its age and hopefully solve for the discrepancies between gyrochronology and isochrones.

\begin{figure*}[t]
\centering
\begin{tabular}{p{.24\textwidth} p{.24\textwidth} p{.24\textwidth}}
\includegraphics[width=.26\textwidth]{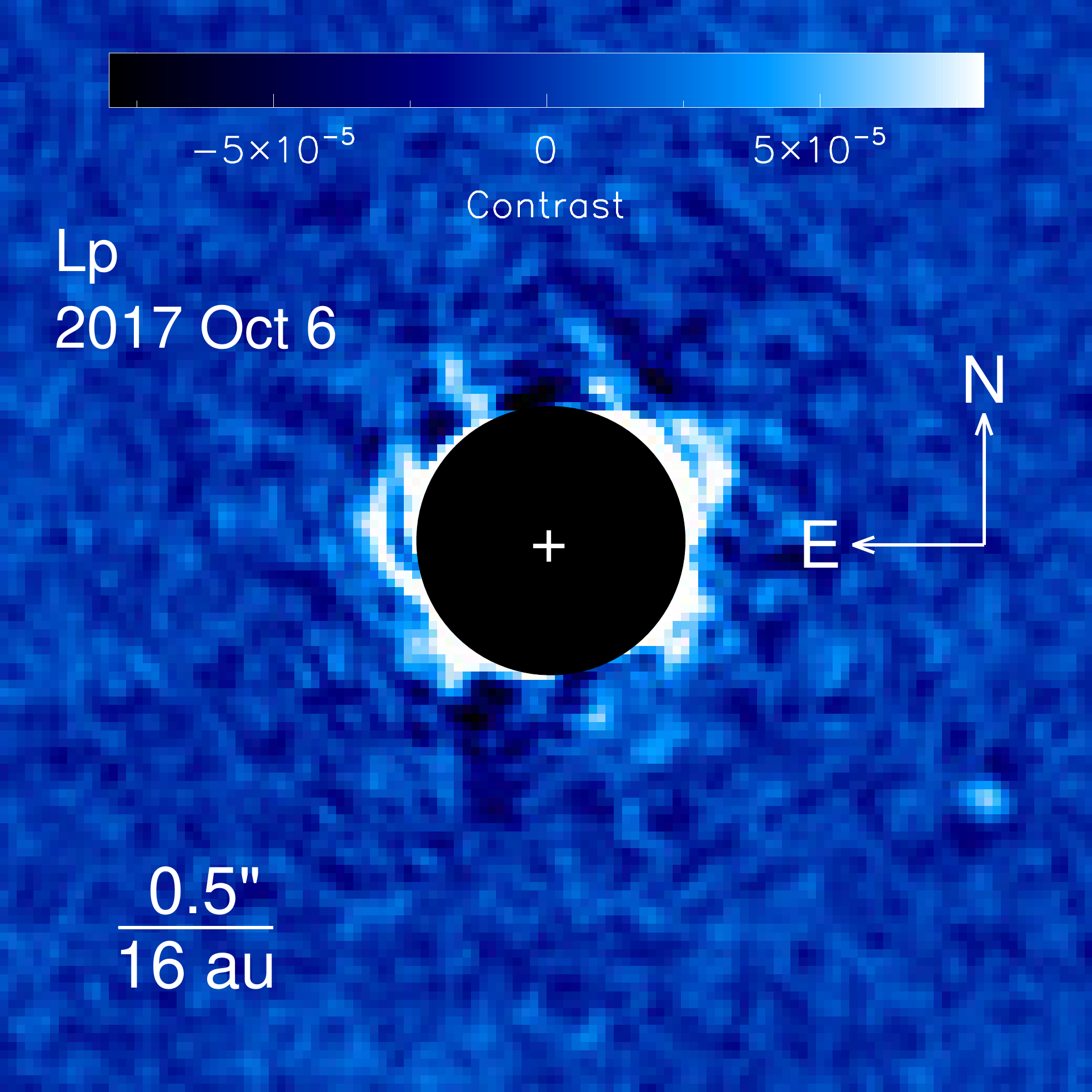} & \includegraphics[width=.26\textwidth]{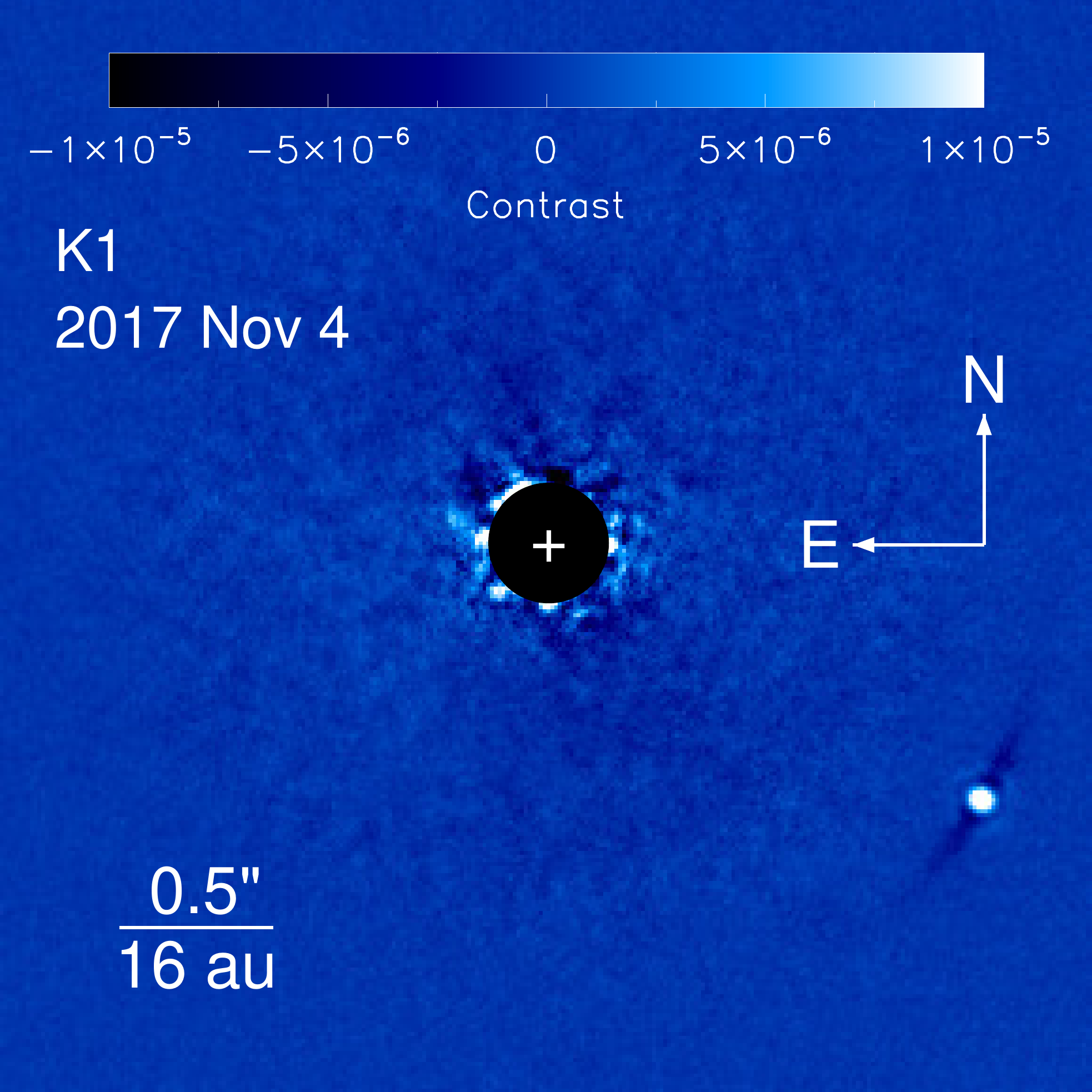} & \includegraphics[width=.26\textwidth]{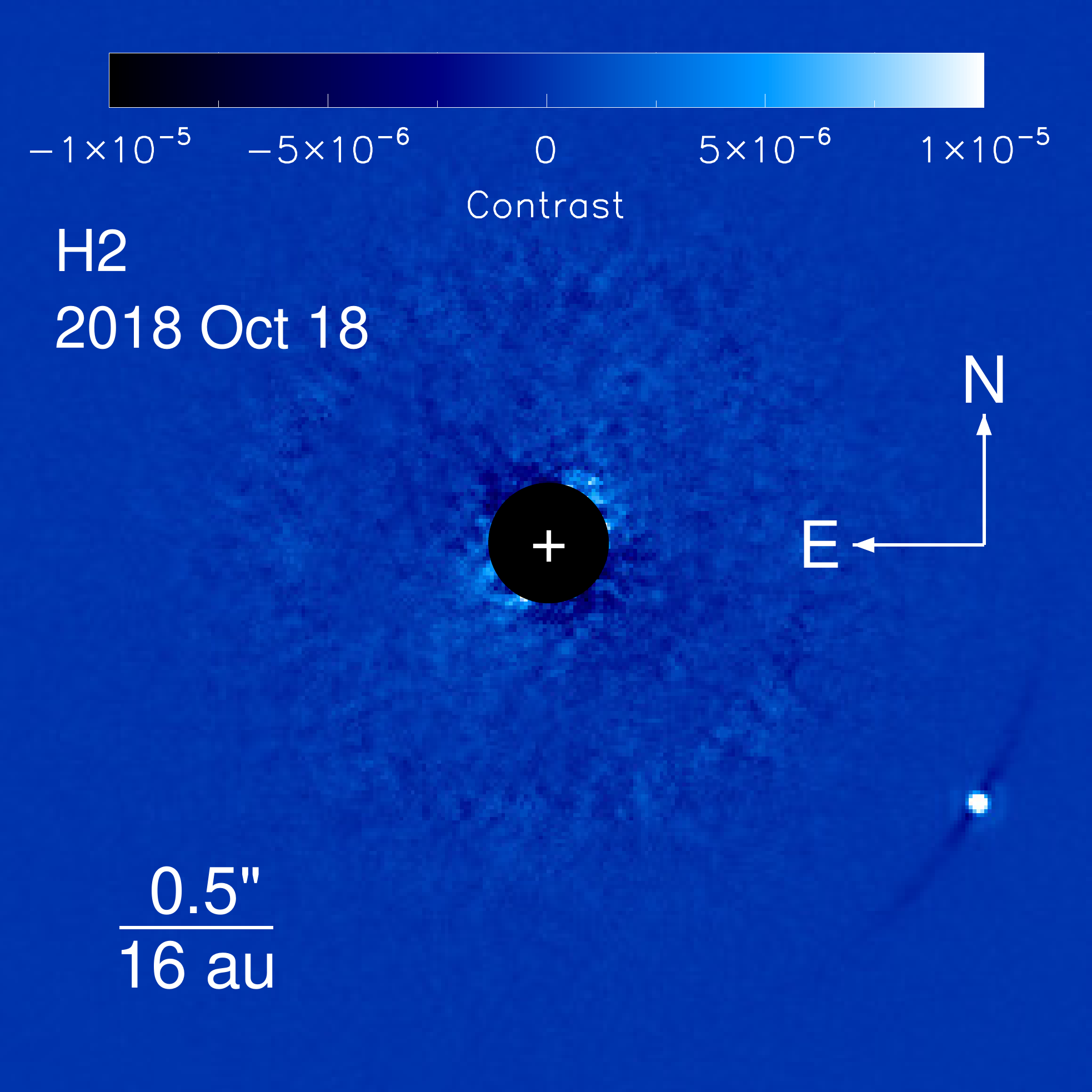} \\
\includegraphics[width=.26\textwidth]{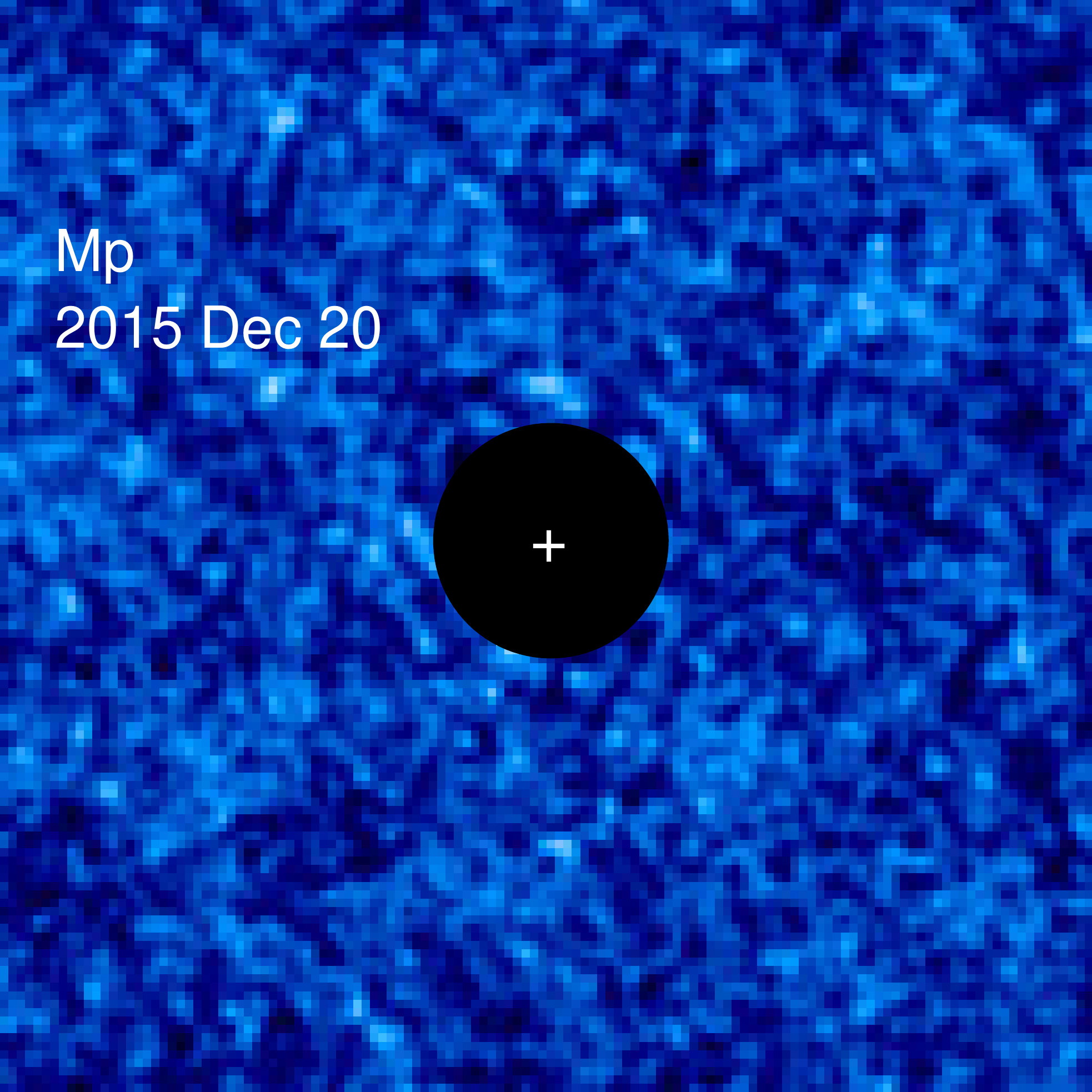} & \includegraphics[width=.26\textwidth]{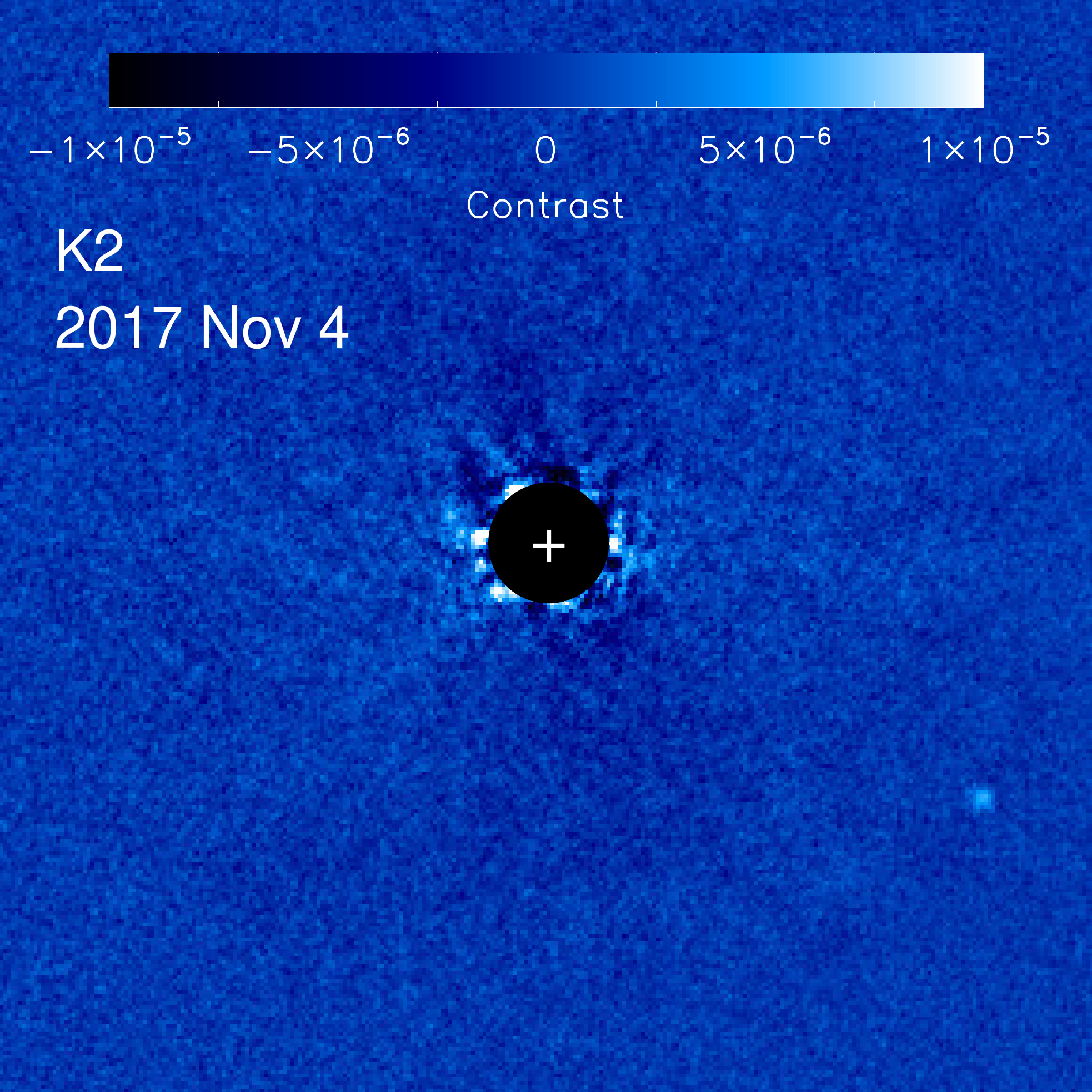} & \includegraphics[width=.26\textwidth]{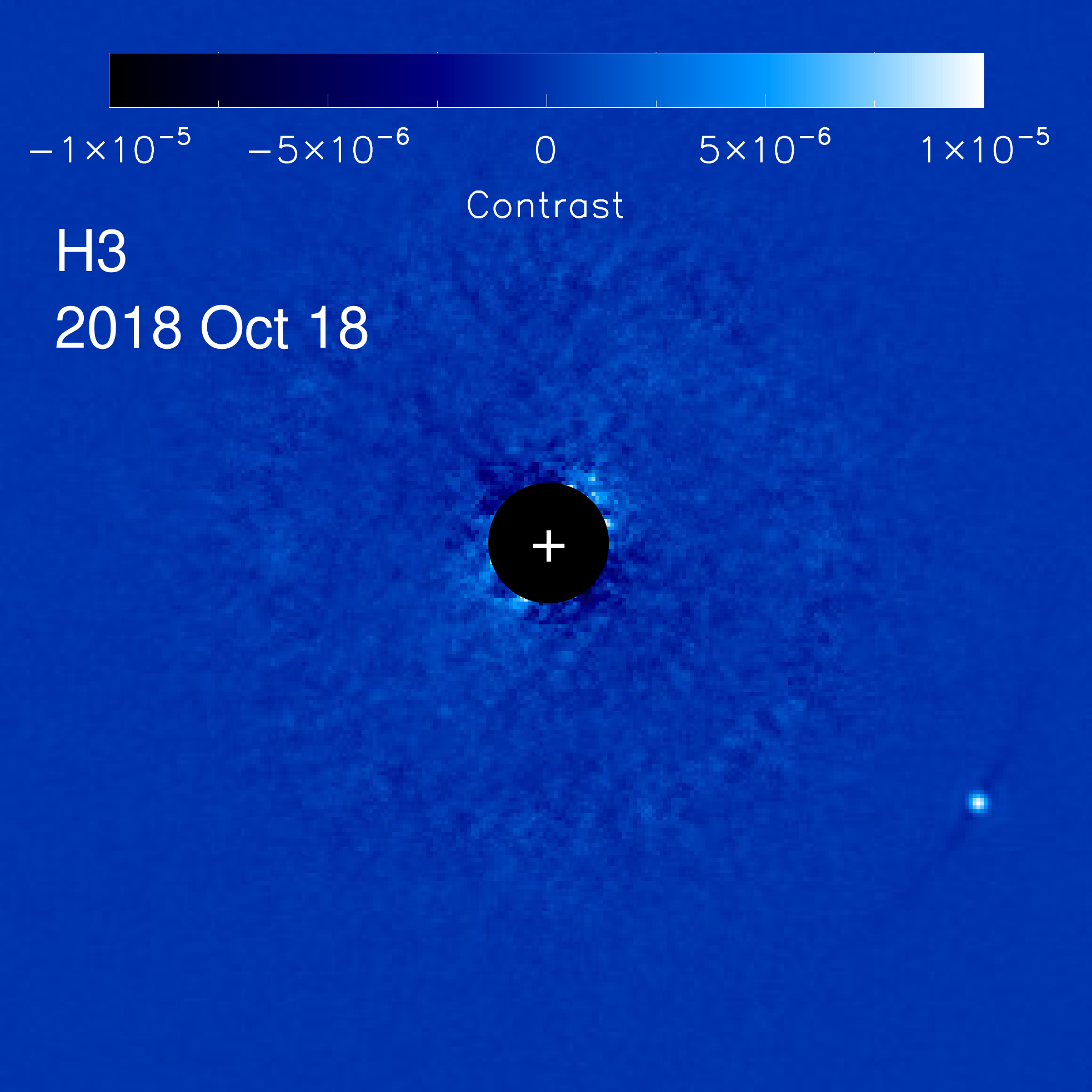} \\
\end{tabular}
\caption{{ADI-processed} images of HD\,19467 obtained with NaCo (\textit{left panels}) and SPHERE (\textit{other panels}). {See text for details.} The central regions of the images {are} numerically masked out to hide bright stellar residuals. The white crosses indicate the location of the star. The brown dwarf companion HD\,19467B is seen in the bottom-right part of all images. The poorer quality of the detection in the $K2$ filter is due to strong methane absorption in the atmosphere of the object.}
\label{fig:images}
\end{figure*}

\section{Observations and data analysis}
\label{sec:data}

\begin{table*}[t]
\caption{Astrometry relative to the star of HD\,19467B.}
\label{tab:astrometry}
\begin{center}
\begin{tabular}{l c c c c c}
\hline\hline
BJD-2\,450\,000 & Filter & $\rho$ & PA & Pixel scale & North correction angle \\
 & & (mas) & ($^{\circ}$) & (mas/pix) & ($^{\circ}$) \\
\hline
8032.3 & $L^{\prime}$ & 1637$\pm$19 & 238.68$\pm$0.47 & 27.20$\pm$0.05 & $-$0.5$\pm$0.1\\
8061.2 & $K1$ & 1636.7$\pm$1.8 & 239.39$\pm$0.13 & 12.267$\pm$0.009 & $-$1.745$\pm$0.053\\
8061.2 & $K2$ & 1634.4$\pm$5.0 & 239.44$\pm$0.21 & 12.263$\pm$0.009 & $-$1.745$\pm$0.053\\
8409.3 & $H2$ & 1631.4$\pm$1.6 & 238.88$\pm$0.12 & 12.255$\pm$0.009 & $-$1.804$\pm$0.043\\
8409.3 & $H3$ & 1631.4$\pm$1.6 & 238.88$\pm$0.12 & 12.251$\pm$0.009 & $-$1.804$\pm$0.043\\
\hline
\end{tabular}
\end{center}
\tablefoot{The astrometric uncertainties {were} derived assuming an error budget including the measurement uncertainties (image post-processing) and the systematic uncertainties (calibration).\\
}
\end{table*}

\subsection{High-contrast imaging}

\subsubsection{SPHERE NIR observations}

HD\,19467 was observed twice with SPHERE in the NIR (Table~\ref{tab:obs}). For the 2017 observation, we only used the {Infra-Red Dual-band Imager and Spectrograph} IRDIS \citep{Dohlen2008a, Vigan2010} in the dual-band imaging mode with the $K12$ filter pair. For the 2018 observation, we used the standard IRDIFS mode, which allows for simultaneous observations with IRDIS with the $H23$ filter pair and the integral field spectrograph IFS \citep{Claudi2008} in the $YJ$ bands.

For both sequences, an apodized pupil Lyot coronagraph \citep{Carbillet2011, Martinez2009} was used. For calibrating the flux of the images, we acquired unsaturated noncoronagraphic images of the star (hereafter reference point-spread function or reference PSF) at the beginning and end of the sequences. To minimize the frame centering uncertainties in the astrometric error budget, the coronagraphic images were recorded with four artificial crosswise replicas of the star \citep{Langlois2013}. Night-time sky background frames were taken and additional daytime calibration performed following the standard procedure at ESO.

\begin{table*}[t]
\caption{Photometry relative to the star of HD\,19467B.}
\label{tab:photometry}
\begin{center}
\begin{tabular}{l c c c c c c c}
\hline\hline
Filter & $\lambda_0$ & $\Delta\lambda$ & $\Delta$mag & {App. mag.} & Abs. mag. & Flux \\
 & ($\mu$m) & ($\mu$m) & (mag) & {(mag)} & (mag) & ($\times$10$^{-16}$ W\,m$^{-2}$\,$\mu$m$^{-1}$) \\
\hline
$H2$ & 1.593 & 0.052 & 11.50$\pm$0.04 & {16.95$\pm$0.05} & 14.42$\pm$0.05 & 1.953$\pm$0.075 \\
$H3$ & 1.667 & 0.054 & 12.43$\pm$0.04 & {17.88$\pm$0.05} & 15.35$\pm$0.05 & 0.734$\pm$0.027 \\
$K1$ & 2.110 & 0.102 & 11.52$\pm$0.07 & {16.92$\pm$0.07} & 14.39$\pm$0.08 & 0.750$\pm$0.049 \\
$K2$ & 2.251 & 0.109 & 13.12$\pm$0.08 & {18.52$\pm$0.08} & 15.99$\pm$0.09 & 0.133$\pm$0.010 \\
$L^{\prime}$ & 3.800 & 0.620 & 10.16$\pm$0.14 & {15.46$\pm$0.17} & 12.93$\pm$0.17 & 0.288$\pm$0.037 \\
$M^{\prime}$ & 4.780 & 0.590 & $>$9.0 & {>14.0} & $>$11.5 &$<$0.332 \\
\hline
\end{tabular}
\end{center}
\tablefoot{The photometric uncertainties {were} derived assuming an error budget including the measurement uncertainties (image post-processing) and the systematic uncertainties (temporal variability of the reference PSF and of the sequence). \\
}
\end{table*}

The data were reduced with the SPHERE Data Center pipeline \citep{Delorme2017b}, which uses the Data Reduction and Handling software \citep[v0.15.0,][]{Pavlov2008} and custom routines. It {corrected} for the cosmetics and instrument distortion, {registered} the frames, and {normalized} their flux. For the IFS data \citep{Mesa2015}, it {also calibrated them spectrally} and {extracted} the image cubes. Subsequently, we sorted the frames using visual inspection to reject poor-quality frames (adaptive optics open loops, low-wind effect) and an automatic criterion to reject frames with low flux in the coronagraphic spot (semi-transparent mask). After this step, we were left with 91\% and 80\% of the frames for the 2017 and 2018 IRDIS data, {respectively.} We kept all the IFS frames. Finally, the data were analyzed with a consortium image processing pipeline \citep{Galicher2018} and with the ANgular DiffeRential Optimal Method Exoplanet Detection Algorithm \citep[ANDROMEDA,][]{Mugnier2009, Cantalloube2015}. Figure~\ref{fig:images} shows the images processed with angular differential imaging \citep[ADI,][]{Marois2006a} with the Template Locally Optimized Combination of Images algorithm \citep[TLOCI,][]{Marois2014} provided in the consortium image processing pipeline.

\subsubsection{NaCo thermal IR observations}

We also obtained high-contrast imaging data in the $L{^\prime}$ band with NaCo (Table~\ref{tab:obs}, program ID: 0100.C-0234, PI. Maire). These observations were performed without a coronagraph, in pupil-tracking mode to take advantage of ADI, and in dithering mode to sample the sky background while maximizing the observing efficiency. The data were reduced using a custom reduction pipeline \citep[cosmetics, frame registering, and frame binning by 380,][]{Mueller2018} and analyzed with the TLOCI and ANDROMEDA high-contrast imaging algorithms. Figure~\ref{fig:images} shows the image processed with TLOCI and smoothed with a Gaussian {with a width of} 2 pixels.

Finally, we analyzed archival NaCo data taken with the $M{^\prime}$ filter (program ID: 096.C-0602, PI. Buenzli). These observations were acquired following the same strategy as for the NaCo/$L{^\prime}$ data. We {processed the data with a custom data reduction and analysis pipeline \citep{Cheetham2019} by} applying a frame binning of 250{. Figure~\ref{fig:images} shows the processed image smoothed with a Gaussian {with a width of} 2 pixels. We find} no significant signal at the expected location of the companion. We {estimated} an upper limit for the companion contrast of 9.0~mag based on the 3$\sigma$ detection limit measured at the expected separation. We verified that fake companions injected into the raw data with this contrast are recovered in the processed image.

\subsubsection{Photometry and astrometry}

\begin{figure*}[t]
\centering
\includegraphics[width=.44\textwidth]{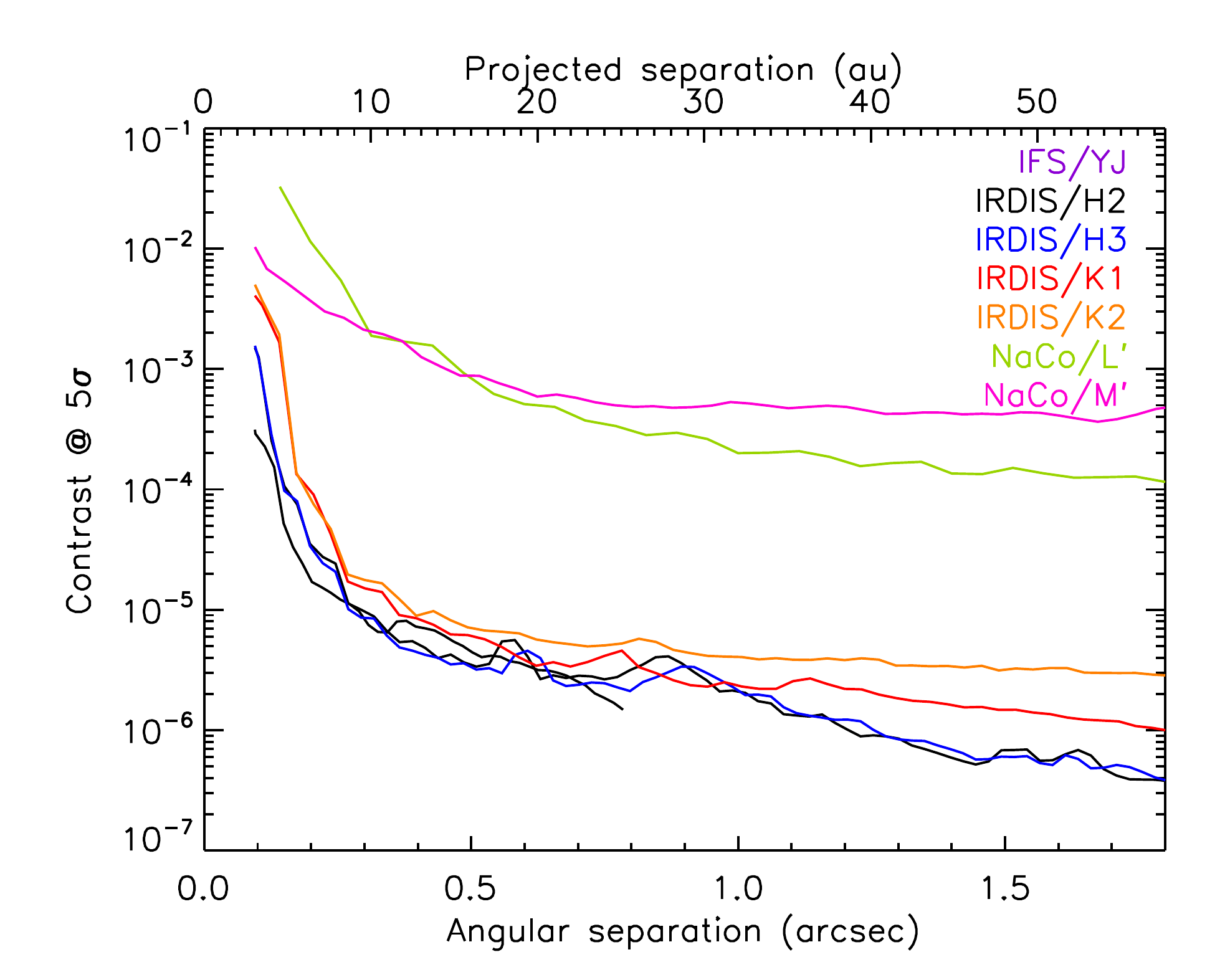}
\includegraphics[width=.44\textwidth]{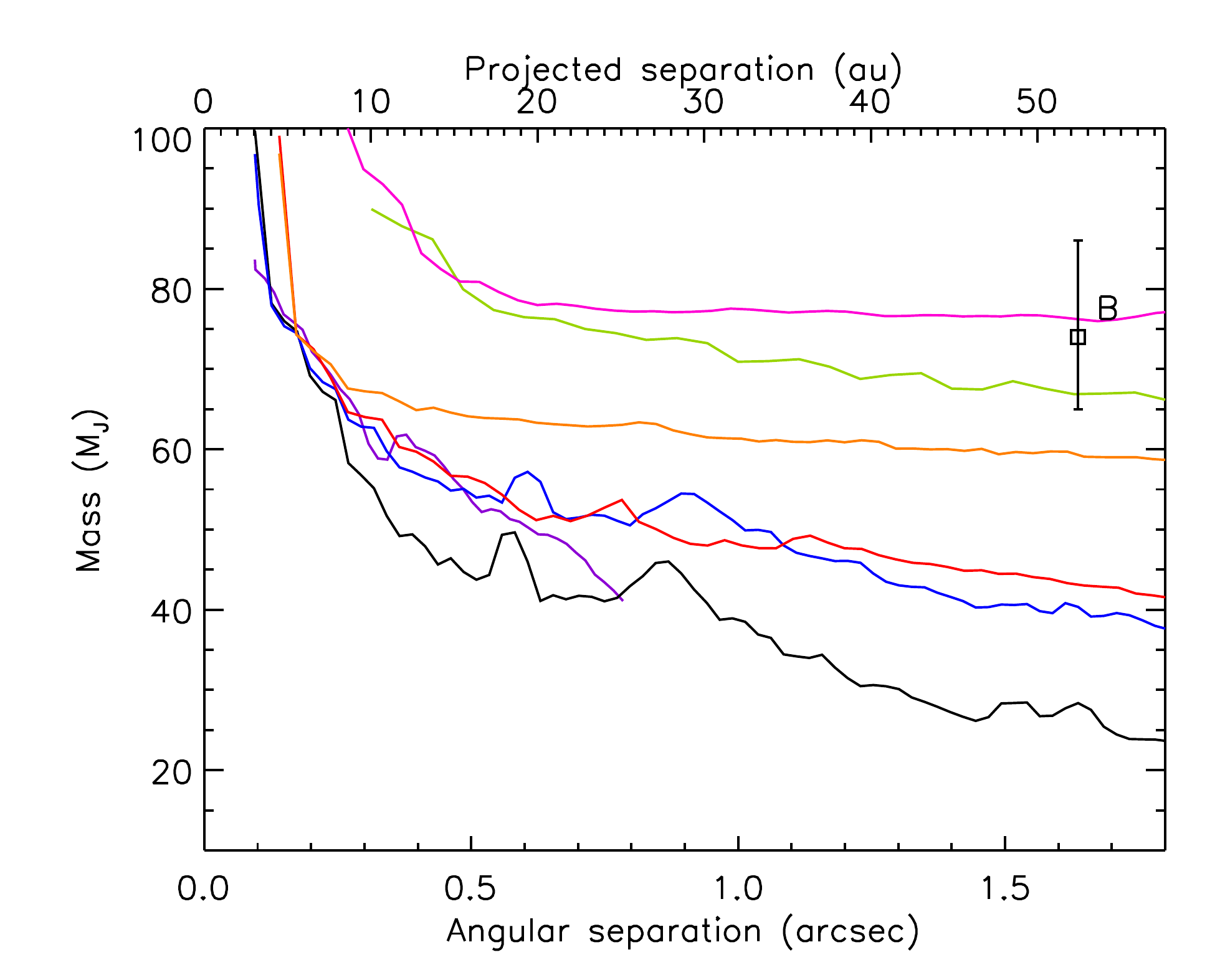}
\caption{{5$\sigma$ detection limits {expressed} in contrast with respect to the star (\textit{left}) and in companion mass (\textit{right}) for the set of instruments (SPHERE IRDIS and IFS, NaCo) and filters (colored curves). In the right panel, we also indicate the location of HD 19467B assuming the mass inferred from the orbital fit (Sect.~\ref{sec:orbit}).} The mass limits achieved at low masses with the $H3$ and $K2$ filters are degraded {compared to those with} the $H2$ and $K1$ filters because these filters match methane absorption bands (see also Fig.~\ref{fig:images}).}
\label{fig:detlimits}
\end{figure*}

For the high-contrast imaging data where the companion could be recovered, the astrometry and photometry listed in Tables~\ref{tab:astrometry} and \ref{tab:photometry} was measured in the TLOCI images using the fit of a model of {the planet image} built from the reference PSF and processed with TLOCI \citep{Galicher2018}. The position and flux of the model of {the planet image} was optimized to minimize the image residuals within a circular region of radius 1.5 full width at half maximum centered on the measured planet location. The astrometry was calibrated following the methods in \citet{Maire2016b} for the SPHERE data and in \citet{Cheetham2019} for the NaCo $L{^\prime}$ data. We compared the TLOCI photometry and astrometry with the ANDROMEDA results and found the values to agree within the TLOCI measurement uncertainties (results not shown). We {use} the TLOCI measurements for the orbital and spectral analyses in the {next} sections, because TLOCI {has been} tested and validated on a larger number of SPHERE datasets to retrieve the astrometry and photometry of detected companions \citep{Galicher2018}. {We note that} the position angle measured with NaCo is smaller by $\sim$0.7${^\circ}$ ($\sim$1.4$\sigma$) {than} the position angle measured with SPHERE in a dataset obtained just a month {after} the NaCo observation.

The absolute magnitudes of the companion at wavelengths shorter than 3~$\mu$m were computed assuming for the stellar magnitudes the 2MASS values \citep{Cutri2003}. The absolute magnitudes of the companion at wavelengths longer than 3~$\mu$m were computed {using} stellar magnitudes in the $L{^\prime}$ band of 5.3$\pm$0.1~mag and in the $M{^\prime}$ band of 5.1$\pm$0.1~mag {estimated} by interpolating the WISE W1 and W2 magnitudes \citep{2013yCat.2328....0C}.

\subsubsection{Detection limits}

The SPHERE and NaCo/$L^{\prime}$ detection limits were computed using the TLOCI-ADI reductions. {The NaCo/$M^{\prime}$ detection limit was computed using the principal component analysis (PCA) algorithm described in \citet{Cheetham2019} with 54 modes (30\% of available modes). The detection limits shown in Fig.~\ref{fig:detlimits}} account for the small sample statistics correction \citep{Mawet2014} and, for the SPHERE datasets, for the coronagraph transmissions \citep{Boccaletti2018}. The conversion from contrast to the star into companion mass {was} computed using a system age of 8~Gyr and ``hot-start'' atmospheric and evolutionary models of \citet{Baraffe2015, Baraffe2003} for the SPHERE data and of \citet{Allard2012} and \citet{Baraffe2003} for the NaCo data. Given the age of the star, we do not expect significant {variations in} the luminosity-mass relation according to the initial conditions assumed in the evolutionary model \citep{Marley2007}. The SPHERE data provide deeper constraints in contrasts and companion masses. Contrasts as deep as 10$^{-5}$ are achieved beyond 0.3$''$ (10~au), which exclude additional companions more massive than $\sim$55~$M_J$. Additional companions more massive than 35~$M_J$ are excluded beyond 1.1$''$ (35~au).

We also show the mass and angular separation of HD\,19467B in the right panel of Fig.~\ref{fig:detlimits} assuming the mass derived in Sect.~\ref{sec:orbit}. If the companion is more massive than $\sim$77~$M_J$, it should have been detected in our NaCo $M^{\prime}$ data.

\subsection{Radial velocity data}
\label{sec:rvdata}

\subsubsection{HIRES}

\begin{figure}[t]
\centering
\includegraphics[width=.4\textwidth]{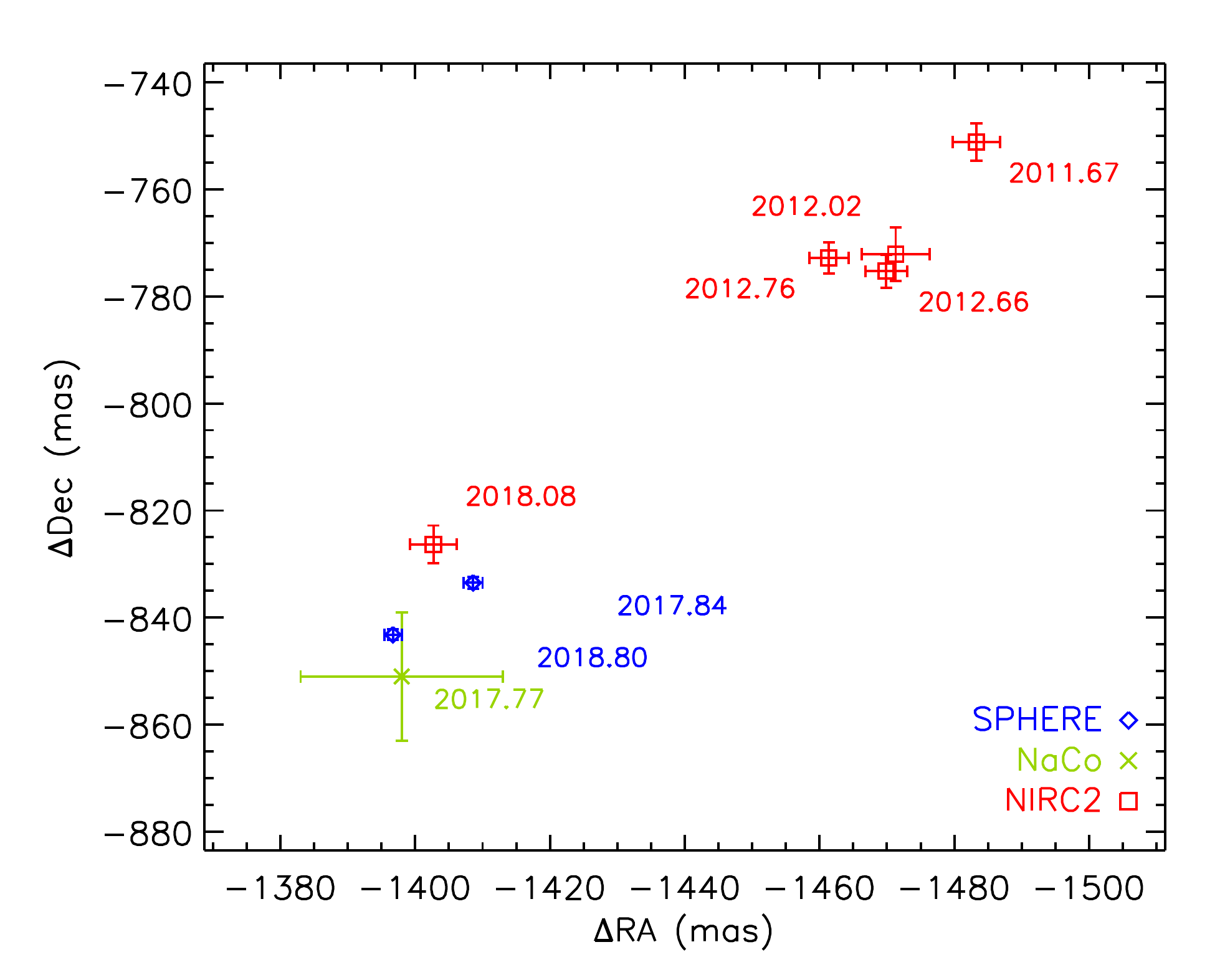}
\caption{{Compilation of the SPHERE, Keck, and NaCo imaging measurements of the position of HD\,19467B relative to the host star in the RA-Dec plane. The orbital motion in the clockwise direction between the first Keck epochs and the SPHERE and NaCo epochs can be seen (see also Fig.~\ref{fig:seppa_time}). The Keck and NaCo measurements {are} not recalibrated (see text).}}
\label{fig:astrometry}
\end{figure}

\begin{figure*}[t]
\centering
\includegraphics[width=.4\textwidth]{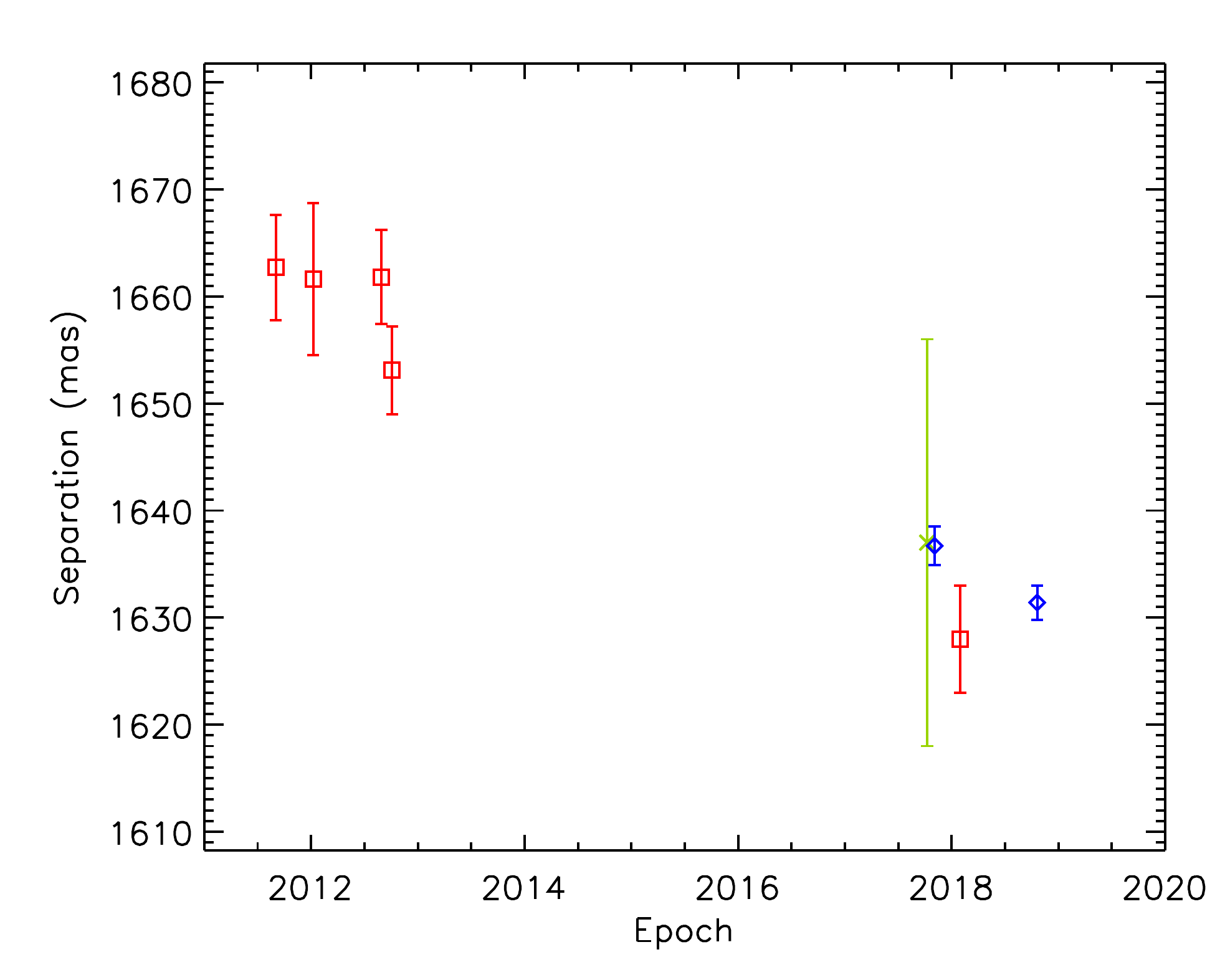}
\includegraphics[width=.4\textwidth]{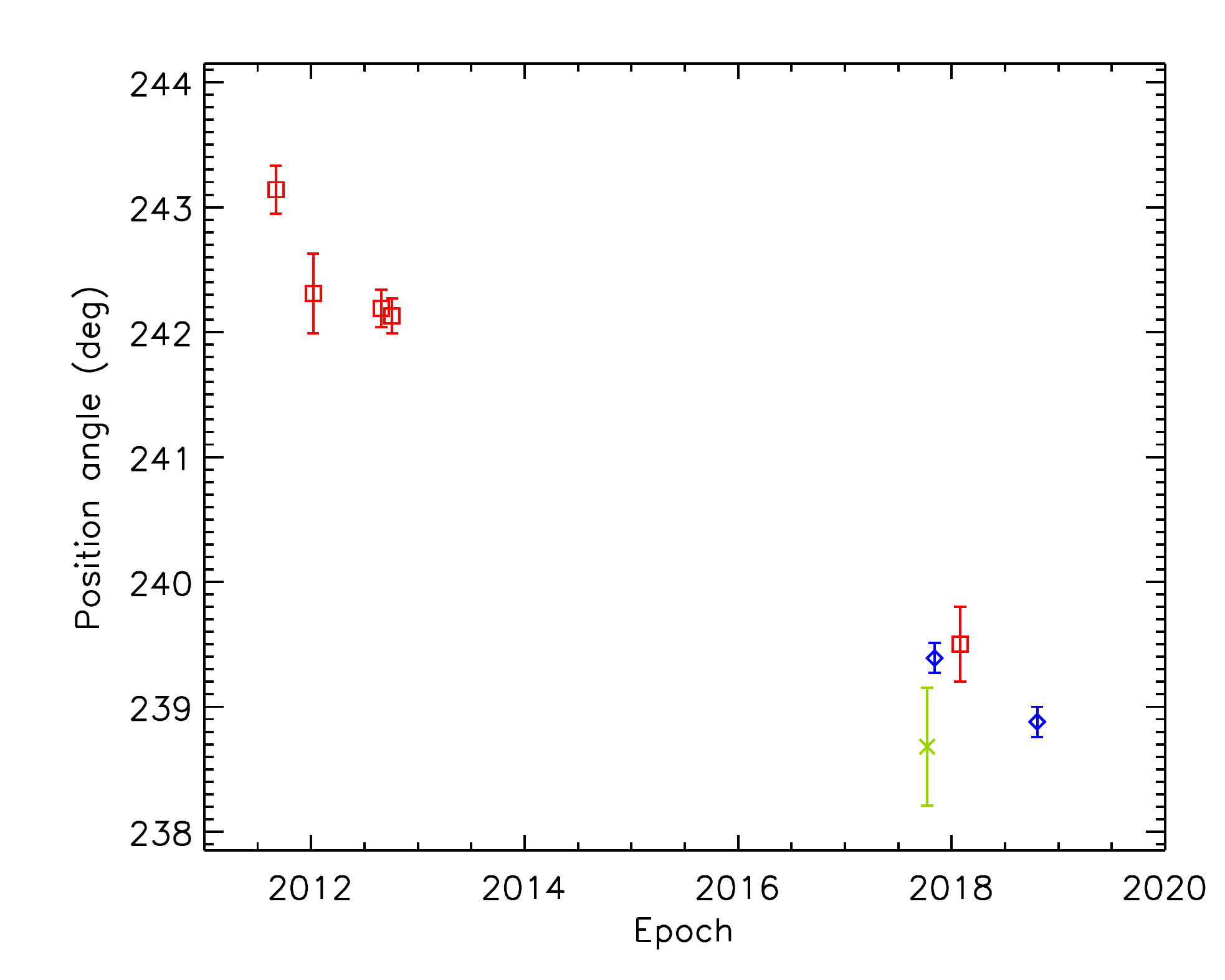}
\caption{{Temporal evolution of the relative separation (\textit{left}) and of the position angle (\textit{right}) of HD\,19467B measured with Keck, SPHERE, and NaCo imaging. The Keck and NaCo measurements {are} not recalibrated (see text).}}
\label{fig:seppa_time}
\end{figure*}

We analyzed the HIRES RV data presented in \citet{Butler2017} following the methods described in \citet{TalOr2019}, which {corrected} in particular for small systematics due to an instrument upgrade in August 2004 (BJD epoch 2453236). {Compared} to the data presented in \citet{Crepp2014}, the data baseline is increased by $\sim$11 months. The data {exhibit} a {decreasing linear} trend (see Appendix~\ref{sec:orbfit_imrv} and Sect.~\ref{sec:orbfit}). We fit a linear trend to the data using the Affine-Invariant Markov Chain Monte Carlo (MCMC) Ensemble Sampler \citep{Goodman2010} provided in the package \texttt{emcee} \citep{ForemanMackey2013} to estimate an acceleration of $-$1.43$\pm$0.04 m\,s$^{-1}$\,yr$^{-1}$ (68\%){. This} is included in the {uncertainties of the value} of $-$1.37$\pm$0.09 m\,s$^{-1}$\,yr$^{-1}$ in \citet{Crepp2014}.

\subsubsection{HARPS}

We also analyzed archival RV data from HARPS taken from 2003 to 2017 (program IDs: 072.C-0488 PI. Mayor, 183.C-0972: PI. Udry, 188.C-0265 PI. Melendez, 192.C-0852 PI. Udry, and 0100.D-0444 PI. Lorenzo de Oliveira). The methods and the data are presented in \citet{Trifonov2020}. Some of the HARPS data were taken after an instrument upgrade in June 2015 \citep[][BJD epoch 2457177]{LoCurto2015}{. They display} an offset {toward larger RVs} with the pre-upgrade data. We note four outlier measurements with small error bars close to BJD epoch 2456500 with measured RVs around $-$20~m\,s$^{-1}$, whereas the other measurements taken around the same epoch show values around $-$12~m\,s$^{-1}$. We excluded these outlier measurements in our analyses (see Appendix~\ref{sec:orbfit_imrv} and Sect.~\ref{sec:orbfit}). We used the MCMC approach applied to the HIRES data to fit a linear trend to the HARPS data. We {derived} an acceleration of $-$1.46$\pm$0.02 m\,s$^{-1}$\,yr$^{-1}$ (68\%), which is within {the uncertainties of} our HIRES acceleration estimate, and a post-upgrade offset of 12.8$\pm$0.3~m\,s$^{-1}$ (68\%).

The systematic offset between the HIRES and HARPS measurements is related to the different zero-points of the instruments. From an MCMC linear fit to the HIRES and HARPS data, we {find} that the HIRES measurements are shifted by $-$4.0$\pm$0.3~m\,s$^{-1}$ (68\%) {compared} to the HARPS data. We used this value as initial guess in the orbital fit (Sect.~\ref{sec:orbit}). For the orbital fit, we also added quadratically to the HIRES and HARPS measurement uncertainties jitter terms with initial guesses of 3.40~m\,s$^{-1}$ and 1.44~m\,s$^{-1}$, respectively. The values were estimated using the statistics of the dispersion of each set of measurements with respect to the predicted values from a robust linear fit. We also verified that they are close to the minimum $\chi^2$ values using the individual RV likelihood terms (Sect.~\ref{sec:orbfit}). \citet{Crepp2014} {note} that given the log\,$R^{\prime}_{HK}$ and $B - V$ color of the star, the expected level of astrophysical noise due to the stellar activity should be 2.4$\pm$0.4~m\,s$^{-1}$.

\section{Orbital analysis}
\label{sec:orbit}

\subsection{Orbital motion}
\label{sec:astrometry}

The astrometry of the brown dwarf is provided in Table~\ref{tab:astrometry}. The data are represented in the RA-Dec plane in Fig.~\ref{fig:astrometry} with the NIRC2 measurements reported by \citet{Crepp2014} and \citet{Bowler2020}. We used a weighted average of two NIRC2 measurements obtained on 2012 January 7 by \citet{Crepp2014}. The SPHERE data confirm the orbital motion of HD\,19467B in the clockwise direction noted by \citet{Crepp2014}, which is inconsistent with the motion expected if it were a stationary background object \citep[HD\,19467 is a high-proper motion star with $\mu_{\mathrm{\alpha}}$~=~$-$8.685$\pm$0.070~mas/yr, $\mu_{\mathrm{\delta}}$~=~$-$260.566$\pm$0.077~mas/yr,][]{GaiaCollaboration2018}. Using the SPHERE data, we {estimated} an orbital motion of 16$\pm$3~mas\,yr$^{-1}$, which is more precise but still within the uncertainties of the estimate of 22$\pm$6~mas\,yr$^{-1}$ in \citet{Crepp2014}. Figure~\ref{fig:seppa_time} shows the temporal evolution of the separation and position angle with time. In $\sim$6.5~yr, the companion got closer to the star by $\sim$28$\pm$5~mas and its position angle decreased by $\sim$3.4$\pm$0.3$^{\circ}$.

When comparing the trends observed for the NIRC2 data and the SPHERE data separately, we note systematic offsets. In particular, the separation measured with NIRC2 in January 2018 (1628$\pm$5~mas) is {shorter} than the separation measured with SPHERE in November 2017 (1636.7$\pm$1.8~mas, Table~\ref{tab:astrometry}). The trends measured for the separation and the position angle from each data series separately agree (we find $-$5.46$\pm$0.96~mas/yr and $-$0.536$\pm$0.053$^{\circ}$/yr using the NIRC2 data){. Then, we used MCMC linear fits to the SPHERE, Keck, and NaCo data to assess potential systematics between the data series {assuming} the SPHERE data series as reference. The fits {confirm} that the separations measured with Keck are {shorter} by a factor of 0.9956$^{+0.0030}_{-0.0031}$ at 68\% and that the position angle mesured with NaCo is offset ($-$0.75$^{+0.48}_{-0.47}$~deg at 68\%). The position angles mesured with Keck may be offset (0.21$\pm$0.30\,$^{\circ}$ at 68\%). For the Bayesian rejection fit to the imaging data only (Appendix~\ref{sec:orbfit_im}), we corrected the NIRC2 and NaCo position angles as well as the NIRC2 separations for the systematics measured above. In the MCMC orbital fits (Sect.~\ref{sec:orbfit} and Appendix~\ref{sec:orbfit_imrv}), we included additional free parameters to account for the systematics. Linear fits to the SPHERE data and the recalibrated NIRC2 and NaCo data give variation rates {of $-$5.48$\pm$0.44~mas/yr for the separation and of $-$0.499$\pm$0.022$^{\circ}$/yr for the position angle}.

When fitting the NIRC2 data with linear fits robust to outliers, we {find} a minimum reduced $\chi^2$ larger than 1 for the position angles{. This implies} that some measured uncertainties are somewhat underestimated. In particular, the position angle measured on 2011 August 30 by \citet{Crepp2014} is deviant from the fit by more than 1$\sigma$. We accordingly increased the uncertainty on this measurement for the orbital fits.

\begin{figure*}[t]
\centering
\includegraphics[width=.99\textwidth, trim = 4mm 4mm 7mm 6mm, clip]{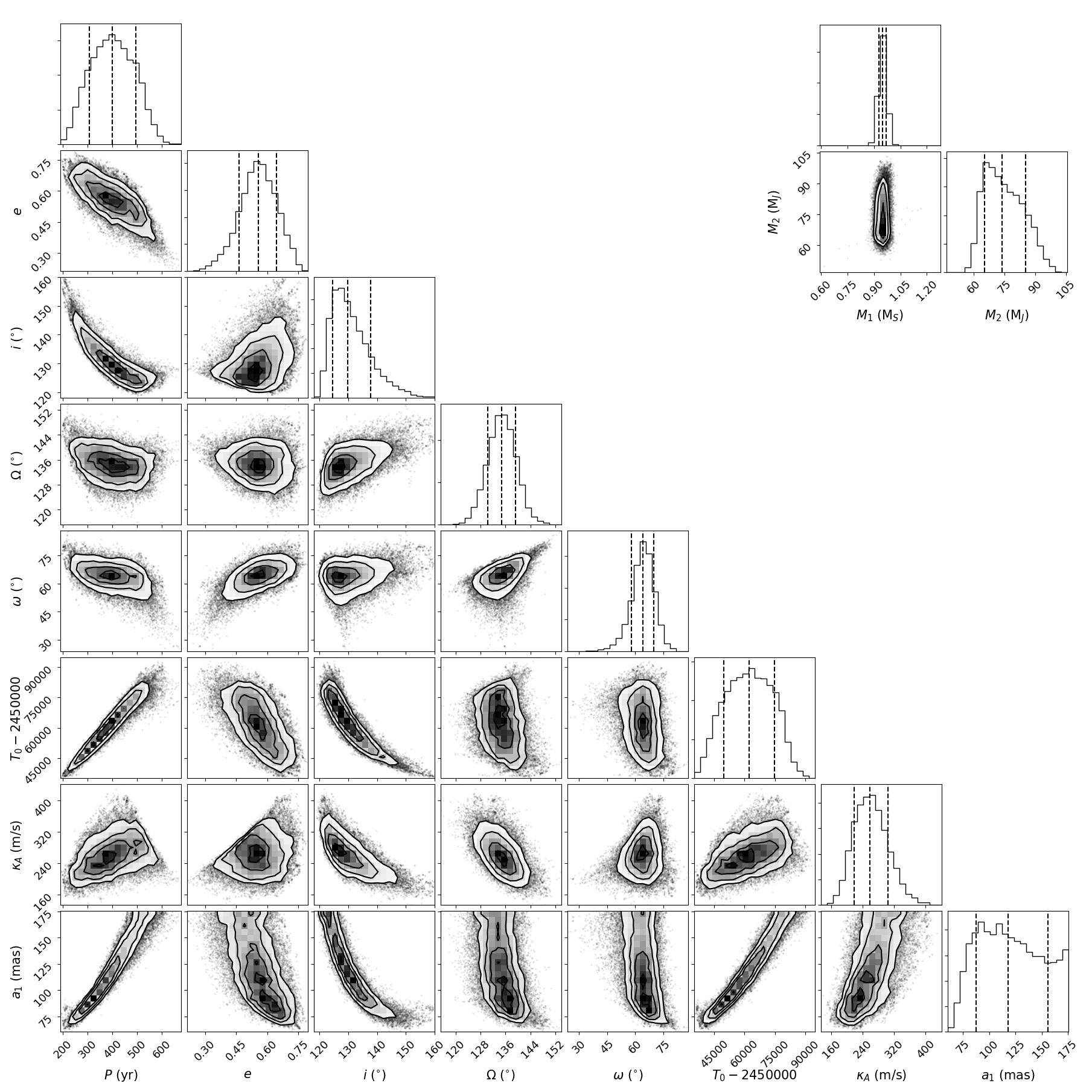}
\caption{{MCMC samples from the posteriors of the orbital parameters (\textit{left}) and of the masses of HD\,19467 A and B (\textit{top right}) obtained by fitting the imaging, RV, and astrometric data.} The diagrams displayed on the diagonal from top left to lower right represent the 1D histogram distributions for the individual elements. The off-diagonal diagrams show the correlations between pairs of orbital elements. In the histograms, the dashed vertical lines indicate the 16\%, 50\%, and 84\% quantiles.}
\label{fig:orbit_imrvpm_corner}
\end{figure*}

\subsection{Minimum dynamical mass of HD\,19467B}

We used the approach in \citet{Torres1999}, \citet{Liu2002}, and \citet{Bowler2016} to assess the minimum dynamical mass of HD\,19467B using our RV acceleration estimated from the HARPS data and the \textit{Gaia} parallax. We assumed Gaussian distributions for the RV acceleration, system distance, and projected separation of the companion. We assumed for the projected separation of the companion a value of 1660$\pm$7~mas using a weighted average of the measurements in \citet{Crepp2014}. We {derived} a value of 60.0$\pm$1.7~$M_J$ (68\% interval), which points toward larger masses than the 51.9$^{+3.6}_{-4.3}$~$M_J$ value in \citet{Crepp2014}. The larger masses that we {derived} are due to the use of the \textit{Gaia} parallax (which is smaller than the \textsc{Hipparcos} parallax) and of our measured acceleration. 

\subsection{Determination of the orbital parameters}
\label{sec:orbfit}

We performed a simultaneous fit of the SPHERE and NIRC2 astrometry with the HIRES and HARPS RV measurements. We also included in the fit astrometric constraints from proper motion measurements of the star from \textsc{Hipparcos} and \textit{Gaia} as done, for example, by \citet{Calissendorff2018} and \citet{Brandt2019b} for the orbital analysis of the T-type brown dwarf GJ\,758B. No clear proper motion anomaly of the \textsc{Hipparcos} and \textit{Gaia} measurements with respect to the long-term proper motion is seen for HD\,19467B in the catalog of \citet{Kervella2019}: pmra\_g\_hg = $-$0.165$\pm$0.088~mas\,yr$^{-1}$, pmdec\_g\_hg = 0.016$\pm$0.094~mas\,yr$^{-1}$, pmra\_h\_hg = 0.640$\pm$0.630~mas\,yr$^{-1}$, pmdec\_h\_hg = $-$0.170$\pm$0.710~mas\,yr$^{-1}$. {The \textsc{Hipparcos} reduction \citep{vanLeeuwen2007} is} well behaved with a goodness-of-fit parameter below 5. {From \citet{Kervella2019}, the} Gaia DR2 record \citep{GaiaCollaboration2018} is well behaved with a renormalized unit weight error below 1.4 \citep{Lindegren2018}. {The proper motion anomaly measurements of \citet{Kervella2019} agree well within the uncertainties with the measurements in \citet{Brandt2018, Brandt2019a} but are slighly better constrained: pmra\_g\_hg = $-$0.164$\pm$0.109~mas\,yr$^{-1}$, pmdec\_g\_hg = 0.030$\pm$0.120~mas\,yr$^{-1}$, pmra\_h\_hg = 0.608$\pm$0.734~mas\,yr$^{-1}$, pmdec\_h\_hg = 0.355$\pm$0.791~mas\,yr$^{-1}$.}

{We fit the measurements simultaneously} using the Parallel-Tempered MCMC algorithm provided in the package \texttt{emcee} \citep{ForemanMackey2013}, which is based on the algorithm described by \citet{Earl2005}. Our implementation follows \citet{Brandt2019b} by and large \citep[see also][]{Maire2020}. We sampled the parameter space of our {17-parameter model} assuming 20 temperatures for the chains and 100 walkers. The first 8 parameters are the semi-major axis $a$, the eccentricity $e$ and {the} argument of {the} periastron passage $\omega$ \citep[parameterized as $\sqrt{e}$\,cos\,$\omega$ and $\sqrt{e}$\,sin\,$\omega$ to mitigate the Lucy-Sweeney bias {toward} low eccentricities,][]{Lucy1971}, the inclination $i$, the longitude of {the} ascending node $\Omega$, the time at {the} periastron passage $T_0$, the RV semi-amplitude of the star $\kappa_A$, and the systemic velocity $\gamma$. We present the results for $\Omega$ and $\omega$ as relative to the companion. The systemic velocity was fixed to zero because HIRES cannot measure it and we subtracted it from the HARPS data {to combine} both datasets.

The initial state of the sampler was set assuming uniform priors in log\,$a$, $\sqrt{e}$\,cos\,$\omega$, $\sqrt{e}$\,sin\,$\omega$, $\Omega$, $T_0$, and $\kappa_A$, as well as a sin\,$i$ prior for $i$. The width of the priors were selected from the results of a preliminary fit to the data. First, we fit the imaging data only (Appendix~\ref{sec:orbfit_im}) to estimate first ranges for the period (log\,$P$=5.0--5.3, with $P$ expressed in days), the eccentricity ($e$=0--0.7), and the inclination ($i$=110--150$^{\circ}$). Then, we employed a least-square Monte Carlo approach \citep{Maire2015, Schlieder2016} to fit the imaging and RV data simultaneously and derive a first range for the RV semi-amplitude (0.08--0.23~km\,s$^{-1}$){. The resulting parameter distributions did not} show multimodality. We present the results of an MCMC fit to the imaging and RV data in Appendix~\ref{sec:orbfit_imrv}.

\begin{figure*}[t]
\centering
\includegraphics[trim = 7mm 0mm 14mm 0mm, clip, height=.28\textwidth]{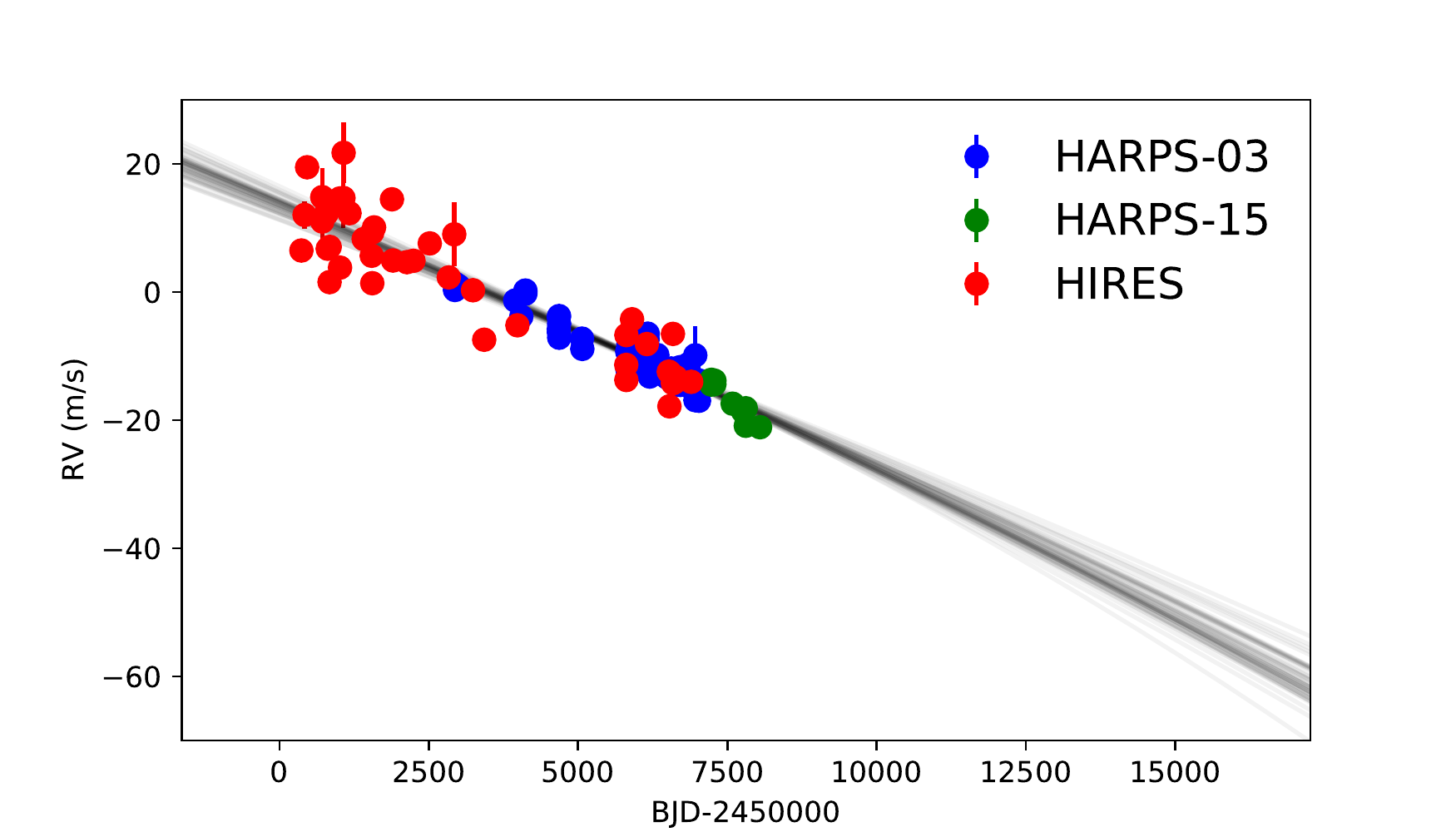}
\includegraphics[trim = 0mm 0mm 10mm 0mm, clip, height=.28\textwidth]{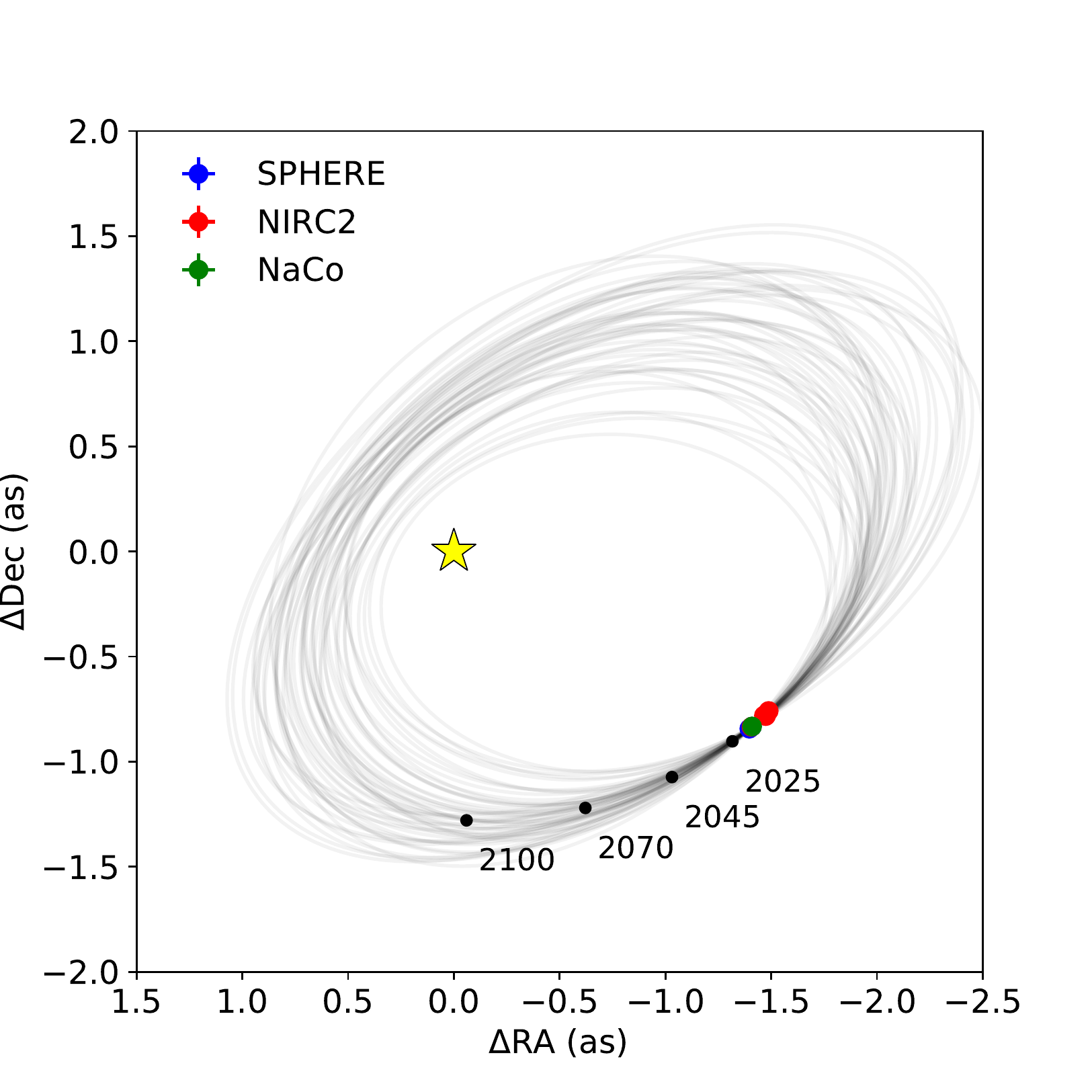}
\includegraphics[trim = 0mm 0mm 7mm 0mm, clip, height=.28\textwidth]{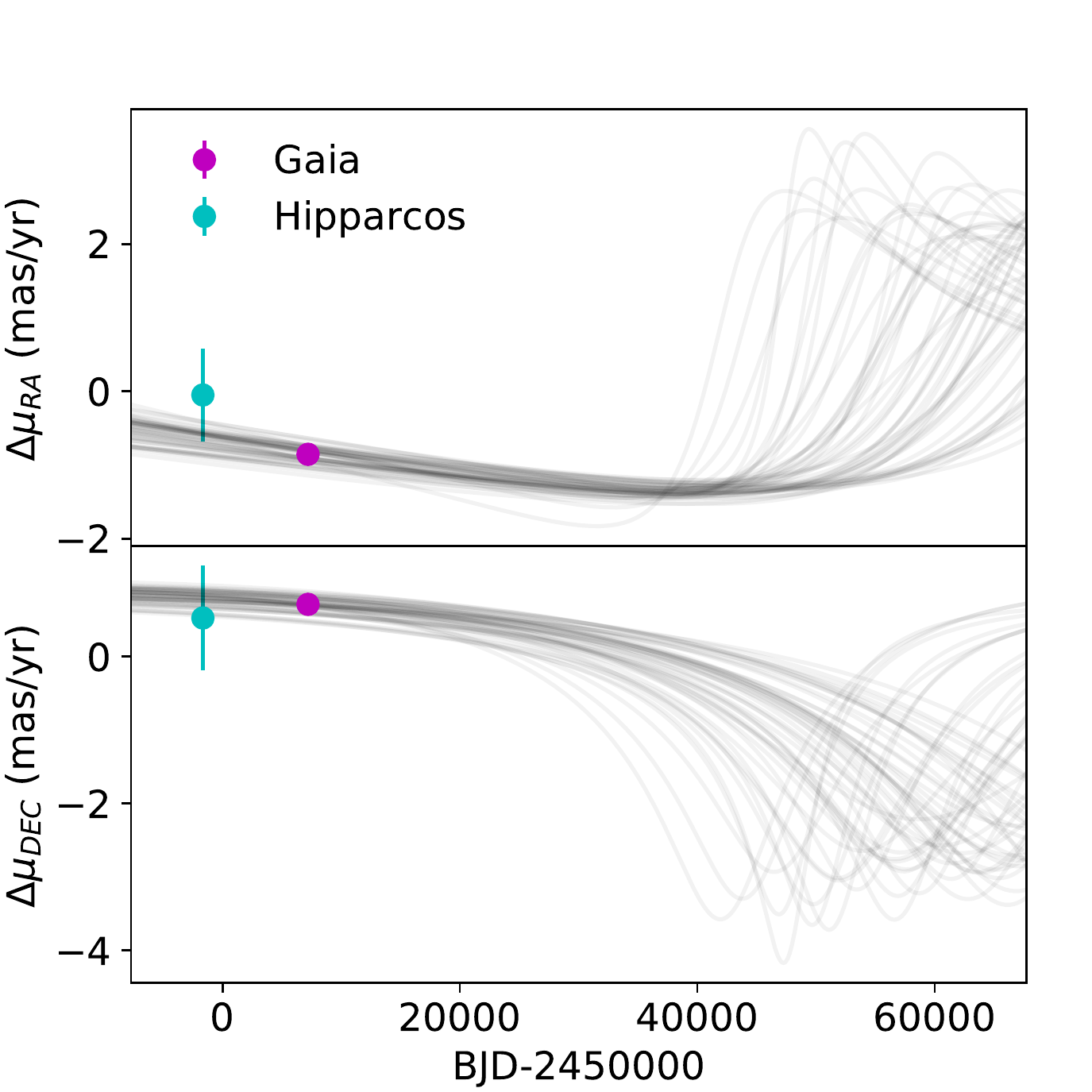}
\caption{{Sample of 50 model orbits (gray curves) fitted on the HD\,19467B data (colored points) from RV (\textit{left}), imaging (\textit{middle}), and astrometry (\textit{right}).} In the \textit{middle} panel, the yellow star marks the location of the star and the black dots show the median predicted position for a few epochs in the future.}
\label{fig:orbit_imrvpm_radec}
\end{figure*}

\begin{table}[t]
\caption{{Orbital parameters and dynamical mass of HD\,19467B.}}
\label{tab:orbparams_rvimpm}
\begin{center}
\begin{tabular}{l c c c}
\hline\hline
Parameter & Unit & Median $\pm$ 1$\sigma$ & Best fit \\
\hline
\multicolumn{4}{c}{Fitted parameters} \\
\hline
Semi-major axis $a$ & mas & 1699$^{+269}_{-277}$ & 1416 \\[3pt]
$\sqrt{e}$\,cos\,$\omega$ & & $-$0.32$\pm$0.06 & $-$0.34 \\[3pt]
$\sqrt{e}$\,sin\,$\omega$ & & $-$0.67$^{+0.09}_{-0.07}$ & $-$0.72\\[3pt]
Inclination $i$ & $^{\circ}$ & 129.8$^{+8.1}_{-5.1}$ & 137.2 \\[3pt]
PA of asc. node $\Omega$ & $^{\circ}$ & 134.8$\pm$4.5 & 134.1 \\[3pt]
Time periastron $T_0$ & BJD & 2512264$^{+12428}_{-12637}$ & 2498964 \\[3pt]
RV semi-ampl. $\kappa_A$ & m\,s$^{-1}$ & 259$^{+46}_{-41}$ & 245 \\[3pt]
Parallax $\pi$ & mas & 31.22$\pm$0.12 & 31.25 \\[1.5pt]
SMA primary $a_1$ & mas & 118$^{+38}_{-30}$ & 89 \\[3pt]
RV offset ZP$_{\rm{HARPS}}$ & m\,s$^{-1}$ & 12.8$\pm$0.7 & 13.1\\[3pt]
RV offset ZP$_{\rm{HIRES}}$ & m\,s$^{-1}$ & $-$4.0$\pm$0.9 & $-$3.7\\[3pt]
RV jitter $\sigma_{\rm{HARPS}}$ & m\,s$^{-1}$ & 1.49$^{+0.18}_{-0.15}$ & 1.39\\[3pt]
RV jitter $\sigma_{\rm{HIRES}}$ & m\,s$^{-1}$ & 3.9$^{+0.6}_{-0.5}$ & 3.9\\[3pt]
Sep. scaling $f\rho_{NIRC2}$ & & 0.9955$^{+0.0034}_{-0.0032}$ & 0.9947\\[3pt]
PA offset $\Delta$PA$_{\rm{NIRC2}}$ & $^{\circ}$ & 0.16$\pm$0.31 & 0.31\\[3pt]
PA offset $\Delta$PA$_{\rm{NaCo}}$ & $^{\circ}$ & $-$0.74$\pm$0.53 & $-$1.23\\[3pt]
\hline
\multicolumn{4}{c}{Computed parameters} \\
\hline
$M_1$ & $M_{\sun}$ & 0.95$\pm$0.02 & 0.94 \\[1.5pt]
$M_2$ & $M_J$ & 74$^{+12}_{-9}$ & 66 \\[1.5pt]
Mass ratio $M_2$/$M_1$ & & 0.074$^{+0.012}_{-0.009}$ & 0.067 \\[3pt]
Period $P$ & yr & 398$^{+95}_{-93}$ & 304 \\[3pt]
Semi-major axis $a$ & au & 54$\pm$9 & 45 \\[1.5pt]
Eccentricity $e$ & & 0.56$\pm$0.09 & 0.64 \\[3pt]
Arg. periastron $\omega$ & $^{\circ}$ & 64.2$^{+5.5}_{-6.3}$ & 64.5 \\[3pt]
\hline
\end{tabular}
\end{center}
\end{table}

The next two parameters are the parallax and semi-major axis of the orbit of the star around the center of mass of the system. For the parallax, we drew the initial guesses around the nominal value measured by \textit{Gaia} assuming a combination of a Gaussian distribution for the measurement uncertainties and a uniform distribution for the potential systematics. We drew the semi-major axis of the star around a guess value computed from its mass (0.95~$M_\sun$), the companion mass (0.065~$M_\sun$), and the total semi-major axis, assuming a log-flat distribution with a half-width of 20~mas. {The last {free parameters in the mode}l are two RV offsets, two RV jitters, one scaling factor for the NIRC2 separation, and two offsets for the NIRC2 and NaCo position angles (Sect.~\ref{sec:astrometry}).} We {assumed} uniform priors for the RV offsets with halfwidths 0.5~m\,s$^{-1}$ and log-flat priors for the RV jitters with halfwidths 0.3~m\,s$^{-1}$. {We {assumed} uniform priors for the scaling factor for the Keck separations and for the offsets for the Keck and NaCo position angles with widths 0.002, 0.05$^{\circ}$, and 0.1$^{\circ}$, respectively.}

We ran the MCMC for 125\,000 iterations and checked the convergence of the chains using the integrated autocorrelation time \citep{ForemanMackey2013, Goodman2010}. The posterior distributions shown in Fig.~\ref{fig:orbit_imrvpm_corner} were obtained after thinning the chains by a factor 100 to mitigate the correlations and discarding the first 75\% of the chains as the burn-in phase.

\subsection{Parameter intervals and correlations}
\label{sec:paramcorr}

\begin{figure*}[t]
\begin{center}
\includegraphics[width=.42\textwidth]{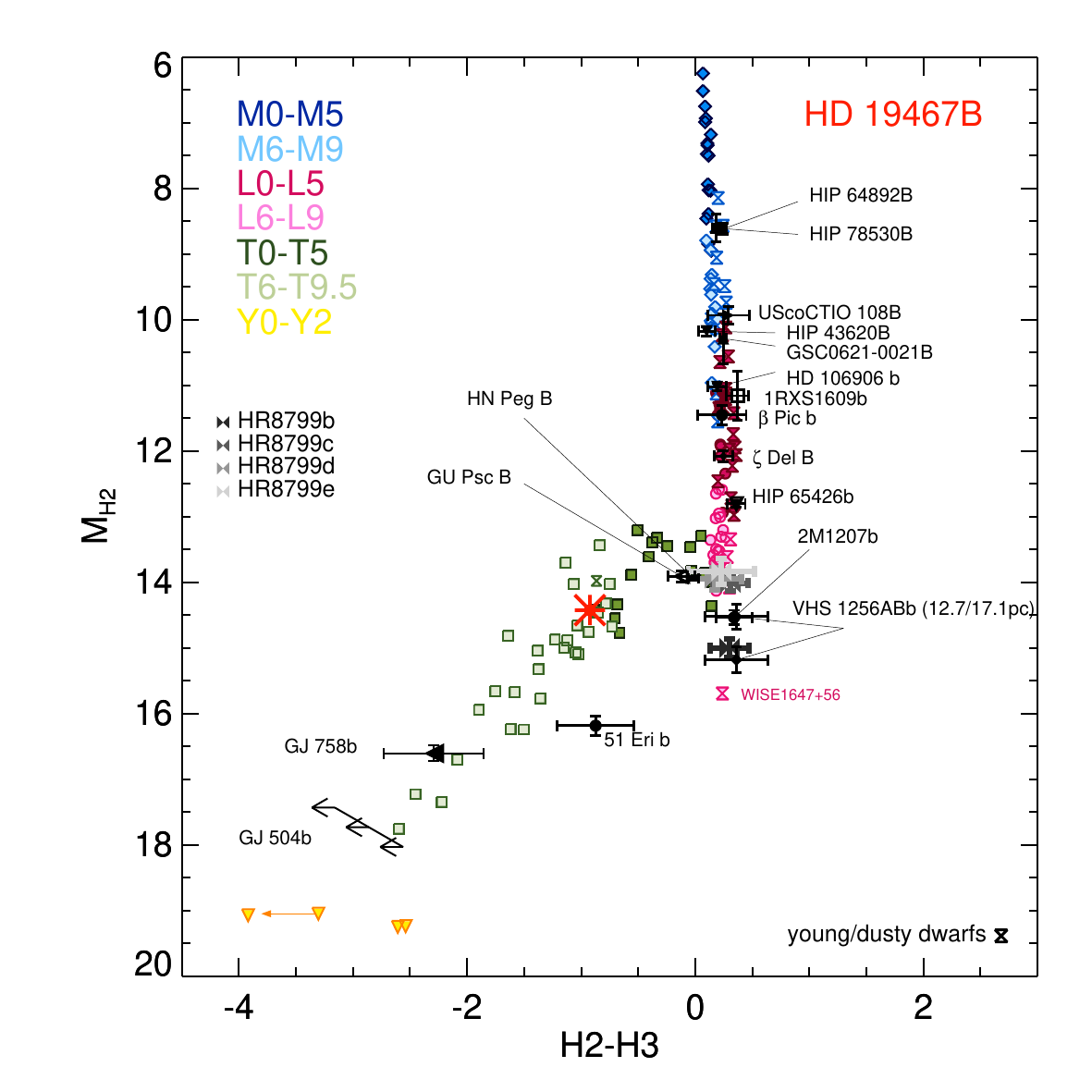}
\includegraphics[width=.42\textwidth]{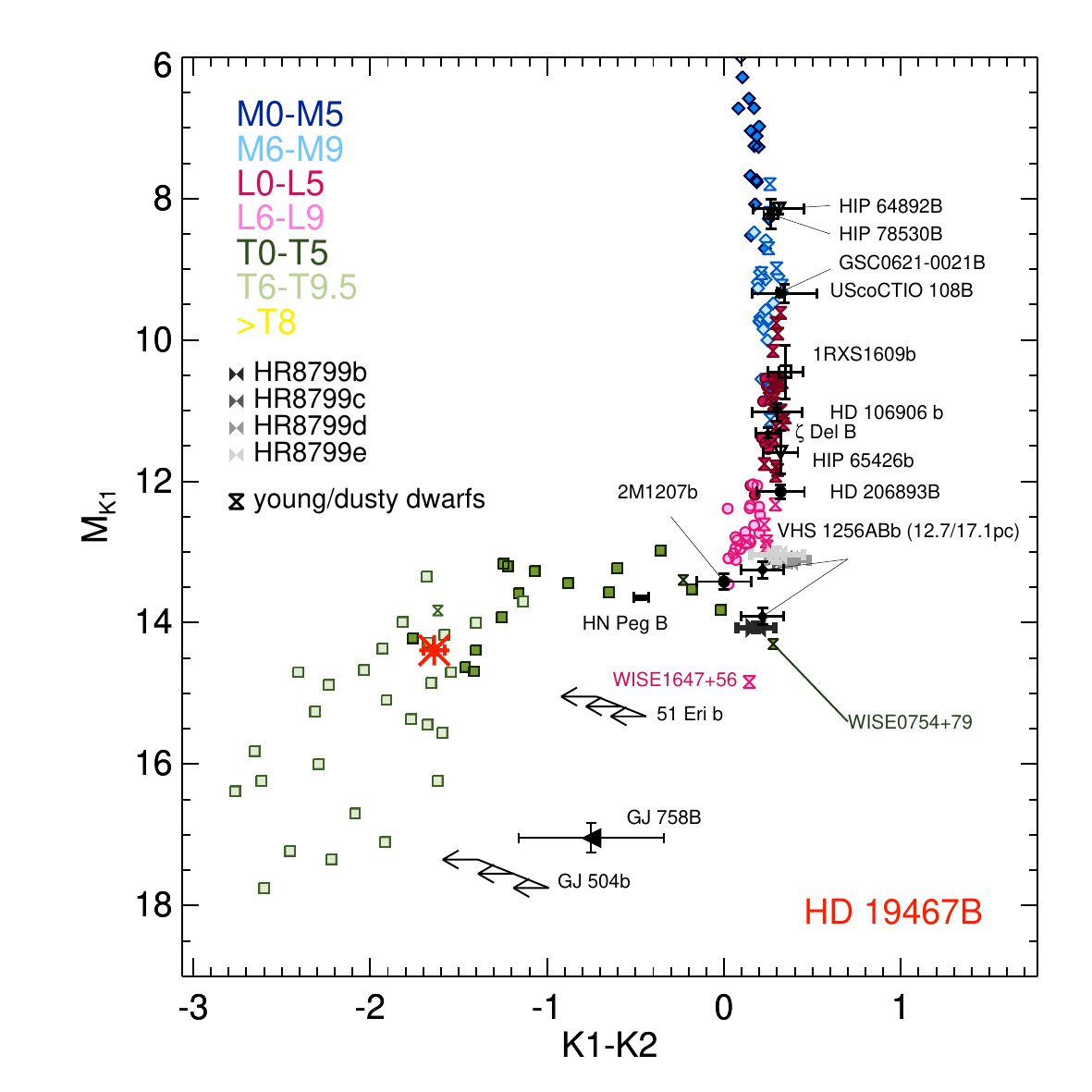}
\includegraphics[width=.42\textwidth]{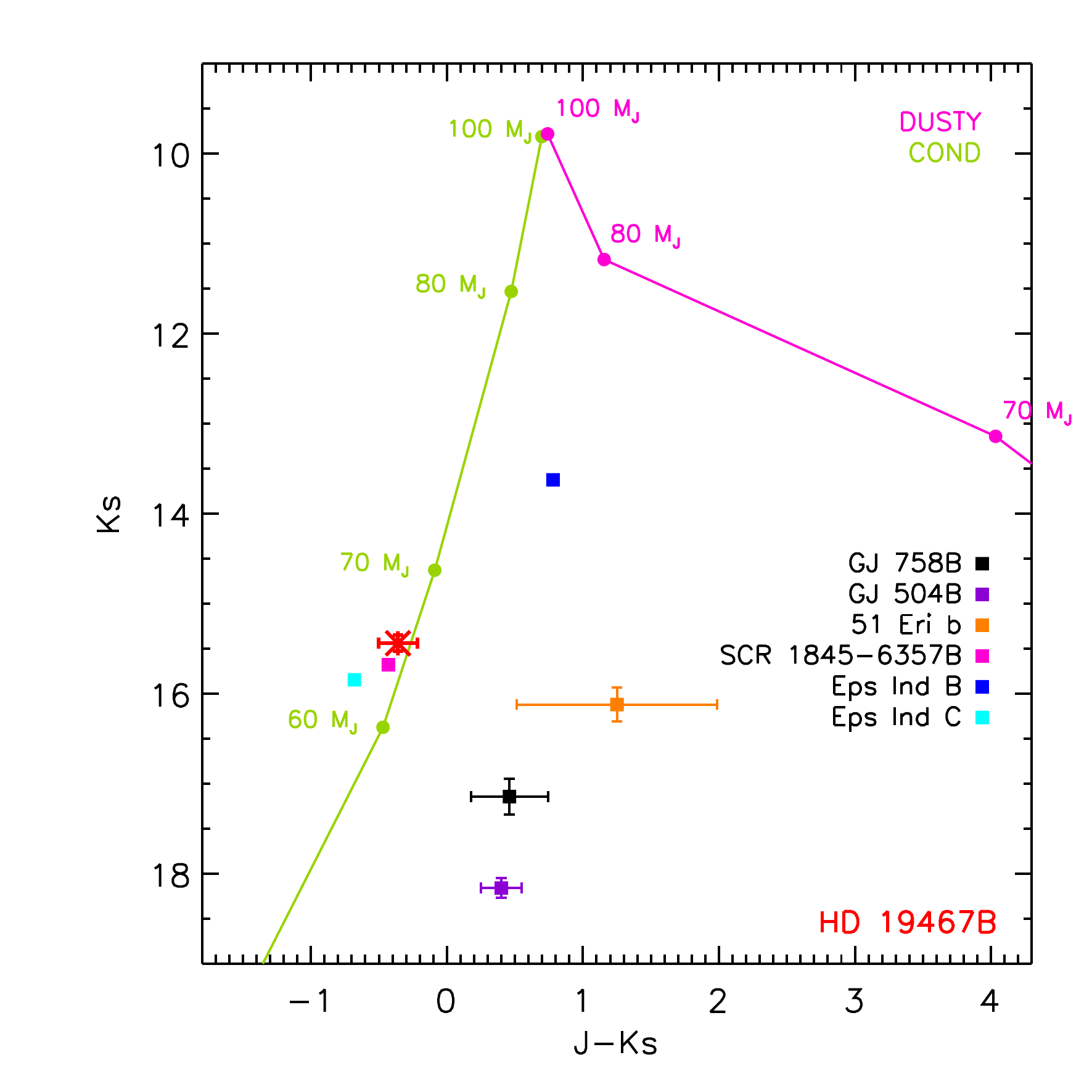}
\includegraphics[width=.42\textwidth]{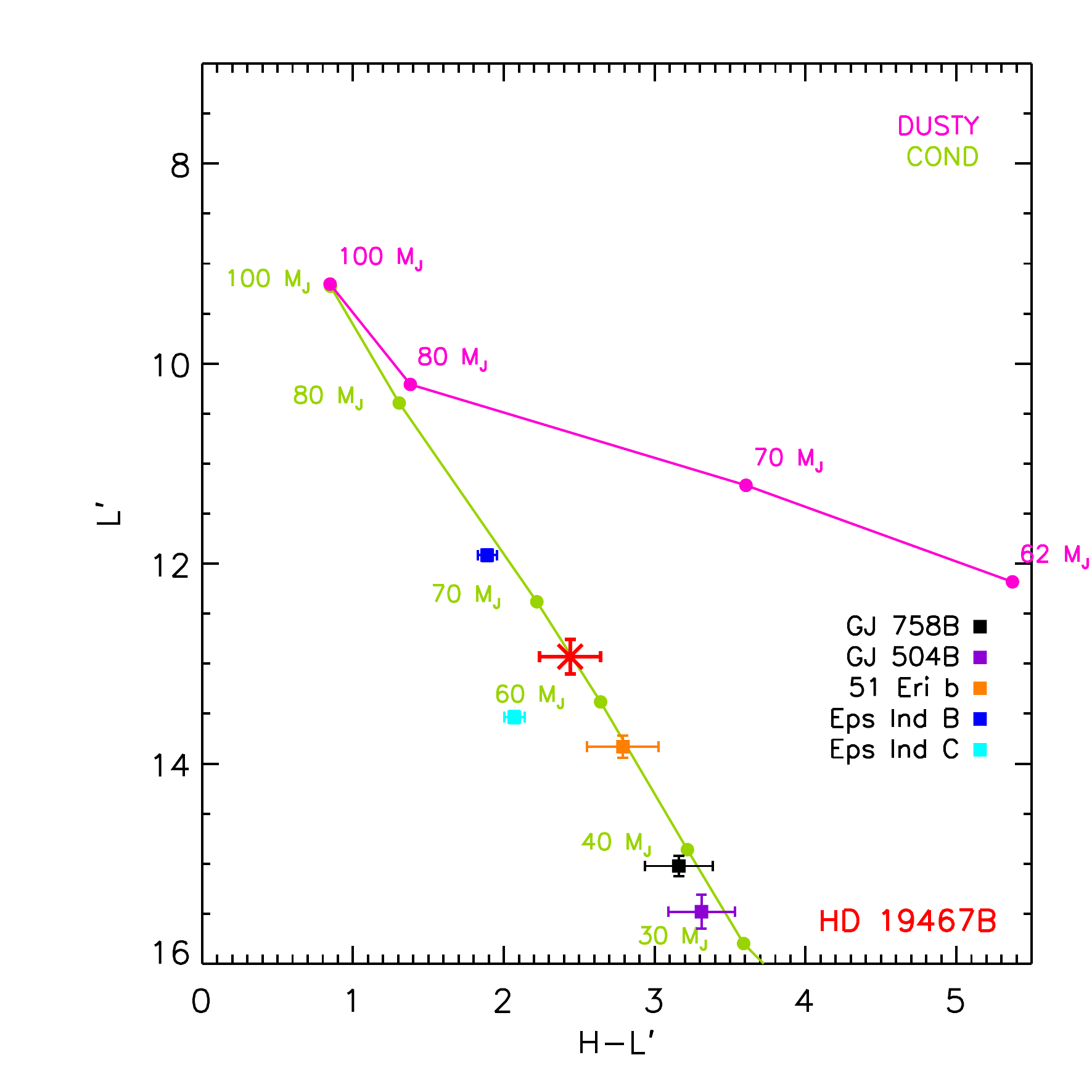}
\caption{\textit{Top}: Color-magnitude diagrams of HD\,19467B (red star) using the SPHERE narrow-band photometry. Template dwarfs (colored points) and a few young low-mass companions (indicated by black labels) are also shown for comparison. \textit{Bottom}: Color-magnitude diagrams of HD\,19467B using the Keck and NaCo broad-band photometry. Evolutionary tracks from the COND model \citep{Baraffe2003} and DUSTY model \citep{Chabrier2000} for an age of 8~Gyr are also indicated with a few known T-type low-mass companions (colored points).}
\label{fig:cmd}
\end{center}
\end{figure*}

Figure~\ref{fig:orbit_imrvpm_corner} provides the histogram distributions of the parameters and the correlation diagrams. Despite the poor orbital coverage of the data and the absence of orbital curvature, most orbital parameters are relatively well constrained except for the semi-major axis, period, and time at {the} periastron passage. The improvements over an orbital fit of the imaging data (Appendix~\ref{sec:orbfit_im}) and an orbital fit of the imaging and RV data (Appendix~\ref{sec:orbfit_imrv}) are noticeable for all parameters in common. In particular, the inclusion of the RV data allows us to break the ambiguity in the longitude of {the} node and {the} argument of {the} periastron inherent to the fit of imaging data only. The semi-major axis of the star with respect to the center of mass of the system is poorly constrained by the current astrometric data (we forced it to stay in the range [57--177]~mas). This results in loose constraints on the mass of the companion with a posterior distribution extending to masses beyond the hydrogen-burning mass limit. We {derived} a 68\% interval of 65--86~$M_J$. {The constraints on the semi-major axis, time at the periastron, companion mass, inclination, and RV semi-amplitude are improved compared to those} derived from {an} RV-imaging fit. {Nevertheless, this behavior is due to the constraints that we imposed on the semi-major axis of the star with respect to the center of mass of the system as seen by the correlations in Fig.~\ref{fig:orbit_imrvpm_corner}.} \citet{Chabrier1997} {computed} hydrogen-burning mass limits of 0.072~$M_{\sun}$ at solar metallicity and 0.083~$M_{\sun}$ at [M/H] =$-$1. If we assume for HD\,19467B the same metallicity as its host star, a linear interpolation gives a mass limit of $\sim$0.074~$M_{\sun}$ or 77~$M_J$. The median values with 1$\sigma$ uncertainties {and} the best-fit values of the parameters are given in Table~\ref{tab:orbparams_rvimpm}. A sample of model orbits are represented in Fig.~\ref{fig:orbit_imrvpm_radec}{. For the} comparison of the companion properties to model predictions (Sect.~\ref{sec:compa_models}), we consider a mass range for the companion of 65--77~$M_J$.

\section{Spectral analysis}
\label{sec:sed}

\subsection{Comparison to color-magnitude diagrams}

We used the IRDIS dual-band photometry of the companion to compute the color-magnitude diagrams shown in the top panels of Fig.~\ref{fig:cmd} \citep[see details in Appendix~\ref{sec:appcmd}, and Appendix C of][]{Bonnefoy2018}. We also used the broad-band photometry of HD\,19467B in \citet{Crepp2014} (we recomputed the absolute magnitudes to account for the new distance estimate from the \textit{Gaia} parallax) and in our analysis to compute the color-magnitude diagrams shown in the bottom panels of Fig.~\ref{fig:cmd}. For these last diagrams, we show for comparison the evolutionary tracks for an age of 8~Gyr from the COND model \citep{Baraffe2003} and the DUSTY model \citep{Chabrier2000}. In all panels, we {indicate} for comparison the T-type substellar companions 51 Eridani b \citep{Macintosh2015, Samland2017, Rajan2017}, GJ\,758B \citep{Thalmann2009, Janson2011a, Vigan2016}, GJ\,504B \citep{Kuzuhara2013, Janson2013, Bonnefoy2018}, and the binary brown dwarf $\epsilon$ Ind BC \citep{King2010}. For the bottom-left panel only, we {indicate} SCR\,1845-6357B \citep{Biller2006, Kasper2007}. We accounted for the new distance estimates from \textit{Gaia} when computing the absolute magnitudes of the companions, except for $\epsilon$ Ind BC for which we used the \textsc{Hipparcos} parallax \citep{vanLeeuwen2007}.

HD\,19467B is located near mid-T template dwarfs in the IRDIS color-magnitude diagrams, which supports its spectral type of T5.5$\pm$1.0 derived in \citet{Crepp2015}. It follows the predictions from the COND model for a mass of $\sim$65~$M_J$ in the color-magnitude diagrams computed from the broad-band photometry. This suggests that atmospheric models with no or very thin clouds should reproduce the spectral properties of HD\,19467B well. The companion is brighter in absolute magnitude {than} 51 Eridani b, GJ\,758B, and GJ\,504B and shows bluer colors in the broad-band color-magnitude diagrams. This could be explained by its larger mass, earlier spectral type, and/or older age. Finally, the companion lies close to SCR\,1845-6357B, which suggests that they share similar spectral properties. SCR\,1845-6357B has a spectral type of T6, $T_{\rm{eff}}$\,=\,950~K, log\,$g$\,=\,5.1~dex, and a mass of 40--50~$M_J$ assuming a system age of 1.8--3.1~Gyr \citep{Biller2006, Kasper2007}. HD\,19467B is also close to $\epsilon$ Ind C, which has a spectral type of T6, $T_{\rm{eff}}$\,=\,880--940~K, log\,$g$\,=\,5.25~dex, and a dynamical mass of 70.1$\pm$0.7~$M_J$ \citep{King2010, Dieterich2018}. \citet{Kasper2009} find for $\epsilon$ Ind C $T_{\rm{eff}}$\,=\,875--925~K and log\,$g$\,=\,4.9--5.1~dex.

\subsection{Atmospheric model fitting}
\label{sec:atmosfits}

We converted the contrast measurements of HD\,19467B reported in \citet{Crepp2014} and in Table~\ref{tab:photometry} into physical fluxes using a model stellar spectrum ($T_{\rm{eff}}$\,=\,5700~K, log\,$g$\,=\,4.5~dex, and [Fe/H]\,=\,0.0~dex) from the BT-NextGen library \citep{Allard2012} and the filter transmission curves. We fit the model spectrum to the stellar spectral energy distribution (SED) over the range 0.3--12~$\mu$m using {the chi-square fitting tool provided in} the Virtual Observatory SED Analyzer \citep{Bayo2008}. {We assumed that the visual extinction is {zero} given the vicinity of the star to the Sun.} The SED was built using data from Tycho \citep{Hog2000}, 2MASS \citep{Cutri2003}, WISE \citep{2013yCat.2328....0C}, and IRAS \citep{Helou1988}, as well as Johnson photometry \citep{Mermilliod2006} and Str\"omgren photometry \citep{Paunzen2015}. We also extracted the normalized P1640 spectrum presented in \citet{Crepp2015} using \texttt{WebPlotDigitizer} \citep{Rohatgi2019} and converted it to physical fluxes using as reference the NIRC2 photometry measured in the $J$ band.


\begin{table*}[t]
\centering
\caption{\label{Tab:atmomodchar} Characteristics of the atmospheric model grids adjusted on the SED of HD\,19467B (see text).}
\begin{tabular}{lllllllllll}
\hline \hline
Model name & $T_{\rm{eff}}$ & $\Delta T_{\rm{eff}}$ & log($g$) & $\Delta$log($g$) & [Fe/H] & $\Delta $[Fe/H] & Clouds & $f_{\rm{sed}}$ or & $\Delta f_{\rm{sed}}$ or \\
	& (K) & (K) & (dex) & (dex) & (dex) & (dex) & & $f_{\rm{sat}}$ & $\Delta f_{\rm{sat}}$ \\
\hline
petitCODE-cloud free & 500--1700 & 50 & 3.0--6.0 & 0.5 & -1.0--1.4 & 0.2 & No & -- & -- \\  
petitCODE-cloudy & 800--1300 & 50 & 1.5--6.0 & 0.5 & -0.4--1.4 & 0.2 & Yes  & 0.5--6.0 & 0.5 \\  
Exo-REM & 500--2000 & 50 & 3.0--6.0 & 0.1 & -0.5--0.5 & 0.5 & No, Yes & 0.1, 0.01 & -- \\  
Morley 2012 & 400--1300 & 50/100 & 4.0--5.5 & 0.5 & 0.0 & -- & Yes & 2.0--5.0 & 1.0 \\  
\hline
\end{tabular}
\tablefoot{The columns give the model name, the range and step for the effective temperature, surface gravity, and metallicity, the type of clouds included in the model, and the range and step for the sedimentation or the saturation parameter for the clouds (see text).}
\end{table*}

\begin{table*}[t]
\centering
\caption{\label{Tab:atmoret} Retrieved HD\,19467B's atmospheric parameters.}
\begin{tabular}{llllllllll}
\hline \hline
Model name & $T_{\rm{eff-cloudy}}$ & $T_{\rm{eff-cloud-free}}$ & log\,$g$ & [Fe/H] & $f_{\rm{sed}}$ & CF & R$_p$ & M$_p$ & {$\chi_\mathrm{min}^2$} \\[3pt] 
 & (K) & (K) & (dex) & (dex) & & & ($R_J$) & ($M_J$) & \\
\hline
Exo-REM-cloud free & -- & 975$\pm$125 & 5.2$\pm$0.1 & {UC} & -- & (1.0) & -- & -- & {231} \\[6pt]  
petitCODE-cloud free & -- & 1186$^{+24}_{-27}$ & 5.61$^{+0.06}_{-0.05}$ & 0.18$^{+0.12}_{-0.11}$ & -- & (0.0) & 0.59$^{+0.03}_{-0.03}$ & 57$^{+7}_{-4}$ & {103.3} \\[4pt]  
petitCODE-cloudy & 1044$^{+12}_{-18}$ & -- & 5.33$^{+0.05}_{-0.05}$ & -0.05$^{+0.07}_{-0.07}$ & 1.02$^{+0.39}_{-0.28}$ & (1.0) & 0.84$^{+0.04}_{-0.02}$ & 63$^{+6}_{-7}$ & {101.3} \\[6pt]  
petitCODE-patchy & 932$^{+66}_{-63}$ & 1291$^{+99}_{-89}$ & 5.34$^{+0.08}_{-0.09}$ & 0.03$^{+0.08}_{-0.08}$ & 1.20$^{+0.79}_{-0.46}$ & 0.79$^{+0.10}_{-0.15}$ & 0.83$^{+0.09}_{-0.06}$ & 60$^{+7}_{-6}$ & {86.6} \\[6pt]  
Morley 2012 & 928$^{+39}_{-42}$ & -- & 5.20$^{+0.09}_{-0.10}$ & (0.0) & 4.07$^{+0.41}_{-0.49}$ & (1.0) & 0.99$^{+0.10}_{-0.09}$ & 63$^{+6}_{-7}$ & {129.7} \\[6pt]  
\hline
\end{tabular}
\tablefoot{Values given in parenthesis are priors {or} assumptions and are not retrieved (see text). {UC: Unconstrained}.}
\end{table*}

\begin{figure*}[t]
\centering
\includegraphics[height=.25\textwidth, trim = 4mm 4mm 4mm 2mm, clip]{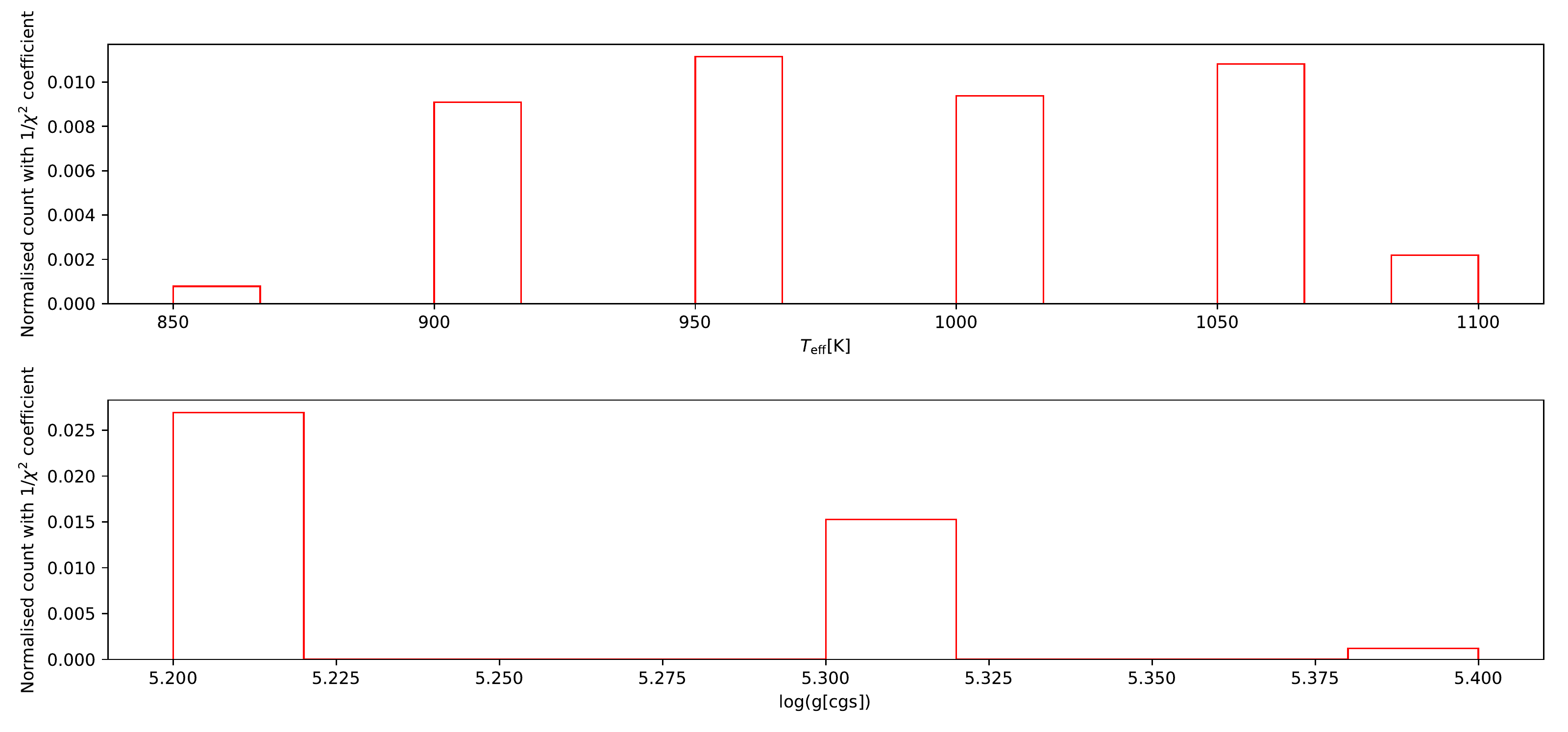}
\includegraphics[height=.25\textwidth]{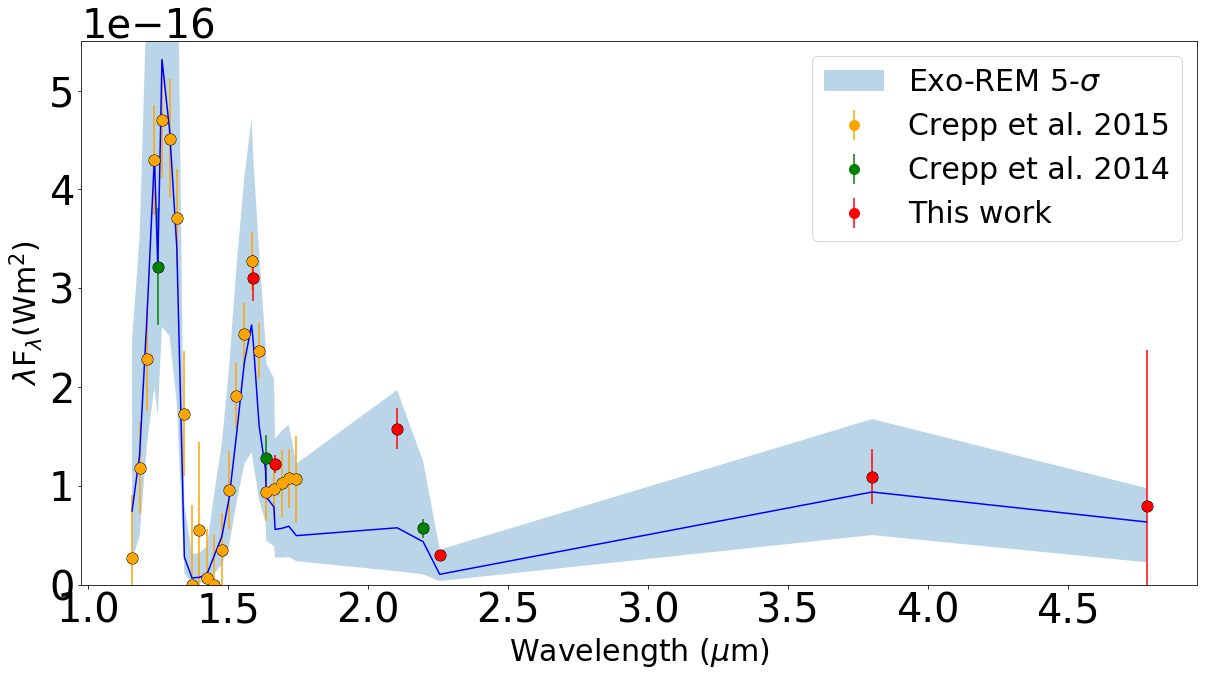}
\caption{{\textit{Left}: Histograms of the effective temperature (top) and surface gravity (bottom) of the models from Exo-REM, reproducing the data, without cloud covering all metallicity values (see text). The number of count is normalized using the invert of the $\chi^2$. \textit{Right}: Comparison of the best-fit model spectra (dark blue line: best fit, light blue area: 5$\sigma$ envelope) and of the mesured SED (colored data points).}}
\label{fig:exorem}
\end{figure*}

\subsubsection{Exo-REM models}

As the first approach to characterize the atmosphere of HD\,19467B, we {fit the SED} with the spectral library Exo-REM \citep[][see Table~\ref{Tab:atmomodchar}]{Baudino2015, Baudino2017, Charnay2018}. The analysis was similar to the one introduced in \citet{Baudino2015}, using $\chi^2$ maps. We explored the $T_{\rm{eff}}$ (between 500 and 2000~K by step of 50~K) and log\,$g$ (between 3 and 6 by step of 0.1) for six cases: metallicity [Fe/H] = -0.5, 0, +0.5, without clouds or with simple microphysics clouds \citep[described in][]{Charnay2018}. The species assumed for the clouds were iron (Fe) and silicates (Mg$_2$SiO$_4$). We included in the analysis a complete research of the {optimal} radius. Usually, we approximate the radius as a shift of the full spectrum \citep{Baudino2015}. For this analysis, we first performed a simple $\chi^2$ minimization as usual. If the radius was outside a given range (0.7--1.3~$R_J$ in this case, coming from evolutionary tracks), we tried to force the radius to decrease or increase to fit in this range. The only rule was to stay in the confidence interval (5-$\sigma$).

In these $\chi^2$ maps, we only kept the results that reproduced the data at less than 5-$\sigma$, with a radius solution between 0.7--1.3~$R_J$ and a mass solution between 52--72~$M_J$ (based on the system's dynamics). Although the adopted mass prior includes smaller values than the actual constraints from the orbital fit (Sect.~\ref{sec:orbfit}), the choice of the bounds {has a} negligible effect on the derivation of the atmospheric parameters. The radius prior has a larger effect. The models fit the data only without clouds (the best fit with clouds is out at more than 10-$\sigma$). We {do not} observe any clues about the metallicity. Figure~\ref{fig:exorem} shows the histograms of the $T_{\rm{eff}}$ and log\,$g$ reproducing the data {{and the comparison} of the best-fit spectra to the measured SED. The count of the histograms} is normalized using the invert of the $\chi^2$ as a coefficient to highlight the {best-fit} cases. The inferred $T_{\rm{eff}}$ is 975$\pm$125~K, the inferred log($g$[cgs])=5.2$\pm$0.1 {(Table~\ref{Tab:atmoret}). We also provide in Table~\ref{Tab:atmoret} the $\chi^2$ values associated with the best-fit solution computed {according to} the definition of \citet{Baudino2015}.}


\subsubsection{petitCODE models}

As the second approach, we used petitCODE \citep{Molliere2015, Molliere2017} to calculate a grid of self-consistent models; assuming both cloud-free and cloudy atmospheres. The characteristics of these models are summarized in Table~\ref{Tab:atmomodchar}. For the cloudy models, the species included are Na$_2$S and KCl. The free parameters in the cloud-free models are the effective temperature, the surface gravity, and the metallicity. For the cloudy models, {the} sedimentation factor ($f_\mathrm{sed}$) is also taken into account as a free parameter.

We performed Bayesian analysis using the \texttt{emcee} tool \citep{ForemanMackey2013} to explore the atmospheric properties of HD\,19467B with the petitCODE models. We considered the statistical treatment of observational uncertainties and explored any underestimation of these uncertainties through a Gaussian Process. Uninformative priors were also assumed for the initialization of the walkers in the MCMC process.

\begin{figure}
\centering
\includegraphics[width=.49\textwidth]{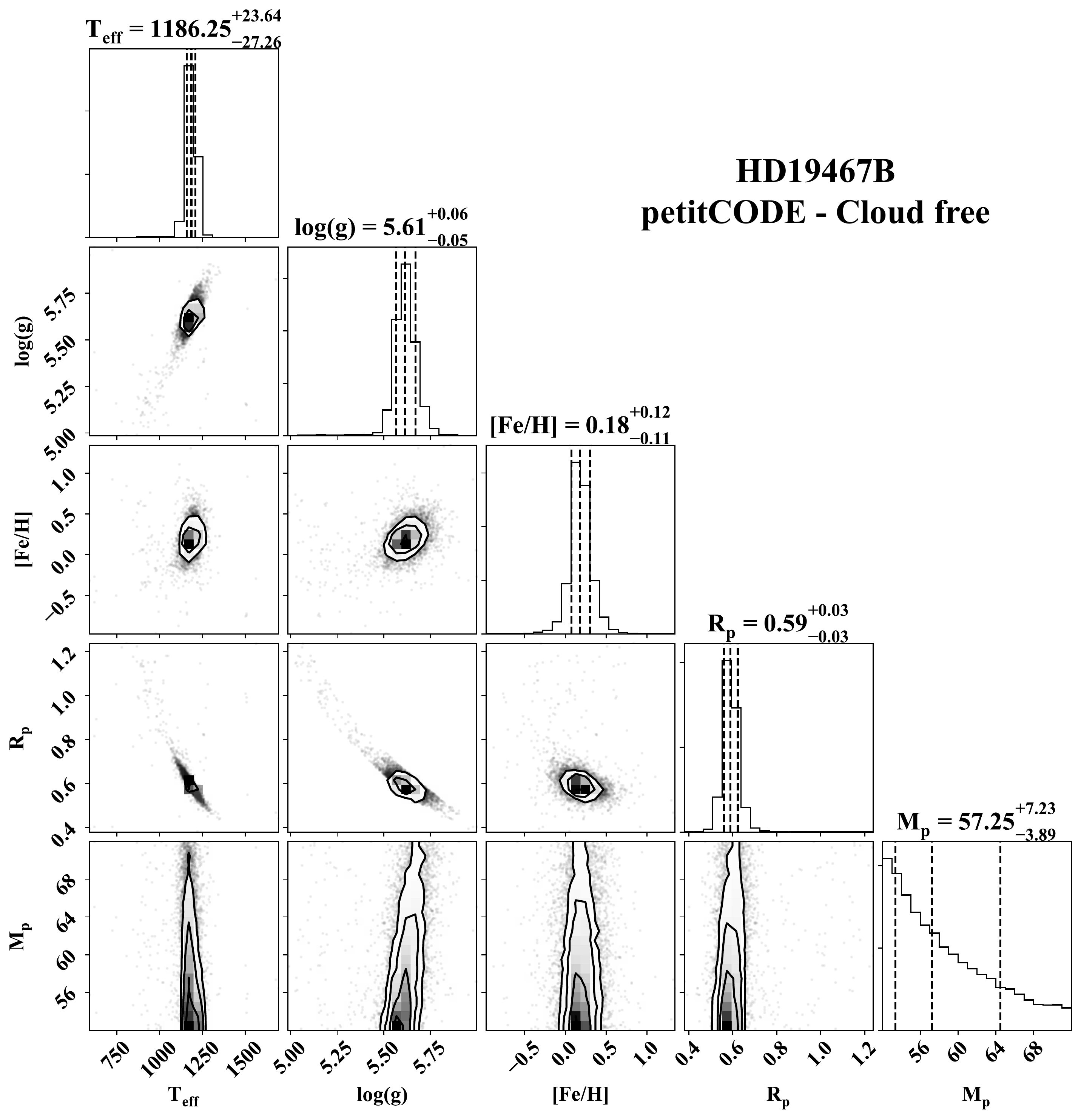}\\
\includegraphics[width=.49\textwidth]{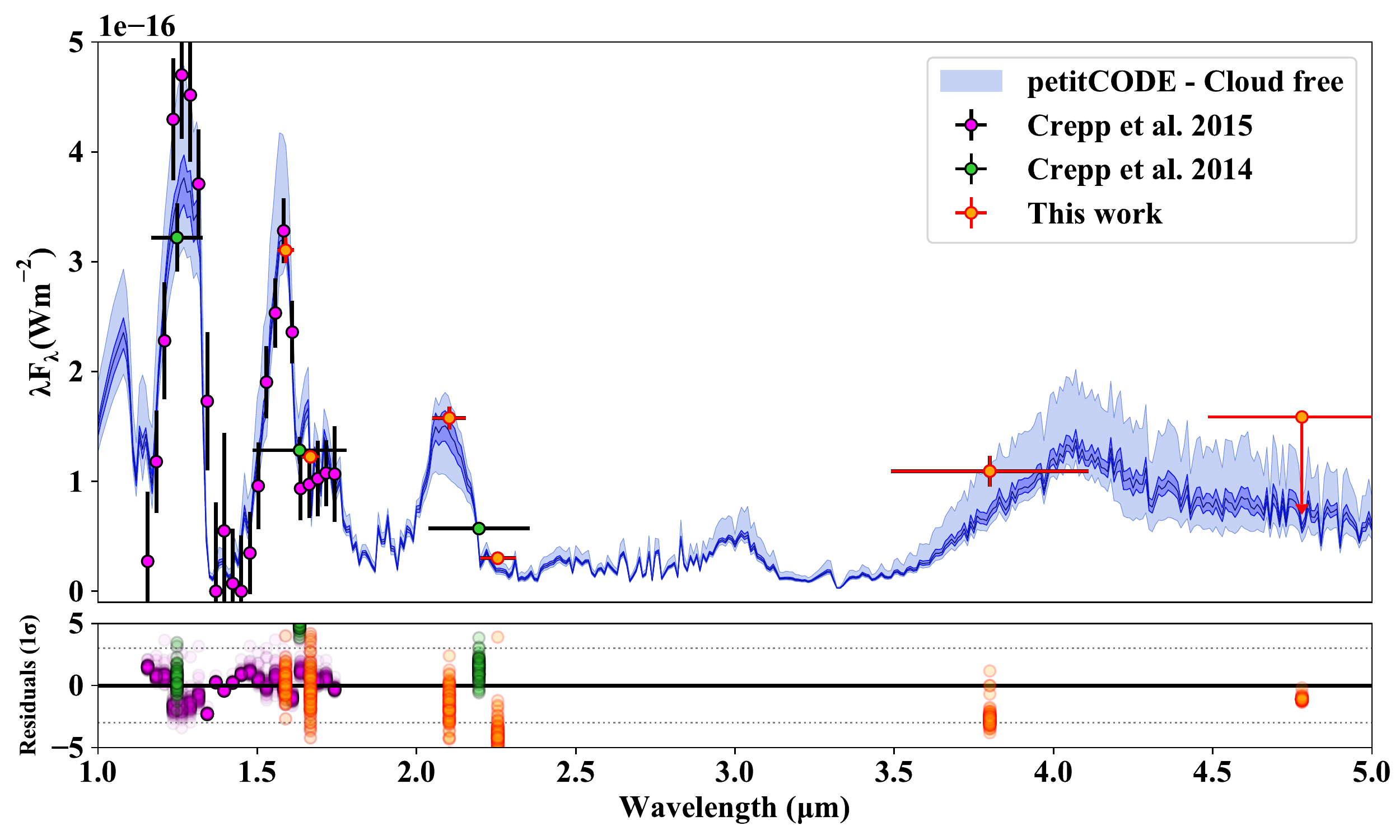}
\caption{{Atmospheric fitting of HD\,19467B with petitCODE cloud-free models.} The \textit{top} panel shows the corner plot of the retrieved atmospheric parameters and the \textit{bottom} panel the comparison of the best-fit model spectra and of the measured SED (colored data points). For the model spectra, the dark blue area corresponds to the region of the posteriors between the 16\% and 84\% quantiles and the light blue area to the region between the 1\% and 99\% quantiles.}
\label{fig:atmo_petitclear}
\end{figure}

\begin{figure}[t]
\centering
\includegraphics[width=.49\textwidth]{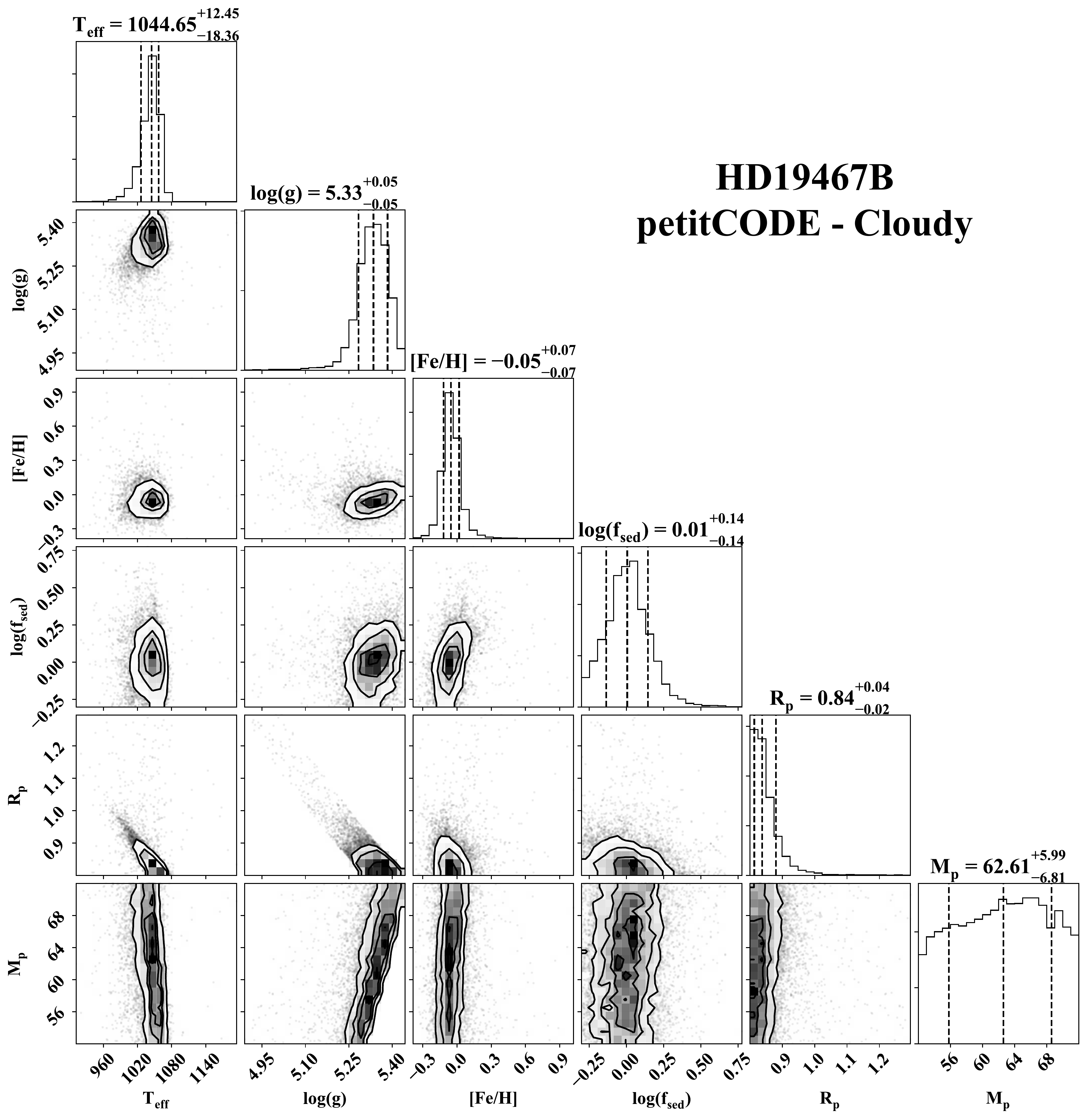}\\
\includegraphics[width=.49\textwidth]{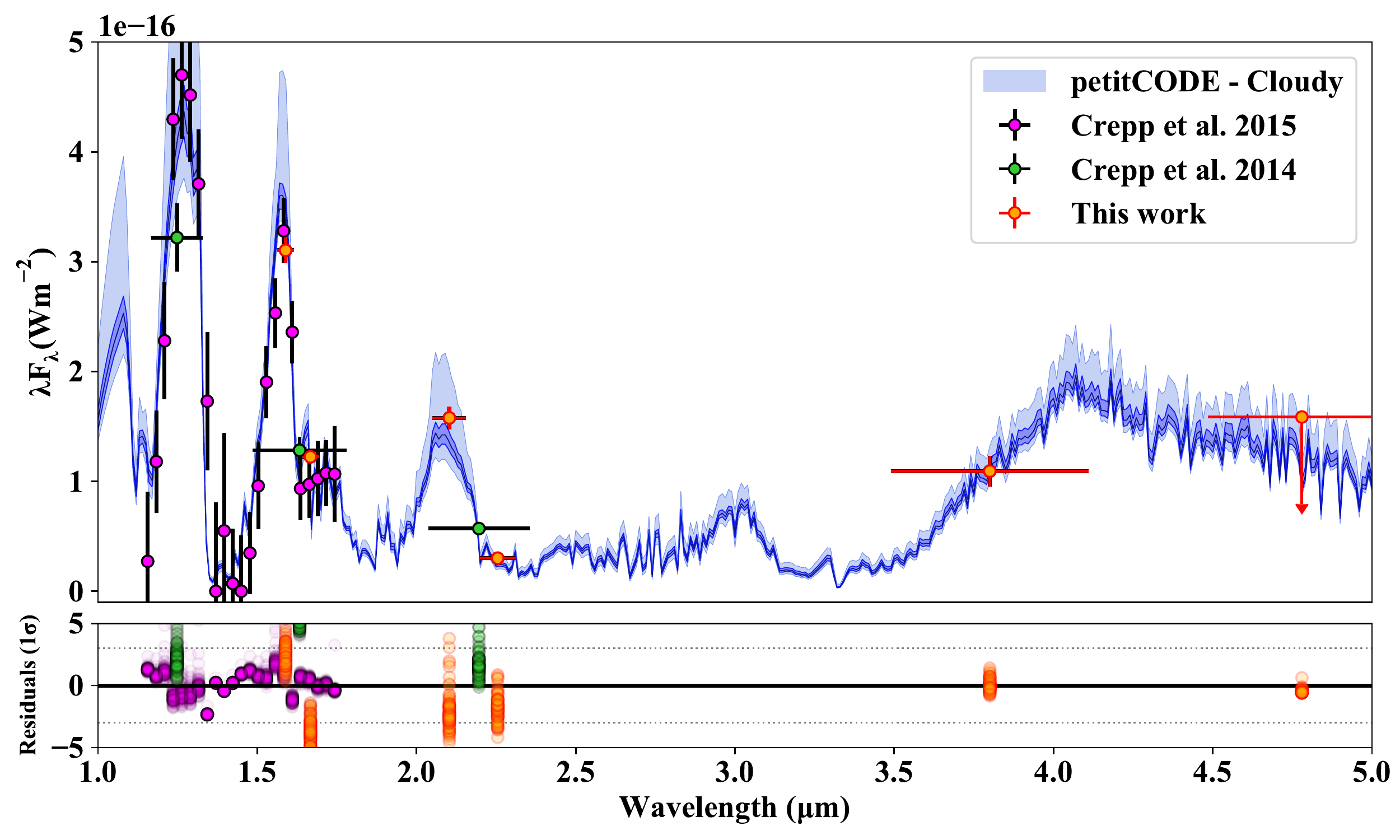}
\caption{{Atmospheric fitting of HD\,19467B with petitCODE cloudy models.} For the radius posterior, values smaller than 0.78~$R_J$ are disfavored.}
\label{fig:atmo_petitcloudy}
\end{figure}

{Firstly, we fit the data with cloud-free models. Figure~\ref{fig:atmo_petitclear} shows the results and the} corner plot of the retrieved parameters. The retrieved properties of HD\,19467B are as follows, assuming a cloud-free atmosphere {(Table~\ref{Tab:atmoret})}: an effective temperature of 1186$^{+24}_{-27}$~K, a surface gravity of 5.61$^{+0.06}_{-0.05}$~dex, and a metallicity of 0.18$^{+0.12}_{-0.11}$~dex. The radius {is 0.59$^{+0.03}_{-0.03}$\;$R_J$ and the mass is 57$^{+7}_{-4}$\;$M_J$}. As discussed, cloud-free models could explain the SED, although tentatively. The photometric points, in particular at 1.633\;$\mu$m, 2.255\;$\mu$m, and 3.8\;$\mu$m, {disagree} with the best-fit model, with the first two by at least 3$\sigma$ (dotted lines in the bottom panel of Fig.~\ref{fig:atmo_petitclear}). In addition, the inferred radius of the companion is significantly smaller than the expected radius from the evolutionary tracks (0.8\;$R_J$). We therefore {examined} cloudy models to improve the fit.


\begin{figure}[t]
\centering
\includegraphics[width=.49\textwidth]{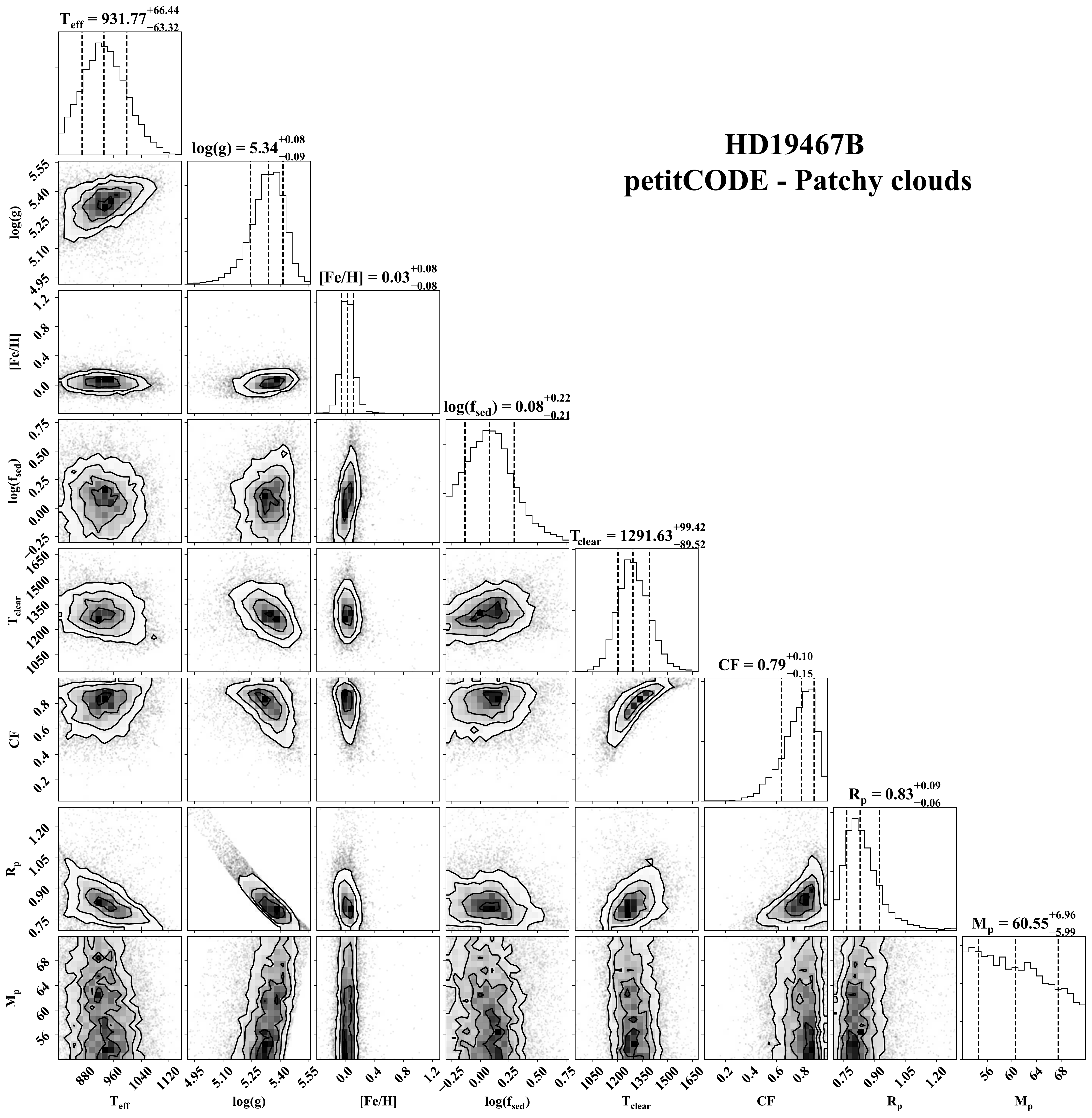}\\
\includegraphics[width=.49\textwidth]{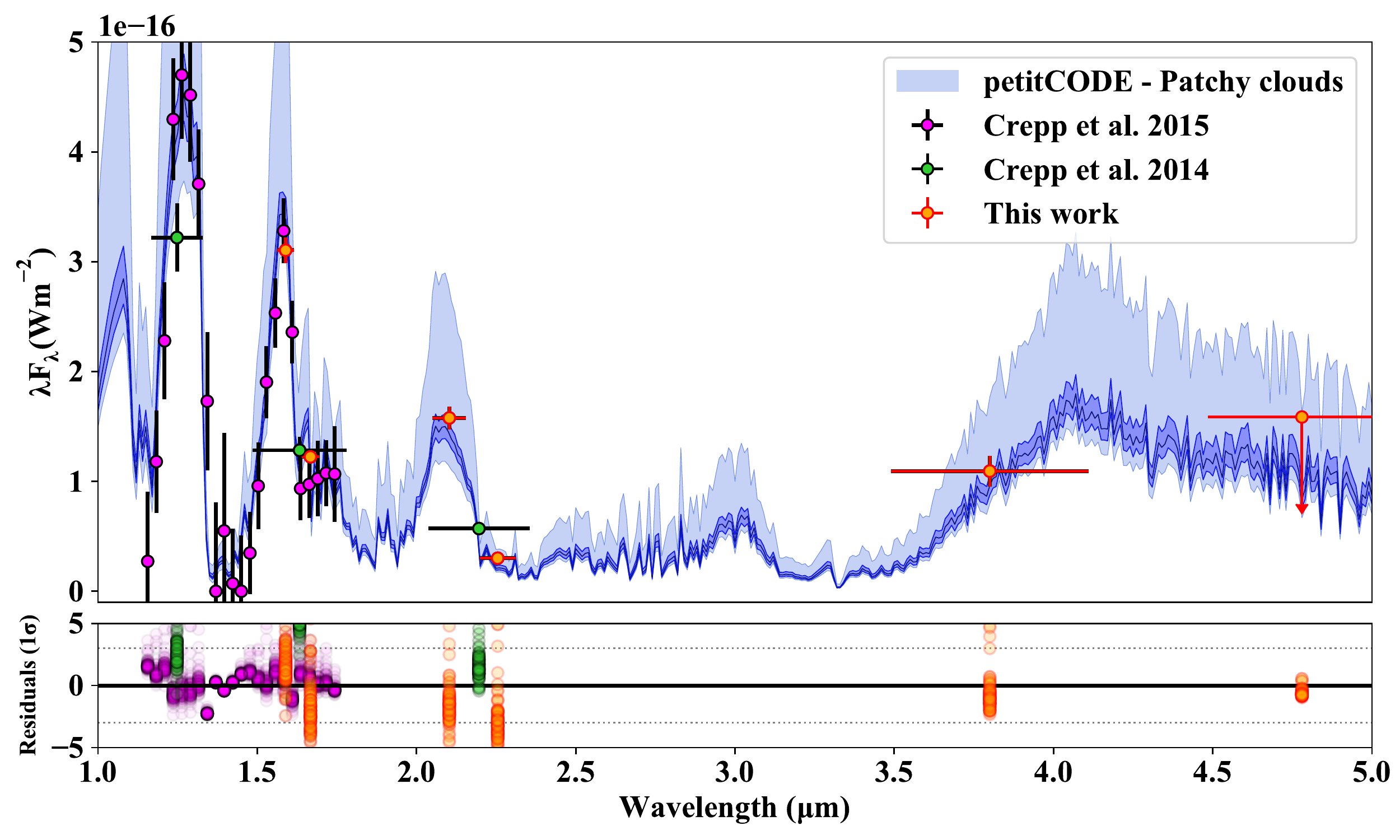}
\caption{{Atmospheric fitting of HD\,19467B with petitCODE patchy cloudy models with temperature constraints (see text).}}
\label{fig:atmo_petitpatchy}
\end{figure}

{Secondly, we fit the data with cloudy models. We assumed} 52$<M_p<$72\;$M_J$ and 0.7$<R_p<$1.3\;$R_J$ as priors. Figure~\ref{fig:atmo_petitcloudy} shows the fitted models to the data and the corner plot of the retrieved parameters. The retrieved atmospheric properties are as follows {(Table~\ref{Tab:atmoret})}: $T_{\rm{eff}}$=1044$^{+12}_{-18}$~K, log\,$g$=5.33$^{+0.05}_{-0.05}$~dex, and [Fe/H]=-0.05$^{+0.07}_{-0.07}$~dex, all have values less than their counterparts when {fitting with} petitCODE cloud-free models. This behavior can be explained by the prior used for the companion radius, which excludes radii smaller than 0.7~$R_J$. The best-fit value for log($f_\mathrm{sed}$) is 0.01$^{+0.14}_{-0.14}$, which corresponds to a sedimentation factor of 1.0. This suggests {that} an active removal of the clouds is required for the clouds to fit the observations. We note that the cloud species considered in these petitCODE cloudy models are Na$_2$S and KCl, which both have a relatively low evaporation temperatures at typical photospheric pressures (i.e., around 1000\;K at 1\;bar). The fitted temperature of $\sim$1050\;K suggests that these species have a reduced contribution to the cloud opacities; supporting an optically thin atmosphere hypothesis. The retrieved radius {is 0.84$^{+0.04}_{-0.02}$\;$R_J$ and the retrieved mass is 63$^{+6}_{-7}$\;$M_J$}. While the best fitted petitCODE’s cloudy models agree with {most data points} within 3$\sigma$, fitting the photometric point at 1.633~$\mu$m demands relaxation of the model.

{Thirdly}, we {examined} the idea of a patchy atmosphere for HD\,19467B, following the method in \citet{Samland2017}. In this approach, we {took} one cloudy model and one cloud-free model and {combined} them linearly as below:

\begin{equation}
    F_{patchy}=CF\cdot F_{cloudy}+(1-CF)F_{cloud-free}
\end{equation}
where $F_{cloudy}$ and $F_{cloud-free}$ are the flux of cloudy and cloud-free models. $CF$ is the cloud fraction, which has a value ranging from 0 (no cloud) to 1 (fully cloudy). We also {imposed} a prior on the temperature of the patches, where the temperature of the cloudy parts {was} assumed to be smaller than the temperature of the cloud-free parts, $T_{cloudy}$<$T_{cloud-free}$. The surface gravity and metallicity of these patches {were} assumed to be the same. Figure~\ref{fig:atmo_petitpatchy} shows the best-fit results and retrieved parameters assuming a patchy atmosphere. While taking this approach {does not} improve the fit significantly, the radius of the companion is constrained. The retrieved atmospheric properties are $T_{eff-cloudy}$=932$^{+66}_{-63}$~K, $T_{eff-cloud-free}$=1291$^{+99}_{-89}$~K, log\,$g$=5.34$^{+0.08}_{-0.09}$~dex, [Fe/H]=0.03$^{+0.08}_{-0.08}$~dex, and log($f_\mathrm{sed}$)=0.08$^{+0.22}_{-0.21}$ {(Table~\ref{Tab:atmoret})}. The cloudy temperature {agrees} with the retrieved temperature by \citet{Crepp2015} and the cloud-free temperature {would agree better} with the expectations given the age and dynamical mass (Sect.~\ref{sec:compa_models}). A cloud fraction of $CF\sim0.8\pm0.1$ hints for an atmosphere to be mostly covered by clouds. Given the cloud-free and cloudy temperatures and the cloud fraction, the global temperature is 1042$^{+77}_{-71}$~K. Nevertheless, the relatively high $f_\mathrm{sed}$ and low evaporation temperatures of the cloud species considered in the cloudy models, as discussed above, call for an optically thin cloud layer. The retrieved radius {is 0.83$^{+0.09}_{-0.06}$\;$R_J$ and the retrieved mass is 60$^{+7}_{-6}$\;$M_J$}. The patchy model constrains the radius of the companion well and in agreement with the evolutionary tracks.


\begin{figure}
\centering
\includegraphics[width=.49\textwidth]{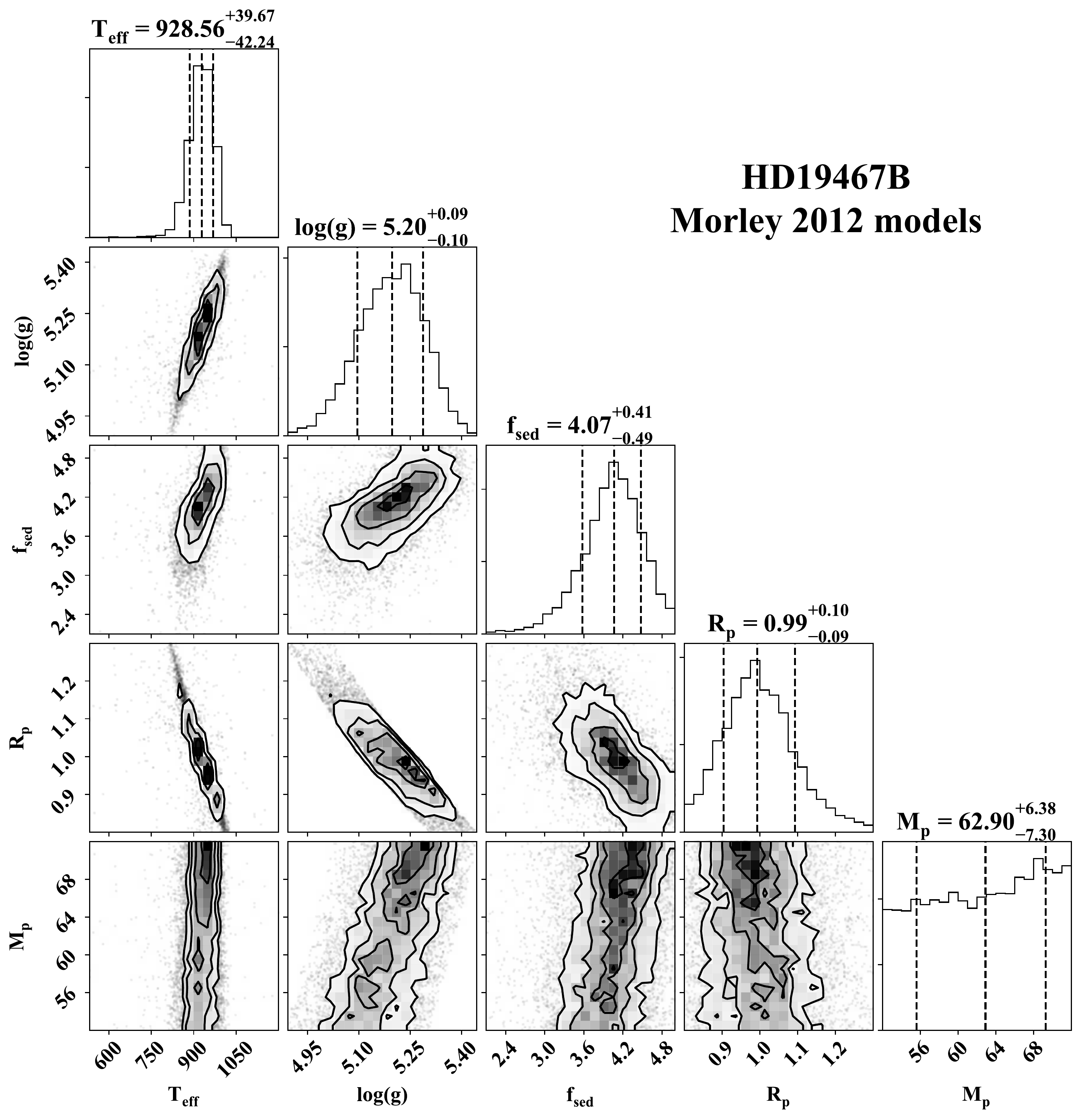}\\
\includegraphics[width=.49\textwidth]{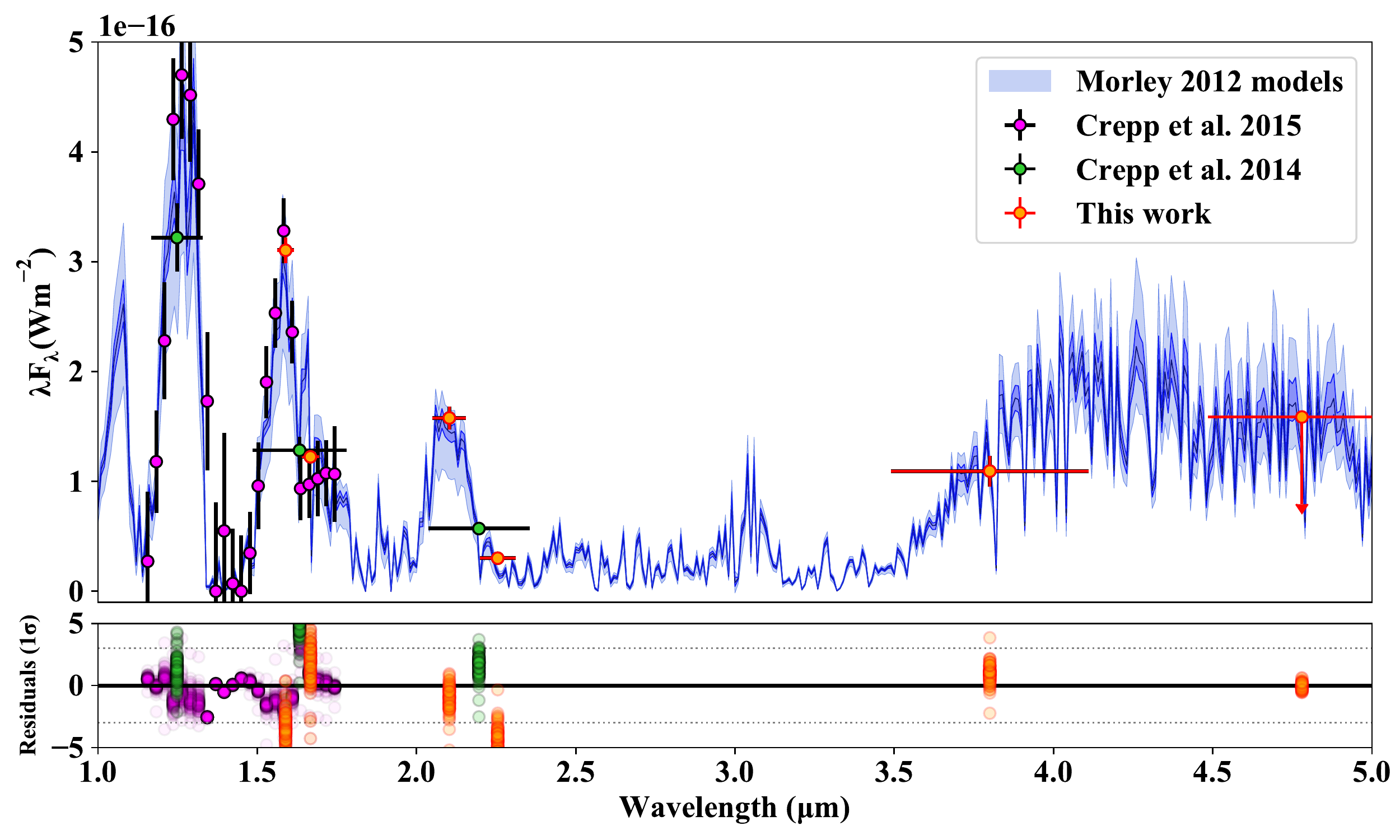}
\caption{{Atmospheric fitting of HD\,19467B with the models of \citet{morley_neglected_2012}.}}
\label{fig:atmo_Morley12}
\end{figure}

\begin{figure*}[t]
\centering
\includegraphics[width=.4\textwidth]{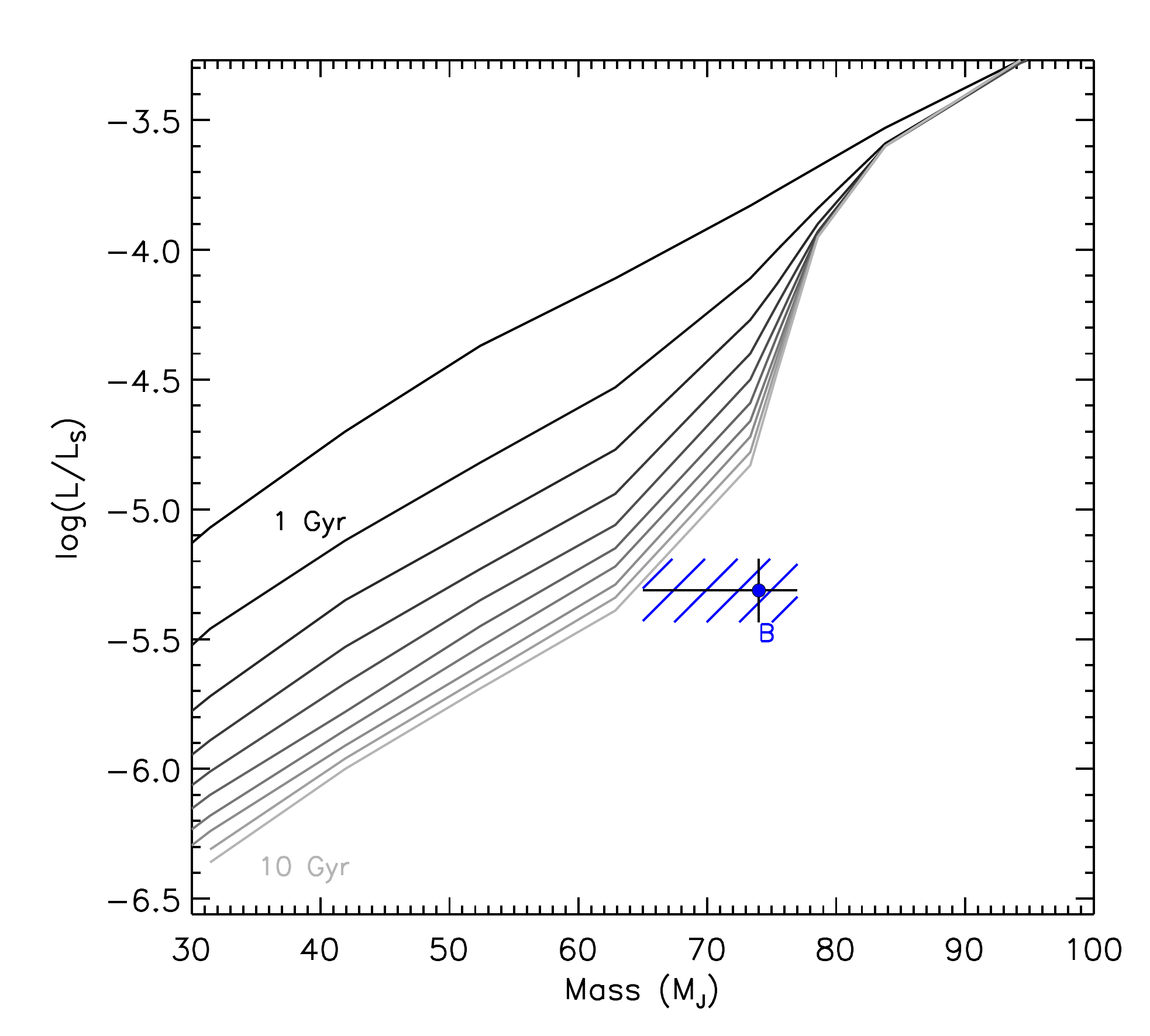}
\includegraphics[width=.4\textwidth]{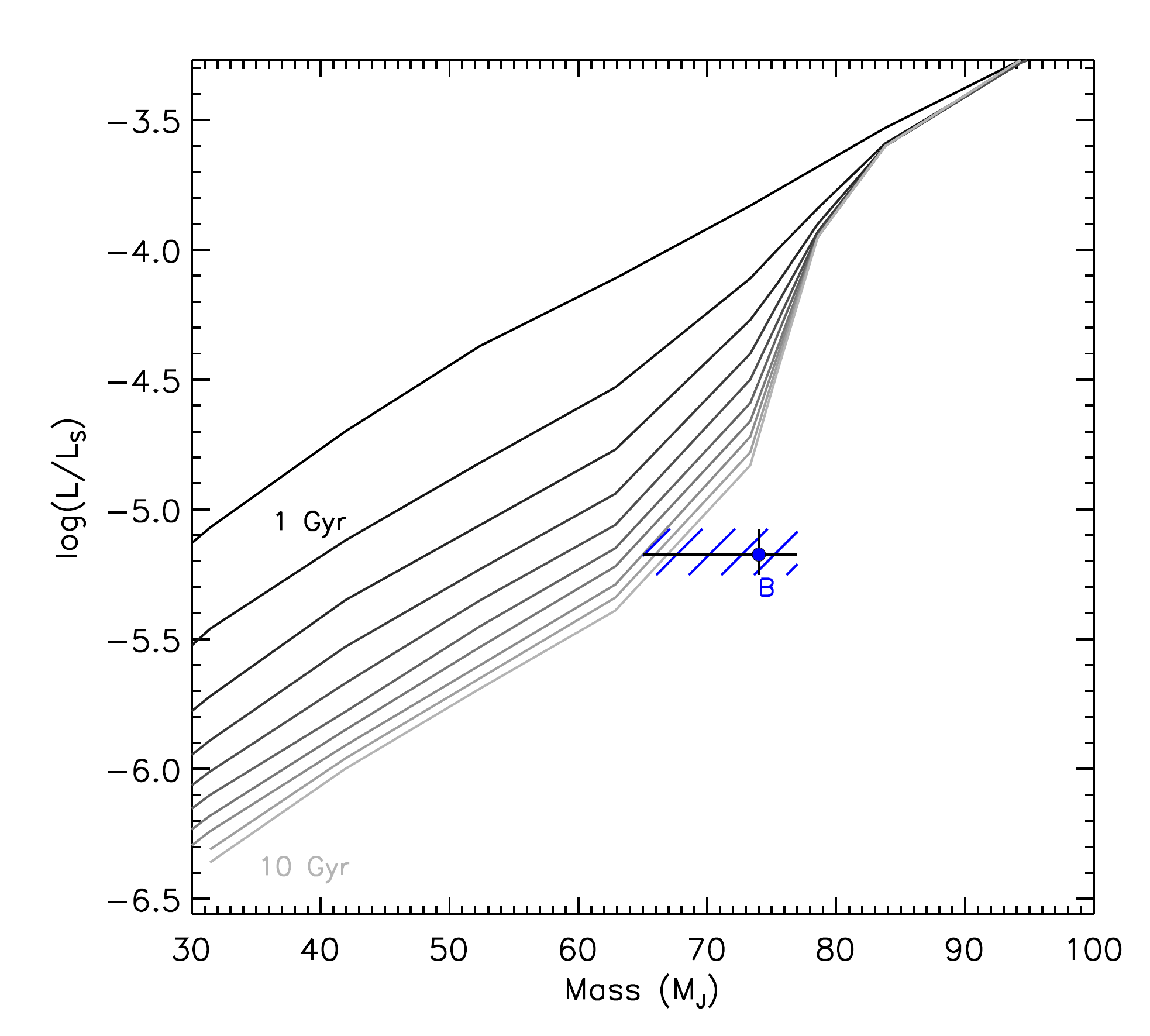}
\caption{{Bolometric luminosity of HD\,19467B computed using the NIRC2 photometry in the $K_s$ band (\textit{left}) and in the $J$ band (\textit{right}) and the relations of \citet{Filippazzo2015} for field dwarfs as a function of the mass estimated from the orbital fit and the theoretical hydrogen-burning mass limit (see text).} For comparison, the model isochrones of \citet{Baraffe2003} are indicated.}
\label{fig:lbol_mass}
\end{figure*}

\subsubsection{Morley 2012 models}

We also {fit the SED} using the grid of models in \citet{morley_neglected_2012}. The properties of their grid are summarized in Table~\ref{Tab:atmomodchar}. The cloud species included are Na$_2$S, KCl, ZnS, MnS, and Cr. We note that they assume some additional cloud species in the models. This results in an abundance of cloud opacities in colder regimes, where more condensates can form to add to the opacity contribution of clouds. A higher retrieved sedimentation factor, 4.07$^{+0.41}_{-0.49}$, is likely a consequence of this treatment of cloud species (Fig.~\ref{fig:atmo_Morley12} {and Table~\ref{Tab:atmoret}}). Other retrieved atmospheric parameters are $T_{\rm{eff}}$=928$^{+39}_{-42}$~K and log\,$g$=5.20$^{+0.09}_{-0.10}$~dex. The {radius is 0.99$^{+0.10}_{-0.09}$\;$R_J$ and the mass is 63$^{+6}_{-7}$\;$M_J$}. While the radius is constrained, the value is larger than the expected value from the evolutionary tracks ($\sim$0.8\;$R_J$).

\subsubsection{{Conclusion and remarks}}

{We conclude that the SED of HD\,19467B is consistent with a patchy atmosphere mostly covered by thin clouds. A summary of the retrieved parameters is given in Table~\ref{Tab:atmoret}. {It is worth highlighting the} relative consistency of the retrieved surface gravity in all the tested atmospheric models{, which suggests} an object with a high surface gravity (greater than 5.1~dex){. A solar metallicity is consistent with} all analyses ($\sim$0.0, for the Exo-REM and petitCODE models). All our retrieved surface gravities are at the high end or larger than the surface gravities of 4.21--5.31~dex inferred by \citet{Crepp2015} using BT-Settl models.}

The effective temperatures retrieved for the fits of the models of \citet{morley_neglected_2012} and of the cloudless Exo-REM models agree well with the expectations from the empirical relations of \citet{Filippazzo2015} for field dwarfs given its measured absolute magnitude in $H$ band ($\sim$875--975~K, see their Fig.~16). The global temperature inferred from the petitCODE patchy model fit (971--1119~K) is slightly higher by $\sim$1$\sigma$. The expected range of effective temperatures from \citet{Filippazzo2015} given the measured spectral type is much wider ($\sim$840--1185~K, see their Fig.~15){. All} our atmospheric fits agree {with it}. Finally, we note that given the age and dynamical mass of the companion, evolutionary models (Sect.~\ref{sec:compa_models}) predict surface gravities higher than 5.3~dex. Only the atmospheric fits with the petitCODE models retrieve such large values.

In all the atmospheric fits, the $H$ broad-band photometric point reported by \citet{Crepp2014} is off by at least $\sim$3$\sigma$. \citet{Mesa2020} {estimated} from a long-slit spectrum an absolute photometry in the $H$ band of 15.84$\pm$0.08~mag assuming the distance derived from the \textit{Gaia} parallax{. This} is fainter by $\sim$3.4$\sigma$ {than} the photometry of 15.37$\pm$0.11~mag {derived} in Sect.~\ref{sec:compa_models} from the apparent magnitude in \citet{Crepp2014}. {The fainter $H$-band magnitude found by \citet{Mesa2020} implies a redder $H$-$L^{\prime}$ color by 0.47~mag in the bottom-right panel of Fig.~\ref{fig:cmd}. It points toward lower effective temperatures of $\sim$800--890~K using the empirical relation of \citet{Filippazzo2015}.} Our absolute magnitude in the $J$ band of 15.08$\pm$0.11~mag recomputed from the apparent magnitude in \citet{Crepp2014} {agrees} with the value of 15.13$\pm$0.02~mag reported by \citet{Mesa2020}.

\section{Comparison of the properties of HD\,19467B to model predictions}
\label{sec:compa_models}

\begin{figure}[!t]
\centering
\includegraphics[width=.4\textwidth]{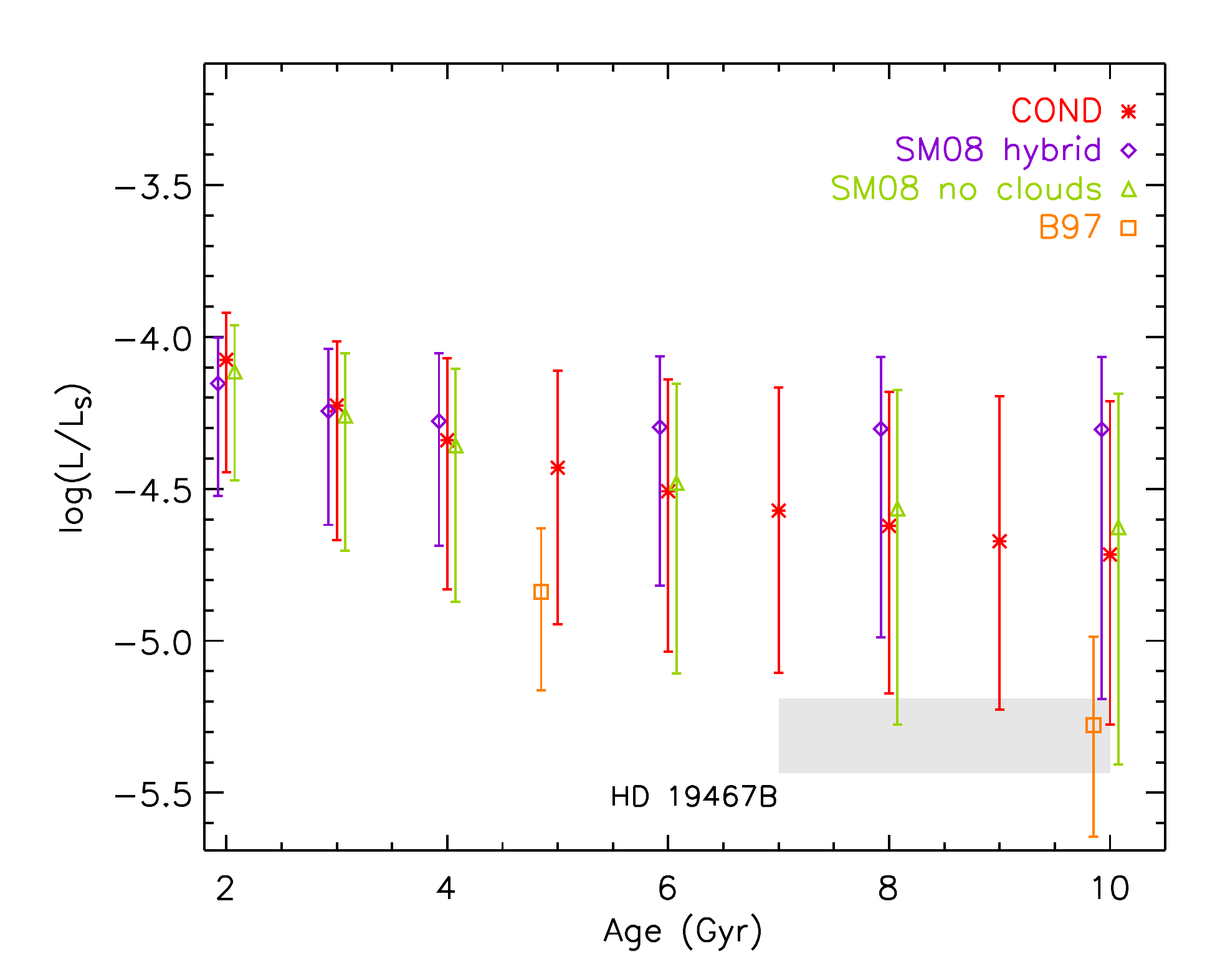}
\includegraphics[width=.4\textwidth]{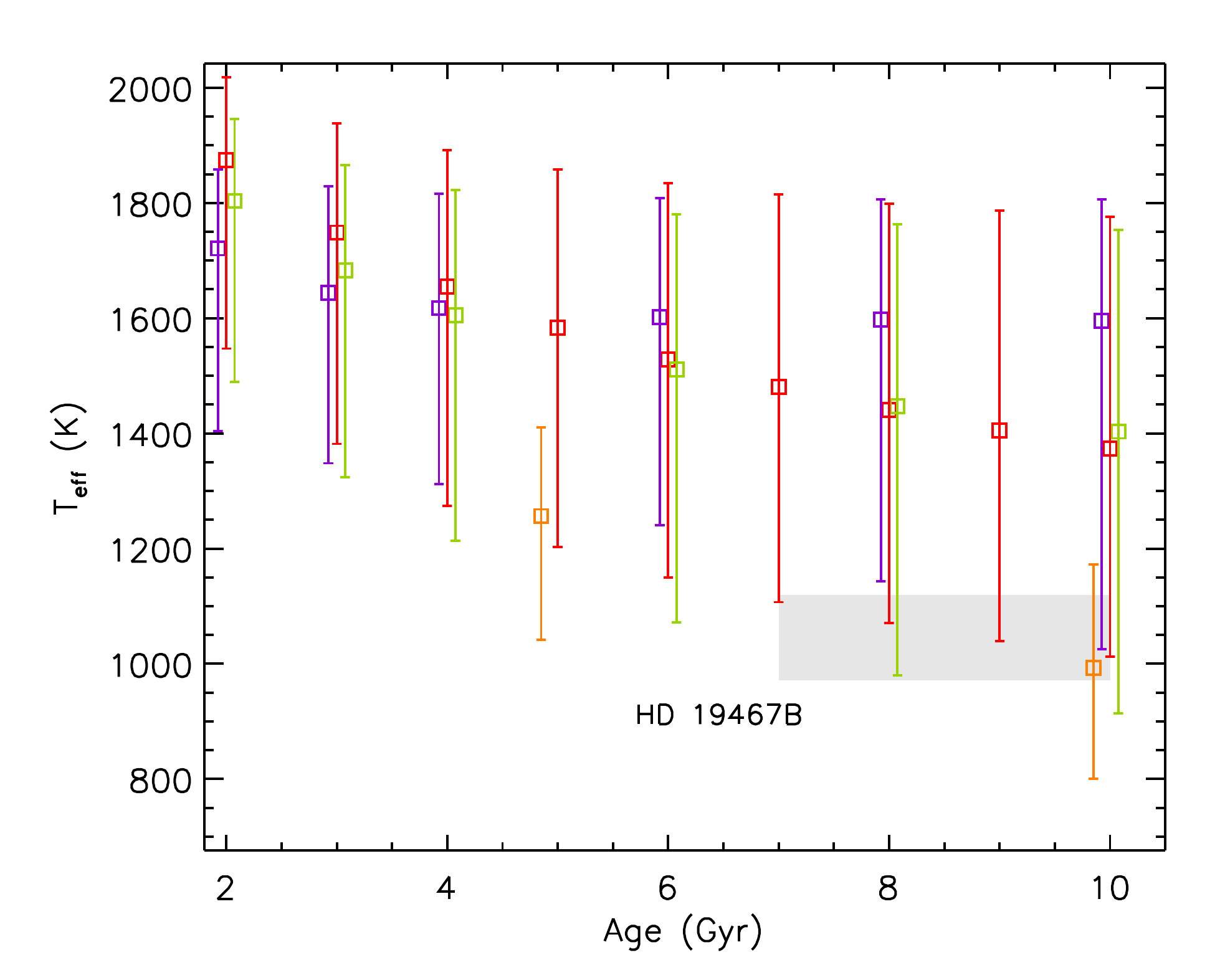}
\caption{{Bolometric luminosity (\textit{top}) and effective temperature (\textit{bottom}) as a function of the age of HD\,19467B (gray area) compared to evolutionary tracks from the models COND \citep{Baraffe2003}, \citet{Saumon2008} (for two treatments of the clouds), and \citet{Burrows1997} assuming the mass for the companion estimated from the orbital fit and the theoretical hydrogen-burning mass limit (data points).} Small horizontal offsets are applied to all models except for COND for clarity.}
\label{fig:lbol_age}
\end{figure}

\begin{figure*}[t]
\centering
\includegraphics[width=.4\textwidth]{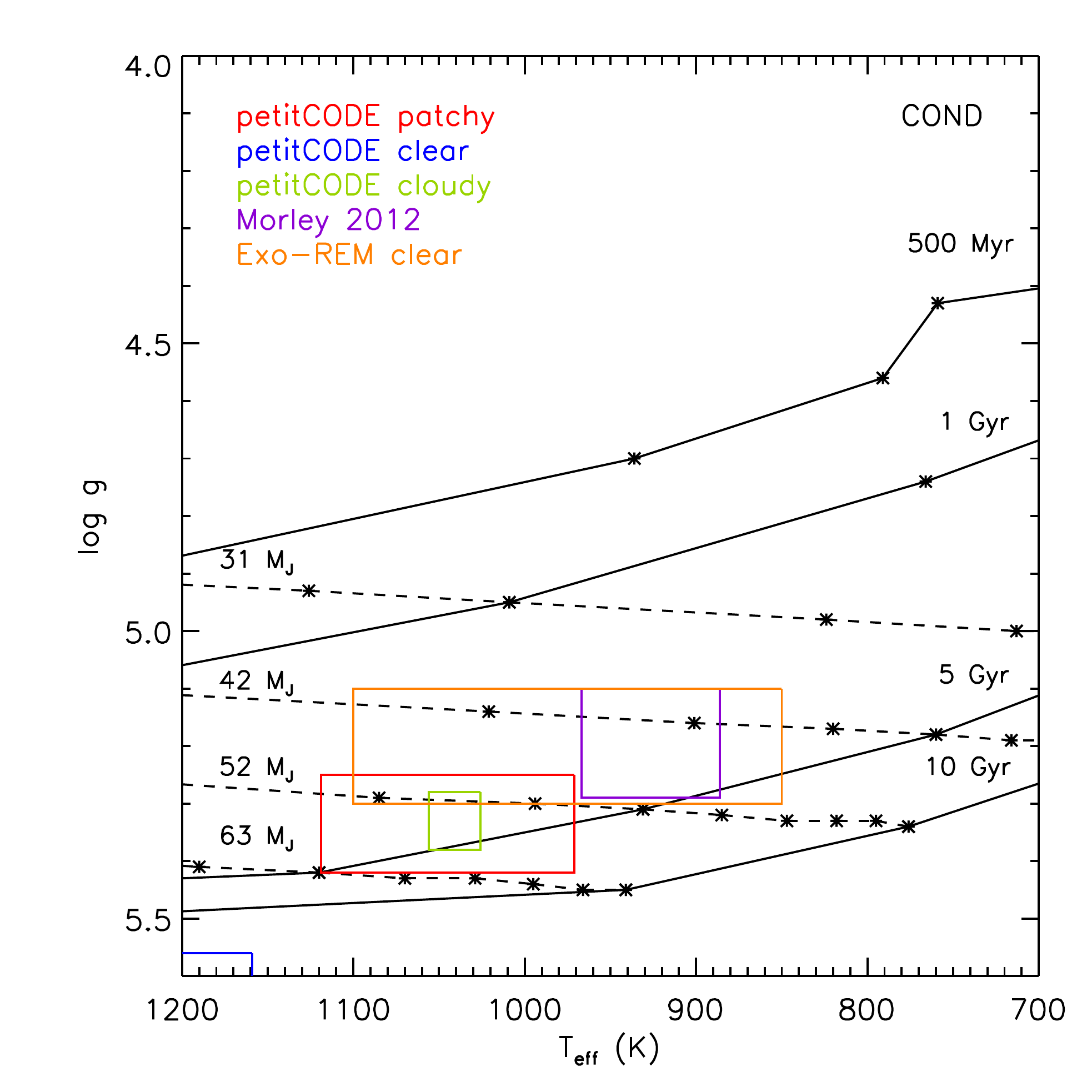}
\includegraphics[width=.4\textwidth]{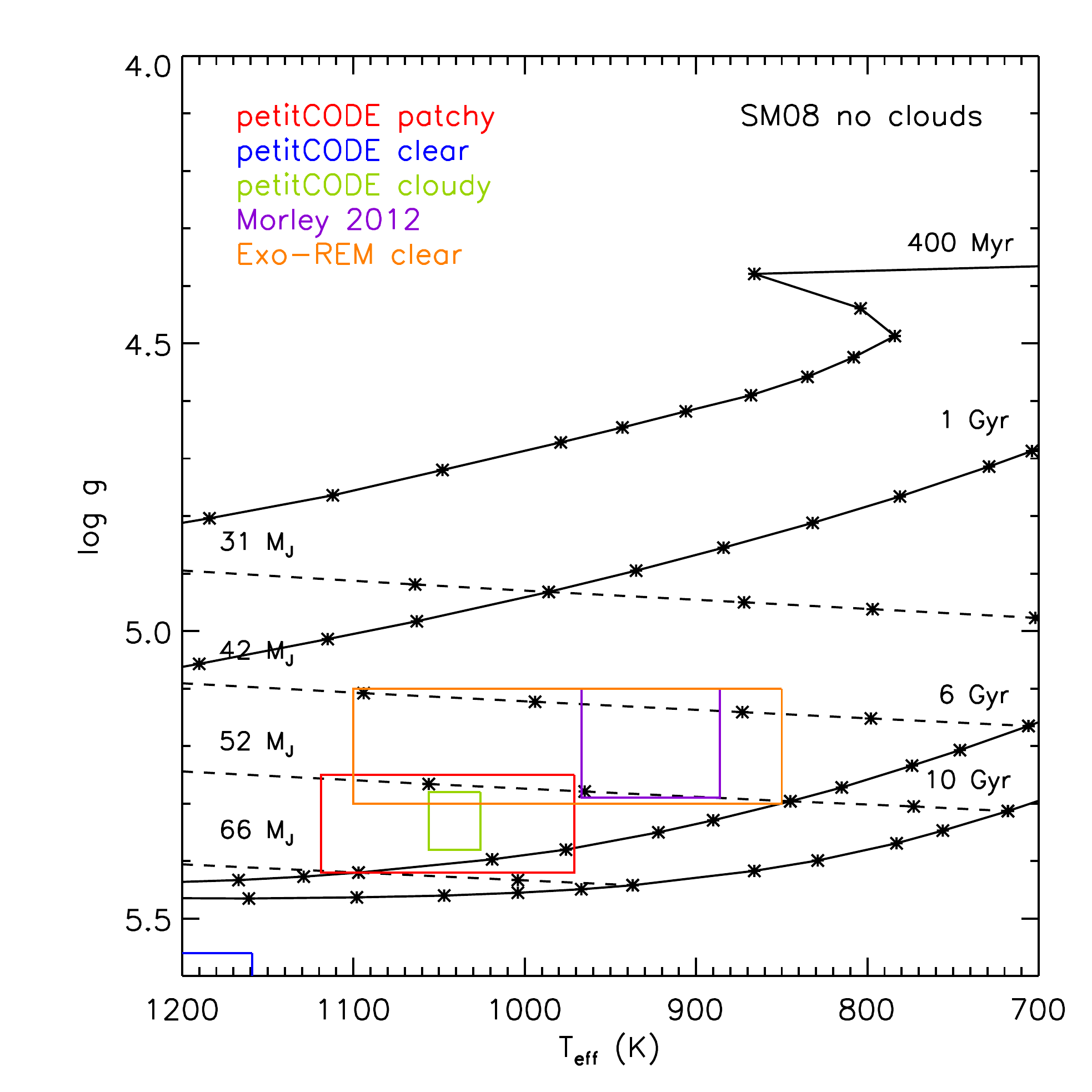}
\includegraphics[width=.4\textwidth]{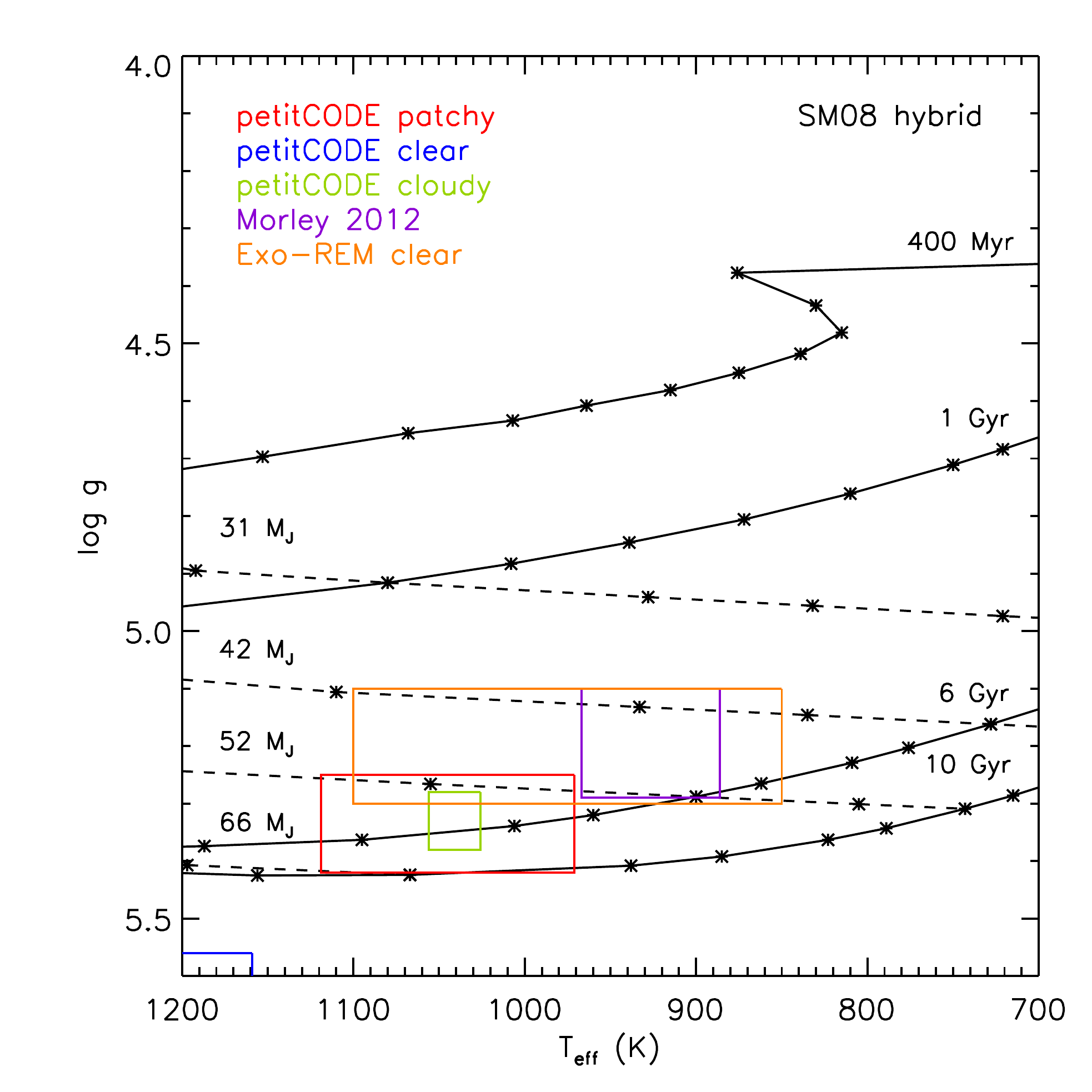}
\includegraphics[width=.4\textwidth]{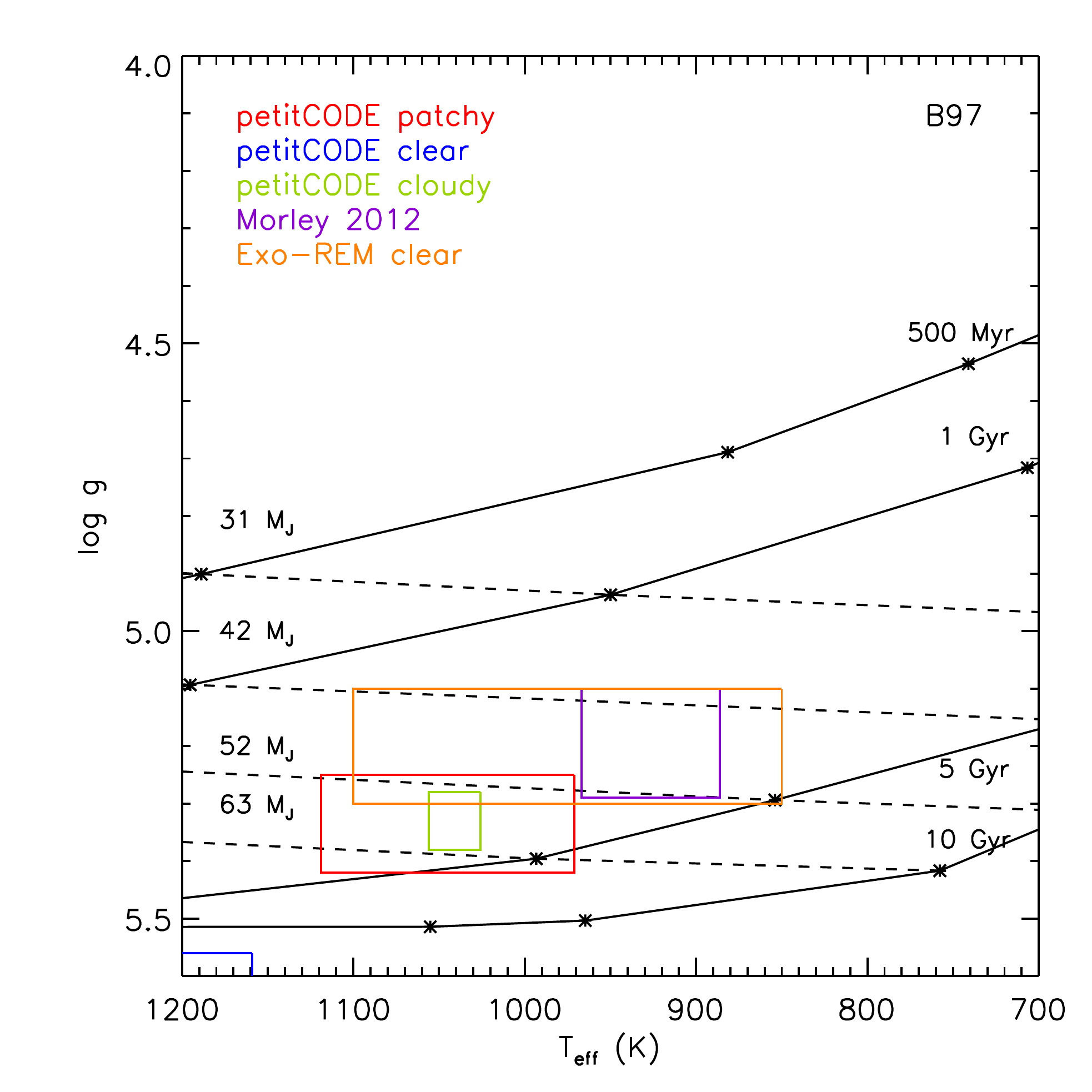}
\caption{Surface gravity as a function of the effective temperature predicted for several ages (black solid curves) and companion masses (dashed curves) by the models COND, the models of \citet{Saumon2008} for two treatments of the clouds, and the models of \citet{Burrows1997}. For comparison, the parameters derived from the atmospheric fits are shown (colored rectangles, Sect.~\ref{sec:atmosfits}).}
\label{fig:plots_teff_logg_mass_age}
\end{figure*}

{Regarding a possible formation mechanism for HD\,19467B, the mass ratio derived in Sect.~\ref{sec:paramcorr} (0.065--0.086 at 68\%) is large and challenging to explain in a disk gravitational instability scenario \citep{Boss1997} without additional mechanisms (non-in situ formation with migration, mass accretion after formation). This would support a stellar-like (or stellar binary-like) formation scenario for the companion.}

Figure~\ref{fig:lbol_mass} compares the bolometric luminosity and mass of HD\,19467B to the model isochrones from \citet{Baraffe2003}. To derive the bolometric luminosities, we used the model relations for field dwarfs of \citet{Filippazzo2015}, a bolometric luminosity for the Sun of 4.74~dex \citep{Prsa2016}, the absolute magnitudes in the $J$ and $K_s$ bands in \citet{Crepp2014} corrected for the new distance estimate from \textit{Gaia} ($M_J$\,=\,15.08$\pm$0.11~mag, $M_H$\,=\,15.37$\pm$0.11~mag, $M_{K_s}$\,=\,15.44$\pm$0.09~mag), and the spectral type of T5.5$\pm$1.0 in \citet{Crepp2015}. We find log($L$/$L_{\sun}$)=$-$5.17$^{+0.10}_{-0.08}$~dex from the $J$-band magnitude and log($L$/$L_{\sun}$)=$-$5.31$\pm$0.12~dex from the $K_s$-band magnitude. Our bolometric luminosity estimate using the $K_s$ magnitude agrees with the estimate of $-$5.19$^{+0.06}_{-0.07}$~dex derived by \citet{Wood2019} based on the absolute $K_s$ magnitude in \citet{Crepp2014} and computed assuming the distance estimated from the \textsc{Hipparcos} parallax. The measured bolometric luminosity and mass of HD\,19467B are compatible with an age older than $\sim$7~Gyr{. These constraints} agree with our age estimate. The models of \citet{Baraffe2003} assume solar metallicity, whereas HD\,19467B could potentially have slightly subsolar metallicity. Very few evolutionary models explore the effects of metallicity. {The models {from} \citet{Saumon2008} have a poor sampling (0.3~dex) and assume cloudless atmospheres}. The Sonora models \citep{Marley2017} should soon allow {for these issues to be alleviated}. However, based on the cloudless models of \citet{Saumon2008} and assuming linear interpolations, we expect small shifts on the predicted bolometric luminosity and effective temperature toward lower values ($\sim$0.03~dex and $\sim$15~K).

\begin{figure*}[t]
\centering
\includegraphics[width=.4\textwidth]{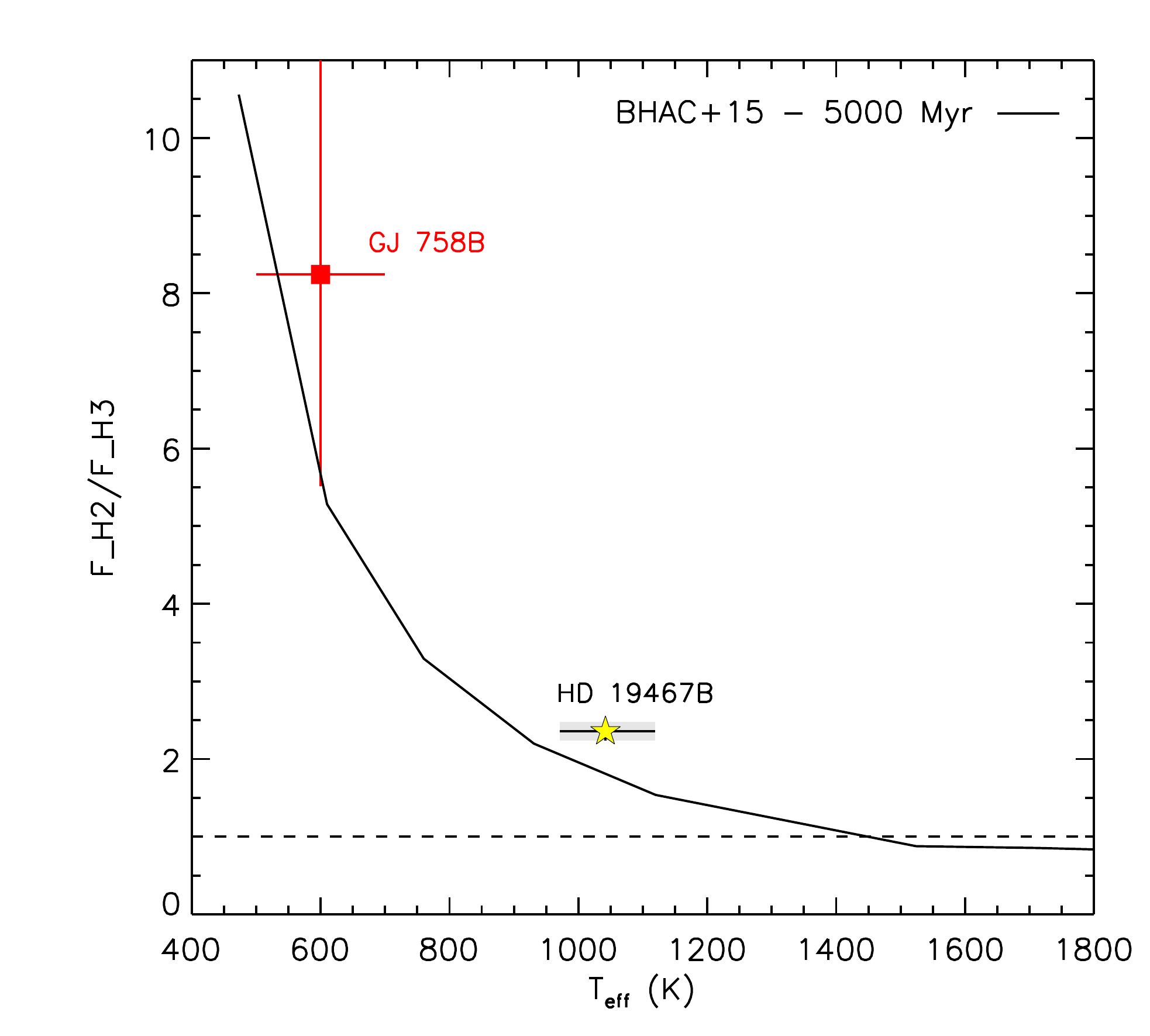}
\includegraphics[width=.4\textwidth]{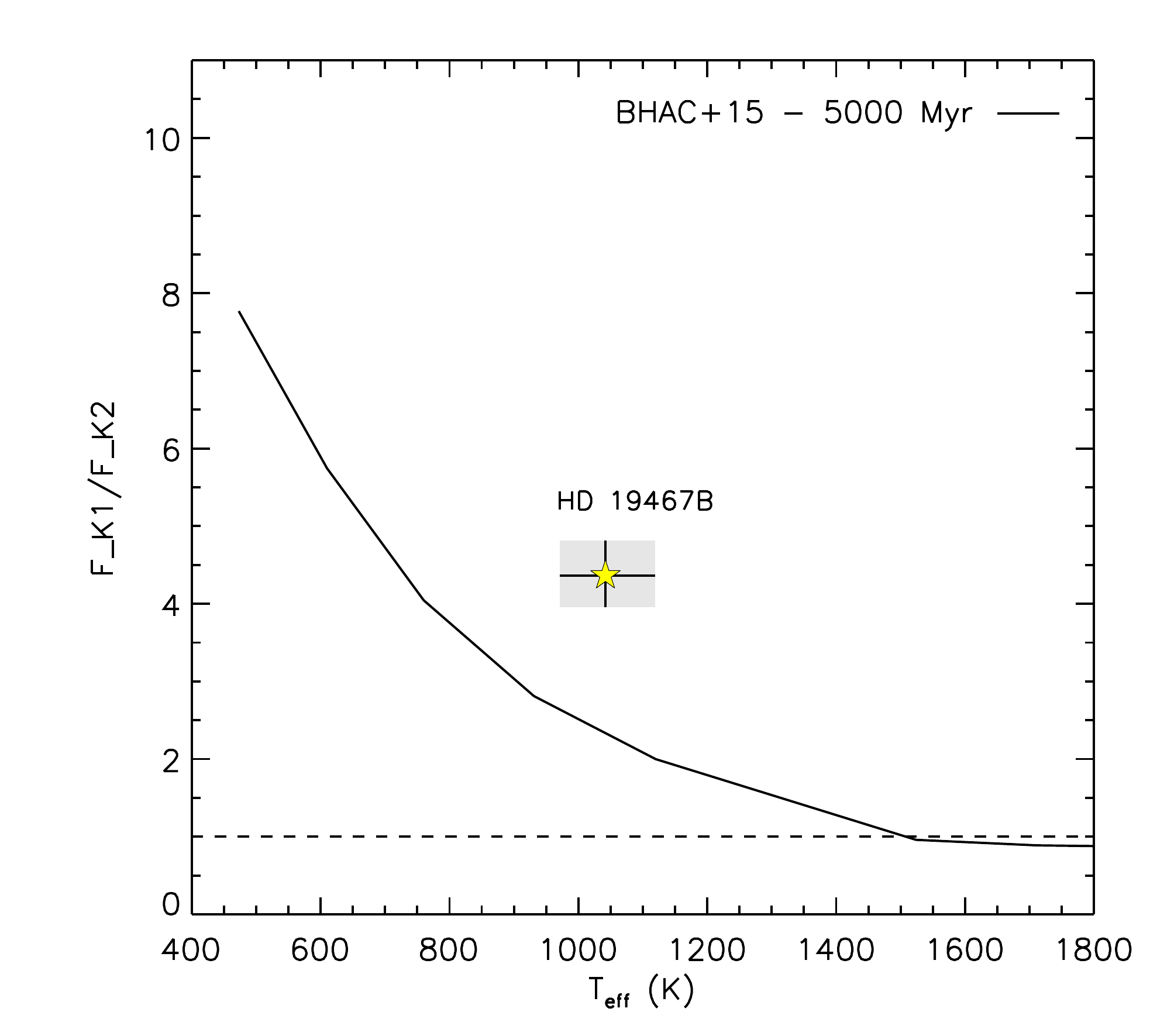}
\caption{Flux ratios outside and inside a methane absorption feature in the $H$ band (\textit{left}) and in the $K_s$ band (\textit{right}) as a function of the effective temperature of HD\,19467B (star symbol) and GJ\,758B (red square). The evolution predicted by the model of \citet{Baraffe2015} for an age of 5~Gyr is also shown (solid curve).}
\label{fig:fluxratio_teff}
\end{figure*}

Figure~\ref{fig:lbol_age} shows the estimated bolometric luminosity (from the $K_s$-band magnitude), effective temperature, and age of HD\,19467B with the predictions from the models COND \citep{Baraffe2003}, of \citet{Saumon2008} (for two treatments of the clouds, hybrid and no clouds), and of \citet{Burrows1997} {assuming for the companion mass 74$^{+3}_{-9}$~$M_J$} (Sect.~\ref{sec:paramcorr}). The hybrid cloudy model of \citet{Saumon2008} intends to model the disappearance of the clouds at the L/T transition by increasing the cloud sedimentation parameter with decreasing $T_{\rm{eff}}$. The evolution model {was} computed assuming for the atmosphere model a combination of cloudless and cloudy atmosphere models. We consider for the effective temperature the constraints from the petitCODE fit with patchy clouds (Sect.~\ref{sec:atmosfits}). We could not test the recent models of \citet{Baraffe2015} {because} they do not extend to effective temperatures below $\sim$1600~K for the age range of HD\,19467B. The measured bolometric luminosity, age, and dynamical mass of the companion are {better} reproduced by the models of \citet{Burrows1997}{. The other models} tend to overestimate its luminosity or equivalently to underestimate its cooling. When considering the effective temperature instead of the bolometric luminosity, the properties of the companion are compatible with more models, but are {better} reproduced by the models of \citet{Burrows1997} and the cloudless models of \citet{Saumon2008}.

\citet{Dieterich2018} {find} that evolutionary models tend to underpredict the cooling rate of $\epsilon$ Ind C and that evolutionary models employing model atmospheres with lower molecular opacities reproduce its measured mass better. \citet{Brandt2019c} {find} that when assuming an age older than 5~Gyr the models of \citet{Burrows1997} reproduce the measured mass of GJ~229B better. \citet{Brandt2019b} {find} for GJ~758B that the models COND, the models of \citet{Burrows1997}, and the models of \citet{Saumon2008} without clouds and a hybrid cloud model are compatible with its measured mass for an age older than 6~Gyr.

\citet{Saumon2008} {discuss} the differences between their models {and} the models COND and of \citet{Burrows1997}. Briefly, the main differences between the cloudless models of \citet{Saumon2008} and COND relevant to the case of an old and massive brown dwarf such as HD\,19467B reside in the surface boundary condition provided by the atmosphere and the noninclusion in the former model of the electron conduction in the core of the object (which is a dominant energy transport mechanism). The noninclusion of the latter effect {produces} lower luminosities. For a 10-Gyr brown dwarf of 0.06~$M_\odot$, \citet{Saumon2008} {find} a difference in bolometric luminosity of $\sim$0.1~dex {compared} to the COND model. This value agrees with the luminosity shift found by \citet{Chabrier2000} when including this effect. The main differences between the cloudless models of \citet{Saumon2008} and the models of \citet{Burrows1997} are the use {in the latter model} of a lower value for the helium abundance \citep[0.25 vs. 0.28~dex; the protosolar value is 0.2741$\pm$0.0120,][]{Lodders2003} and a less opaque atmosphere. Both a lower helium abundance and a less opaque atmosphere result in lower luminosities.

Figure~\ref{fig:plots_teff_logg_mass_age} compares the results from our atmospheric fits to the predictions of the four evolutionary models tested above in the effective temperature vs. surface gravity plane. We show model relations between these two parameters for several ages and companion masses. Only the atmospheric parameters derived from the petitCODE patchy fit are consistent with an object of the age of HD\,19467B, when assuming the models COND and of \citet{Saumon2008}. For the models of \citet{Burrows1997}, the predicted ages are too young, because for given age and $T_{\rm{eff}}$ the surface gravities predicted by this model are higher {than those by} the other models. The Exo-REM and Morley 2012 fits suggest too young ages and too low masses, whereas the petitCODE cloudy and clear fits suggest ages which are too young and too old, respectively. However, the temperature and surface gravity derived from the petitCODE patchy fit indicate a mass range slightly smaller ($\sim$57--66~$M_J$) {than} the mass range suggested by the orbital fit.

Finally, we {compare} in Fig.~\ref{fig:fluxratio_teff} the measured CH$_4$ flux ratios in the IRDIS narrow-band filters and the estimated effective temperature to the expectations from the model of \citet{Baraffe2015}. We selected the model curve for an age of 5~Gyr, but we checked that the model curve for an age of 10~Gyr is very similar for the temperature range of HD\,19467B. We {computed} flux ratios $F_{H2}/F_{H3}$=2.36$^{+0.13}_{-0.12}$ and $F_{K1}/F_{K2}$=4.37$^{+0.45}_{-0.41}$. The CH$_4$ flux ratio in the $K_s$ band is $\sim$1.9 times larger than the CH$_4$ flux ratio in the $H$ band. The measured CH$_4$ flux ratio in the $H$ band {is close} to the predictions given the effective temperature of HD\,19467B estimated in our spectral analysis (Sect.~\ref{sec:atmosfits}), whereas the measured CH$_4$ flux ratio in the $K_s$ band is larger than predicted.

The underpredicted bolometric luminosity by $\sim$0.5~dex of evolutionary models with respect to the measured bolometric luminosity found in \citet{Wood2019} is due to a combination of slightly older age, brighter bolometric luminosity, and smaller dynamical mass estimated from the RV acceleration {compared} to our results.

\section{Conclusions}

We {have presented} VLT/SPHERE and VLT/NaCo observations of the benchmark T-type brown dwarf HD\,19467B to further characterize its orbital and spectral properties. We {have also refined} the properties of the host star using archival data from ASAS, HARPS, and UVES. Our direct rotation period measurement indicates a gyrochronological age of 5.6$\pm$0.8~Gyr, which is older than the {range of 3.1--5.3~Gyr} derived in \citet{Crepp2014} from an indirect rotation period estimate from chromospheric activity indicators. Our isochronal analysis suggests an older age of 9.3$\pm$1.6~Gyr. The chemical abundances and kinematics of the star suggest an age younger than 10~Gyr and a possible membership to the thin disk population, which would set an upper age limit of $\sim$8~Gyr. Considering potential biases in the gyrochronological and isochronal methods at low metallicities and/or ages older than the Sun, we {estimated} an age of 8.0$^{+2.0}_{-1.0}$~Gyr. By fitting the SPHERE and NaCo data, archival RV data from HARPS and HIRES, literature imaging measurements from Keck/NIRC2, and \textsc{Hipparcos}-\textit{Gaia} data, we {derived} constraints on the orbital parameters of HD\,19467B and a {dynamical mass of 65--86~$M_J$. We {have further constrained} the latter to 65--77~$M_J$ using a theoretical limit on the hydrogen-burning mass limit.} Our new photometric data extend the SED of the companion to the $K$ and $L^{\prime}$ bands and confirm that the companion has a cool atmosphere. The spectrophotometric data of the companion are best fitted with model spectra of atmospheres with no clouds or very thin clouds for temperatures of 971--1118~K and high surface gravities of 5.25--5.42~dex. Finally, we {have found} that the measured bolometric luminosity and dynamical mass of HD\,19467B are better reproduced by the evolutionary models of \citet{Burrows1997}, whereas the models of \citet{Baraffe2003} and the models of \citet{Saumon2008} tend to underestimate the cooling of the companion.

Further precise monitoring of the companion with both HARPS and high-contrast imaging in the coming years will be critical to measure at high significance an orbital curvature and place more robust {upper limits} on its dynamical mass. Spectral measurements at higher resolutions and/or at longer wavelengths will help to better constrain its atmospheric properties and chemical abundances. Finally, a more precise age estimate from asteroseismology will improve the comparison of the companion properties to model predictions and help to better distinguish them.

\begin{acknowledgements}
     We thank an anonymous referee for a constructive report that helped to improve the manuscript. The authors thank the ESO Paranal Staff for support in conducting the observations {and Eric Lagadec}, and Nad\`ege Meunier (SPHERE Data Centre) for their help with the data reduction. We thank Justin Crepp for sending us the P1640 spectrum of HD\,19467B during the revision of the manuscript for cross-checks. We thank Hans-Walter Rix for discussions regarding the use of stellar abundances for age determination. This work has made use of recipes from the IDL libraries: IDL Astronomy Users's Library \citep{Landsman1993}, Coyote (\url{http://www.idlcoyote.com/index.html}), EXOFAST \citep{Eastman2013}, JBIU (\url{http://www.simulated-galaxies.ua.edu/jbiu/}), and the s3drs package (\url{http://www.heliodocs.com/xdoc/index.html}). It has also made use of the Python packages: NumPy \citep{Oliphant2006}, emcee \citep{ForemanMackey2013}, corner \citep{Foreman2016}, Matplotlib \citep{Hunter2007}, Astropy \citep{AstropyCollaboration2013, AstropyCollaboration2018}, and dateutil (\url{http://dateutil.readthedocs.io/}). We acknowledge financial support from the Programme National de Planétologie (PNP) and the Programme National de Physique Stellaire (PNPS) of CNRS-INSU. This work has also been supported by a grant from the French Labex OSUG@2020 (Investissements d'avenir -- ANR10 LABX56). The project is supported by CNRS, by the Agence Nationale de la Recherche (ANR-14-CE33-0018). This work has made use of the SPHERE Data Centre, jointly operated by OSUG/IPAG (Grenoble), PYTHEAS/LAM/CeSAM (Marseille), OCA/Lagrange (Nice), Observatoire de Paris/LESIA (Paris), and Observatoire de Lyon, also supported by a grant from Labex OSUG@2020 (Investissements d’avenir – ANR10 LABX56). T.H. acknowledges support from the European Research Council under the Horizon 2020 Framework Program via the ERC Advanced Grant Origins 83 24 28. This publication has made use of VOSA, developed under the Spanish Virtual Observatory project supported by the Spanish MINECO through grant AyA2017-84089. VOSA has been partially updated by using funding from the European Union's Horizon 2020 Research and Innovation Programme, under Grant Agreement nº 776403 (EXOPLANETS-A). This research has made use of the SIMBAD database and the VizieR Catalogue access tool, both operated at the CDS, Strasbourg, France. The original descriptions of the SIMBAD and VizieR services were published in \citet{Wenger2000} and \citet{Ochsenbein2000}. This research has made use of NASA's Astrophysics Data System Bibliographic Services. SPHERE is an instrument designed and built by a consortium consisting of IPAG (Grenoble, France), MPIA (Heidelberg, Germany), LAM (Marseille, France), LESIA (Paris, France), Laboratoire Lagrange (Nice, France), INAF -- Osservatorio di Padova (Italy), Observatoire de Genève (Switzerland), ETH Zurich (Switzerland), NOVA (Netherlands), ONERA (France), and ASTRON (Netherlands), in collaboration with ESO. SPHERE was funded by ESO, with additional contributions from CNRS (France), MPIA (Germany), INAF (Italy), FINES (Switzerland), and NOVA (Netherlands). SPHERE also received funding from the European Commission Sixth and Seventh Framework Programs as part of the Optical Infrared Coordination Network for Astronomy (OPTICON) under grant number RII3-Ct-2004-001566 for FP6 (2004--2008), grant number 226604 for FP7 (2009--2012), and grant number 312430 for FP7 (2013--2016).    

\end{acknowledgements}

%
   \bibliographystyle{aa} 
   \bibliography{../../biblio} 
%

\begin{appendix}

\section{Orbital fit on the imaging data}
\label{sec:orbfit_im}

We fit the imaging data of HD\,19467B using a custom implementation of the Bayesian rejection sampling approach Orbits For The Impatient \citep{Blunt2017, Maire2019}. {We corrected the Keck and NaCo data for the systematics measured with respect to the SPHERE data in Sect.~\ref{sec:orbfit}.} We assumed uniform distributions in $e$, cos\,$i$, $\omega$, and $T_0$. Figure~\ref{fig:orbit_im_radec} shows a sample of fitted orbits. Figure~\ref{fig:orbit_im} and Table~\ref{tab:orbit_im} show the derived orbital parameters based on the statistics of {27\,374 fitted orbits}. {The longitude of {the} node and the argument of {the} periastron are restrained} to the interval [0,180) deg to account for the ambiguity on the longitude of {the} node inherent to the fitting of imaging data only.

Compared to the constraints derived in \citet{Bowler2020}, our constraints {agree} given the uncertainties but {are} broader. We confirm that the orbital eccentricity of the companion is below 0.8. The most significant difference is for the longitude of {the} node. Our distribution for this parameter extends to values smaller than 60$^{\circ}$, whereas no such values are found in \citet{Bowler2020}.

\begin{figure}[h]
\centering
\includegraphics[width=.42\textwidth]{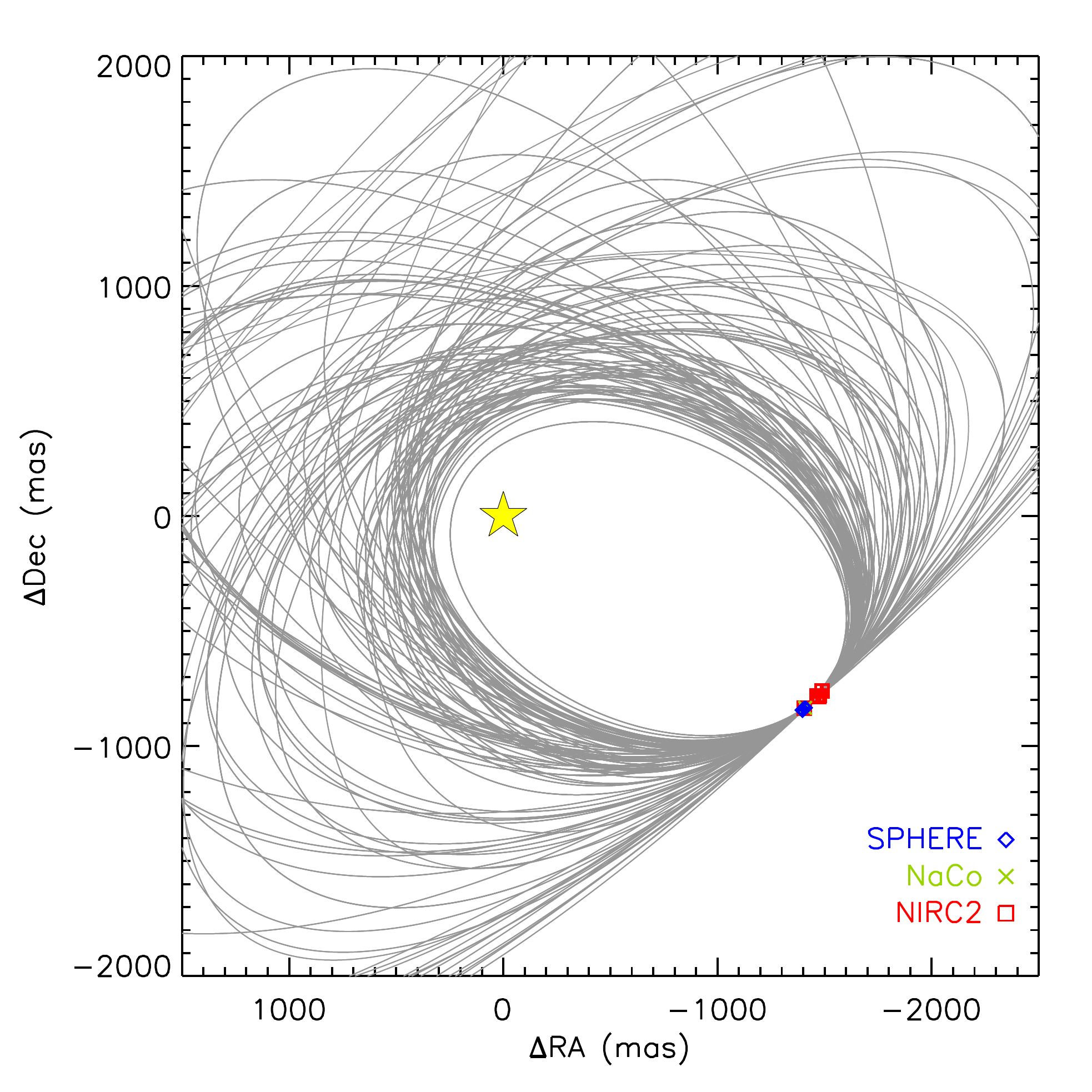}
\caption{{Sample of 100 orbits (gray curves) fitted on the imaging data (colored data points).} The yellow star indicates the position of the star.}
\label{fig:orbit_im_radec}
\end{figure}

\begin{table}[h]
\caption{{Orbital parameters derived using the imaging data.}}
\label{tab:orbit_im}
\begin{center}
\begin{tabular}{l c c c c c}
\hline\hline
Parameter & Unit & Median $\pm$ 1$\sigma$ & $\chi^2_{\rm{min}}$ \\
\hline
$P$ & yr & 390$^{+397}_{-154}$ & 1324 \\[3pt]
$a$ & au & 52$^{+31}_{-15}$ & 119 \\[3pt]
$e$ & & 0.43$^{+0.21}_{-0.23}$ & 0.20 \\[3pt]
$i$ & $^{\circ}$ & 127$^{+17}_{-8}$ & 112 \\[3pt]
$\Omega$ & $^{\circ}$ & 88$^{+44}_{-33}$ & 120 \\[3pt]
$\omega$ & $^{\circ}$ & 74$^{+65}_{-36}$ & 142 \\[3pt]
$T_0$ & AD & 2096$^{+58}_{-231}$ & 1674 \\[3pt]
\hline
\end{tabular}
\end{center}
\end{table}

\begin{figure*}[t]
\centering
\includegraphics[width=.99\textwidth]{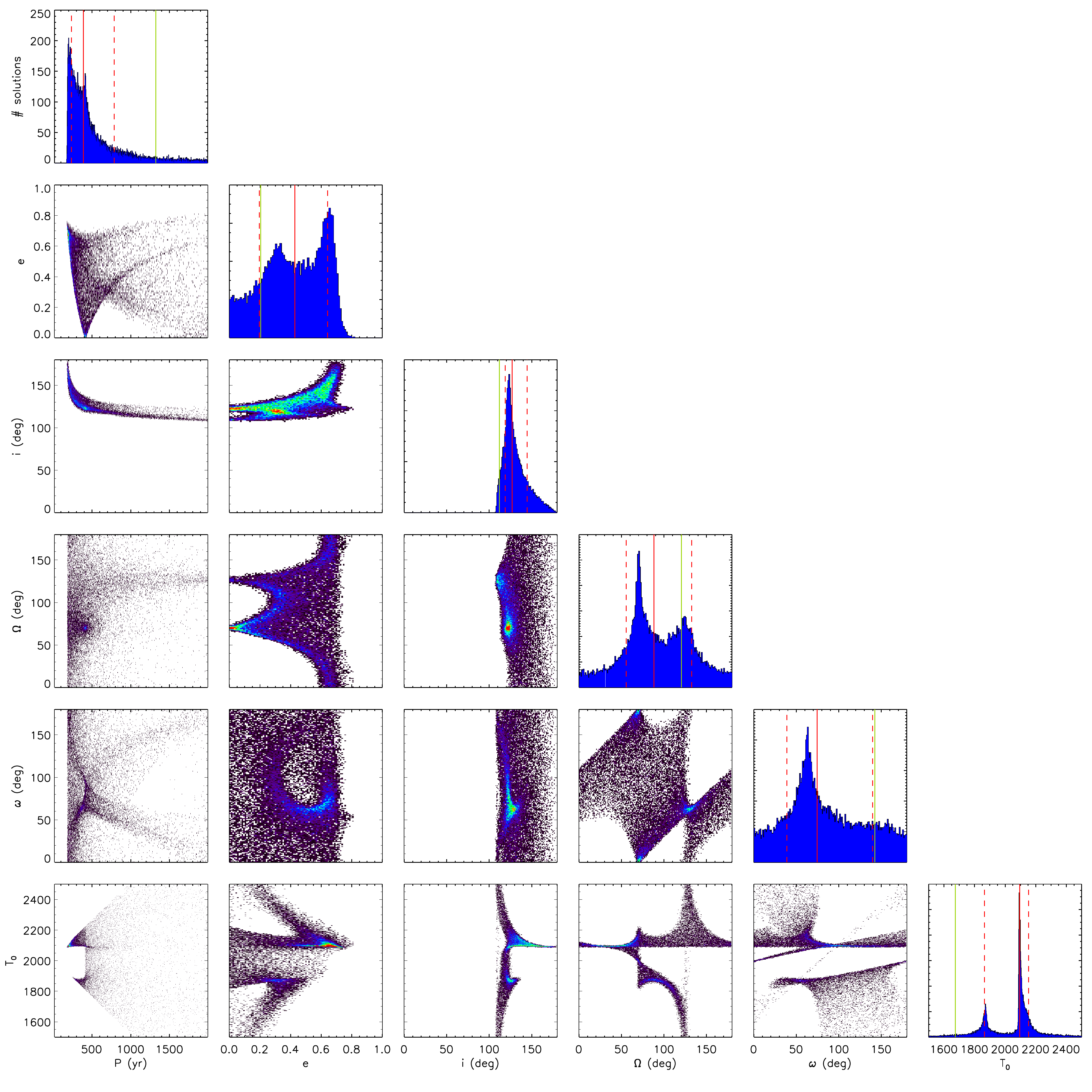}
\caption{{Posterior distributions of the orbital parameters obtained by fitting the imaging data.} The diagrams displayed on the diagonal from top left to lower right represent the 1D histogram distributions for the individual elements. The off-diagonal diagrams show the correlations between pairs of orbital elements. The linear color-scale in the correlation plots accounts for the relative local density of orbital solutions. In the histograms, the green solid line indicates the best $\chi^2$ fitted solution, the red solid line shows the 50\% percentile value, and the red dashed lines represent the interval at 68\%.}
\label{fig:orbit_im}
\end{figure*}

\section{Orbital fit on the imaging and RV data}
\label{sec:orbfit_imrv}

\begin{table}
\caption{{Orbital parameters and dynamical mass of HD\,19467B from the imaging-RV fit.}}
\label{tab:orbparams}
\begin{center}
\begin{tabular}{l c c c}
\hline\hline
Parameter & Unit & Median $\pm$ 1$\sigma$ & Best fit \\
\hline
\multicolumn{4}{c}{Fitted parameters} \\
\hline
$a$ & $''$ & 1652$^{+516}_{-354}$ & 1376\\[3pt]
$\sqrt{e}$\,cos\,$\omega$ & & $-$0.32$\pm$0.06 & $-$0.19\\[3pt]
$\sqrt{e}$\,sin\,$\omega$ & & $-$0.69$^{+0.11}_{-0.08}$ & $-$0.75\\[3pt]
$i$ & $^{\circ}$ & 130$^{+12}_{-9}$ & 140\\[3pt]
$\Omega$ & $^{\circ}$ & 135$\pm$5 & 145\\[3pt]
$T_0$ & BJD & 2510135$^{+25480}_{-15949}$ & 2500799\\[3pt]
$\kappa_A$ & m\,s$^{-1}$ & 263$^{+69}_{-51}$ & 210\\[3pt]
$\pi$ & mas & 31.23$\pm$0.12 & 31.05\\[3pt]
System mass $M_{\rm{tot}}$ & $M_{\sun}$ & 1.024$^{+0.030}_{-0.026}$ & 1.046\\[3pt]
ZP$_{\rm{HARPS}}$ & m\,s$^{-1}$ & 12.8$\pm$0.7 & 12.8\\[3pt]
ZP$_{\rm{HIRES}}$ & m\,s$^{-1}$ & $-$4.0$\pm$0.9 & $-$3.7\\[3pt]
$\sigma_{\rm{HARPS}}$ & m\,s$^{-1}$ & 1.49$^{+0.18}_{-0.15}$ & 1.39\\[3pt]
$\sigma_{\rm{HIRES}}$ & m\,s$^{-1}$ & 3.9$^{+0.6}_{-0.5}$ & 3.5\\[3pt]
Sep. scaling $f\rho_{NIRC2}$ & & 0.9955$^{+0.0034}_{-0.0035}$ & 1.0023\\[3pt]
PA offset $\Delta$PA$_{\rm{NIRC2}}$ & $^{\circ}$ & 0.22$^{+0.35}_{-0.34}$ & 0.16\\[3pt]
PA offset $\Delta$PA$_{\rm{NaCo}}$ & $^{\circ}$ & $-$0.73$^{+0.54}_{-0.55}$ & $-$0.90\\[3pt]
\hline
\multicolumn{4}{c}{Computed parameters} \\
\hline
$M_1$ & $M_{\sun}$ & 0.95$\pm$0.02 & 0.99\\[3pt]
$M_2$ & $M_J$ & 74$^{+23}_{-9}$ & 63\\[3pt]
$M_2$/$M_1$ & & 0.074$^{+0.023}_{-0.010}$ & 0.061\\[3pt]
$P$ & yr & 381$^{+187}_{-114}$ & 288\\[3pt]
$a$ & au & 52$^{+16}_{-11}$ & 43\\[3pt]
$e$ & & 0.58$^{+0.11}_{-0.13}$ & 0.60\\[3pt]
$\omega$ & $^{\circ}$ & 65$^{+6}_{-7}$ & 76\\[3pt]
\hline
\end{tabular}
\end{center}
\end{table}

We fit the imaging and RV data of HD\,19467B using a similar MCMC approach to Sect.~\ref{sec:orbfit}. We sampled the parameter space of our {17-parameter model} assuming 20 temperatures for the chains and 100 walkers. The first 8 parameters are the same as for the imaging-RV-astrometry fit in Sect.~\ref{sec:orbfit}{. We} assumed similar priors. The next two parameters are the parallax and total mass of the system. We used the same prior on the parallax as in Sect.~\ref{sec:orbfit}. We drew the system mass around a guess value of 1.015~$M_{\sun}$ considering a host star mass of 0.95~$M_{\sun}$ (Sect.~\ref{sec:star_ppties}) and a companion mass of 0.065~$M_{\sun}$ and assuming a Gaussian distribution with a half width at half maximum of 0.025~$M_{\sun}$. We included the prior information on the host star mass (0.95$\pm$0.02~$M_{\sun}$) in the likelihood function instead of the system mass (by computing the difference between the fitted system mass and the companion mass derived from the binary mass function using the fitted orbital parameters). The remaining parameters and the associated priors are the same as in the imaging-RV fit.

We ran the MCMC analysis for 125\,000 iterations and verified the convergence of the chains with the integrated autocorrelation time. Figure~\ref{fig:cornerplot} shows the posteriors on the parameters obtained after thinning the chains by a factor 100 and discarding the first 75\% of the chains as the burn-in phase. Table~\ref{tab:orbparams} gives the median values with 1$\sigma$ uncertainties and the best-fit values. Figure~\ref{fig:modelorbits} shows a sample of fitted orbits. 

\begin{figure*}[!t]
\centering
\includegraphics[width=.99\textwidth, trim = 5mm 5mm 8mm 6mm, clip]{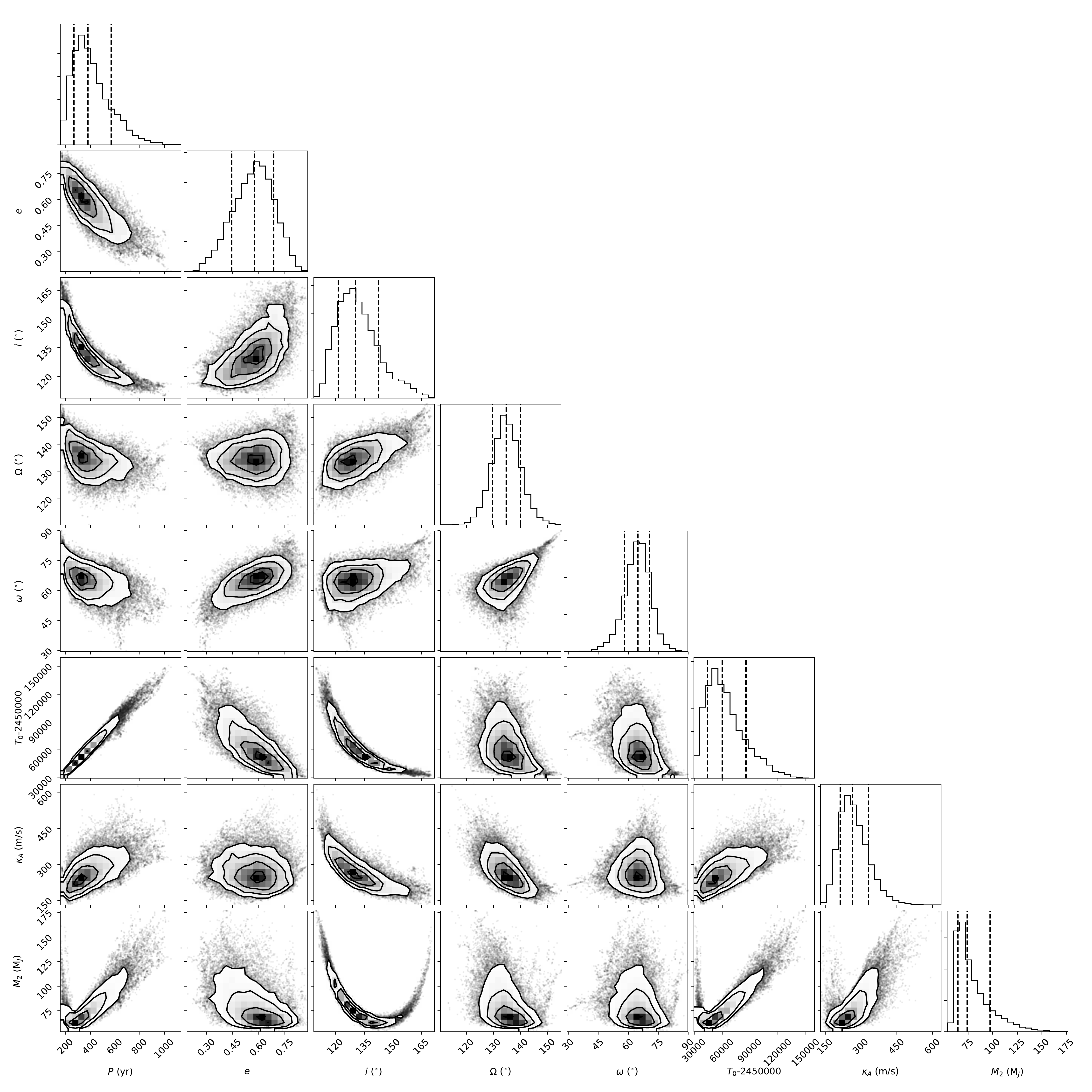}
\caption{{MCMC samples from the posteriors of the orbital parameters and of the mass of HD\,19467B from the imaging-RV fit.} See also Fig.~\ref{fig:orbit_imrvpm_corner}.}
\label{fig:cornerplot}
\end{figure*}

{Compared} to a fit on the imaging data only (Appendix~\ref{sec:orbfit_im}), we note significant improvements on the derived parameters, especially the longitude of {the} ascending node, argument of {the} periastron, and eccentricity. For the eccentricity, values smaller than {$\sim$0.19} are excluded, whereas for the imaging fit circular orbits are possible. The longitude of {the} ascending node and the time at {the} periastron do not show bimodal distributions. The longitude of {the} ascending node is restrained to values of {130--140$^{\circ}$} at 68\%. The argument of {the} periastron is also better constrained to values of {58--71$^{\circ}$} at 68\%. We also note correlations between parameters, with longer periods associated with smaller eccentricities, lower inclinations with respect to the line of sight, and larger RV semi-amplitudes.

\begin{figure*}[t]
\centering
$\vcenter{\hbox{\includegraphics[trim = 0mm 6mm 0mm 15mm, clip, width=.4\textwidth]{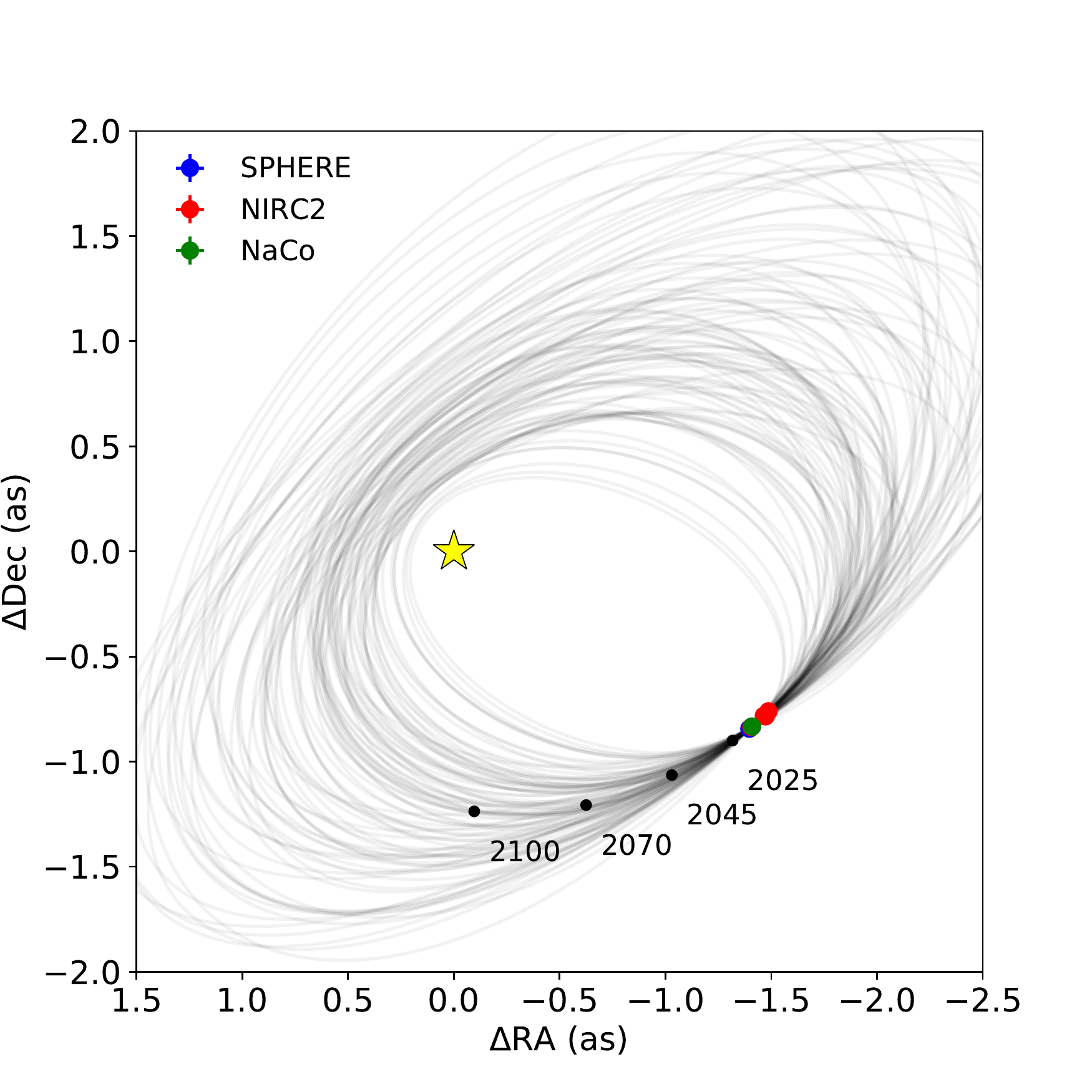}}}$
\hspace*{.1in}
$\vcenter{\hbox{\includegraphics[trim = 9mm 1mm 19mm 12mm, clip, width=.52\textwidth]{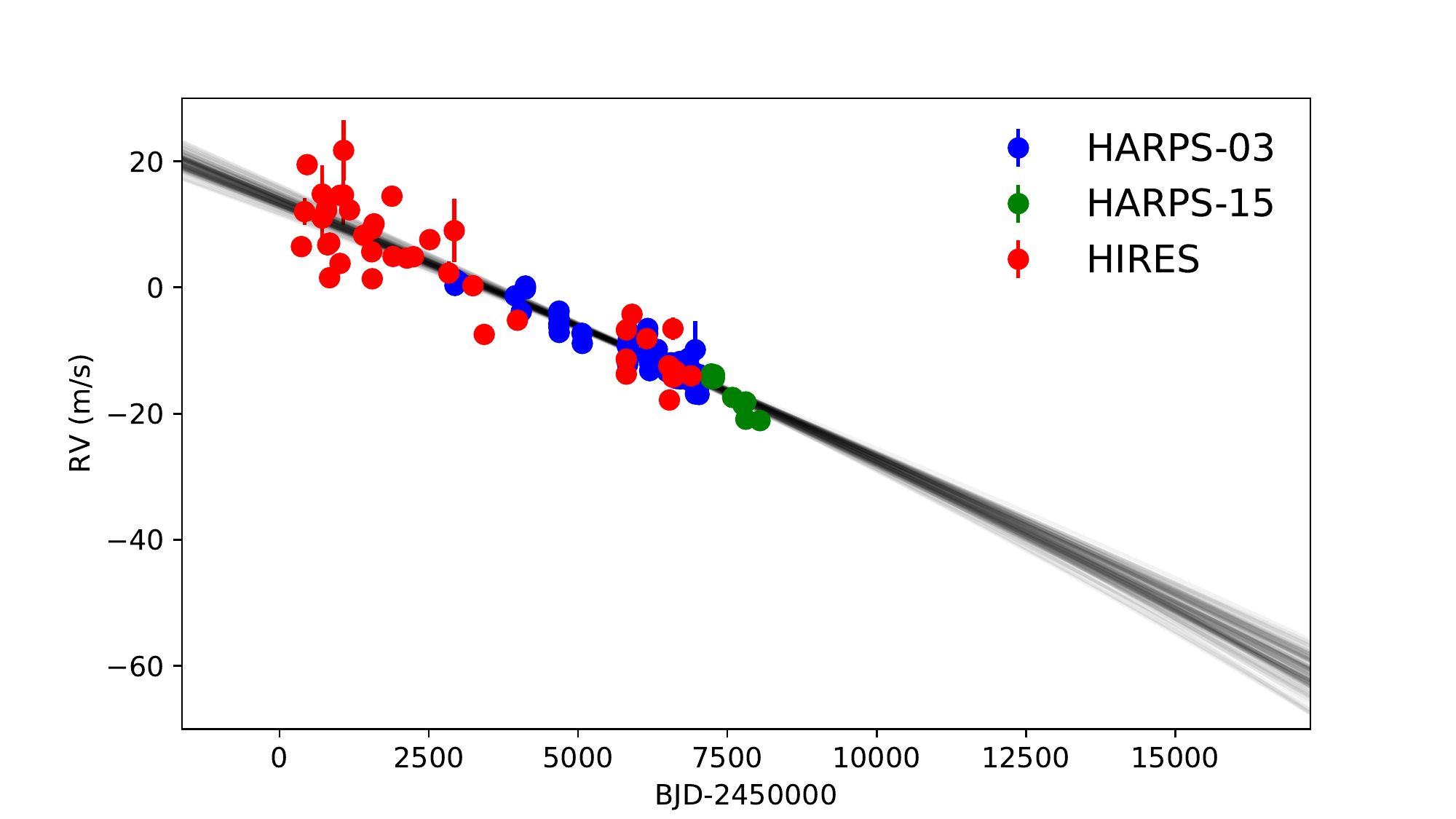}}}$
\caption{{Sample of 100 model orbits (gray curves) fitted on the HD\,19467B data points (colors) from imaging (\textit{left}) and RV (\textit{right}).} In the left panel, the yellow star marks the position of the star and the black dots show the median predicted position for a few epochs in the future.}
\label{fig:modelorbits}
\end{figure*}

Figure~\ref{fig:cornerplot_masses_rvoffjit} shows the posterior distributions for the masses of HD\,19467 A and B as well as for the RV offsets and jitters. The mass posterior for HD\,19467B exhibits a tail toward unphysically large masses beyond the hydrogen-burning mass limit, because the current data do not show a clear curvature.

\begin{figure*}[t]
\centering
$\vcenter{\hbox{\includegraphics[width=.3\textwidth, trim = 5mm 5mm 8mm 6mm, clip]{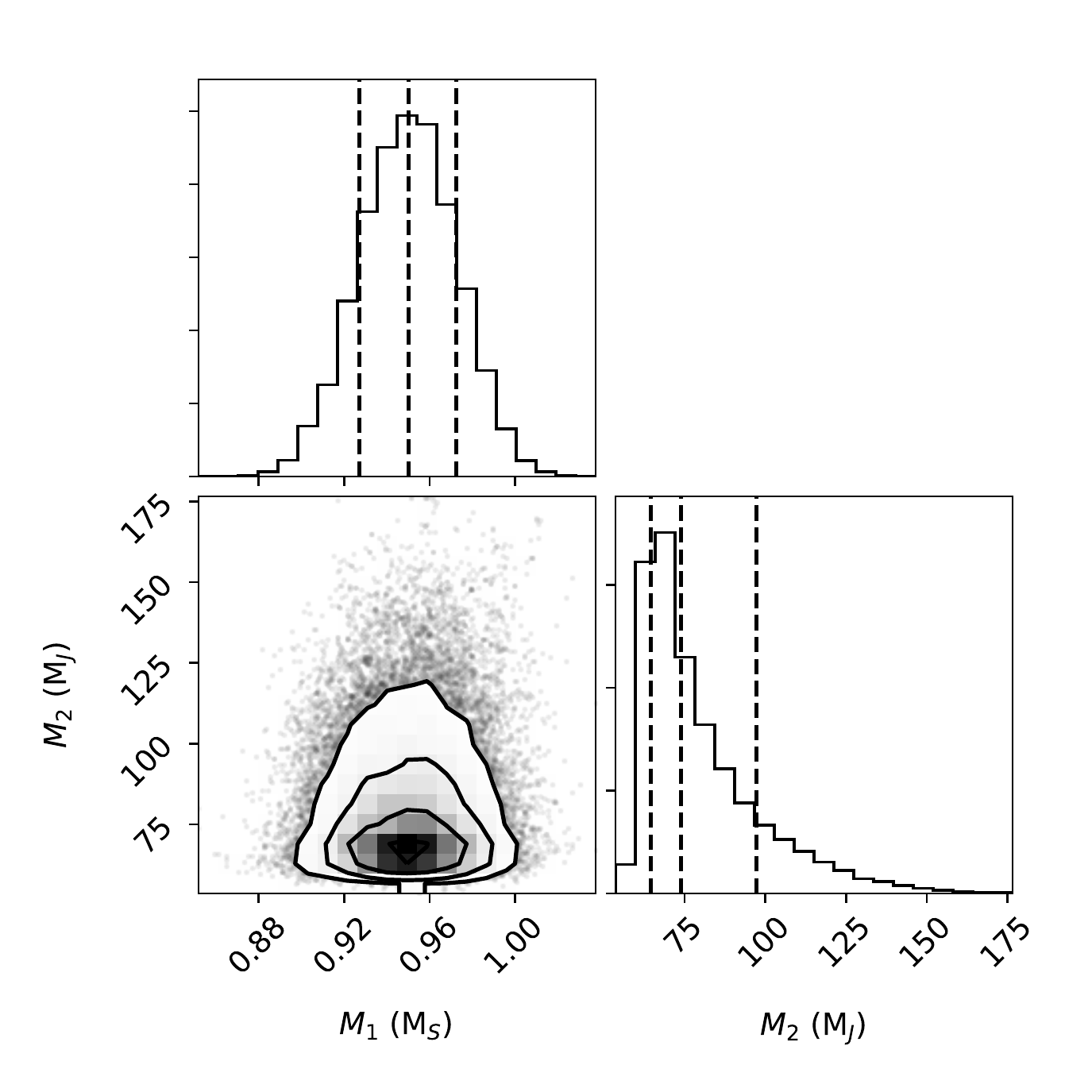}}}$
\hspace*{.2in}
$\vcenter{\hbox{\includegraphics[width=.66\textwidth,trim = 6mm 5mm 10mm 6mm, clip]{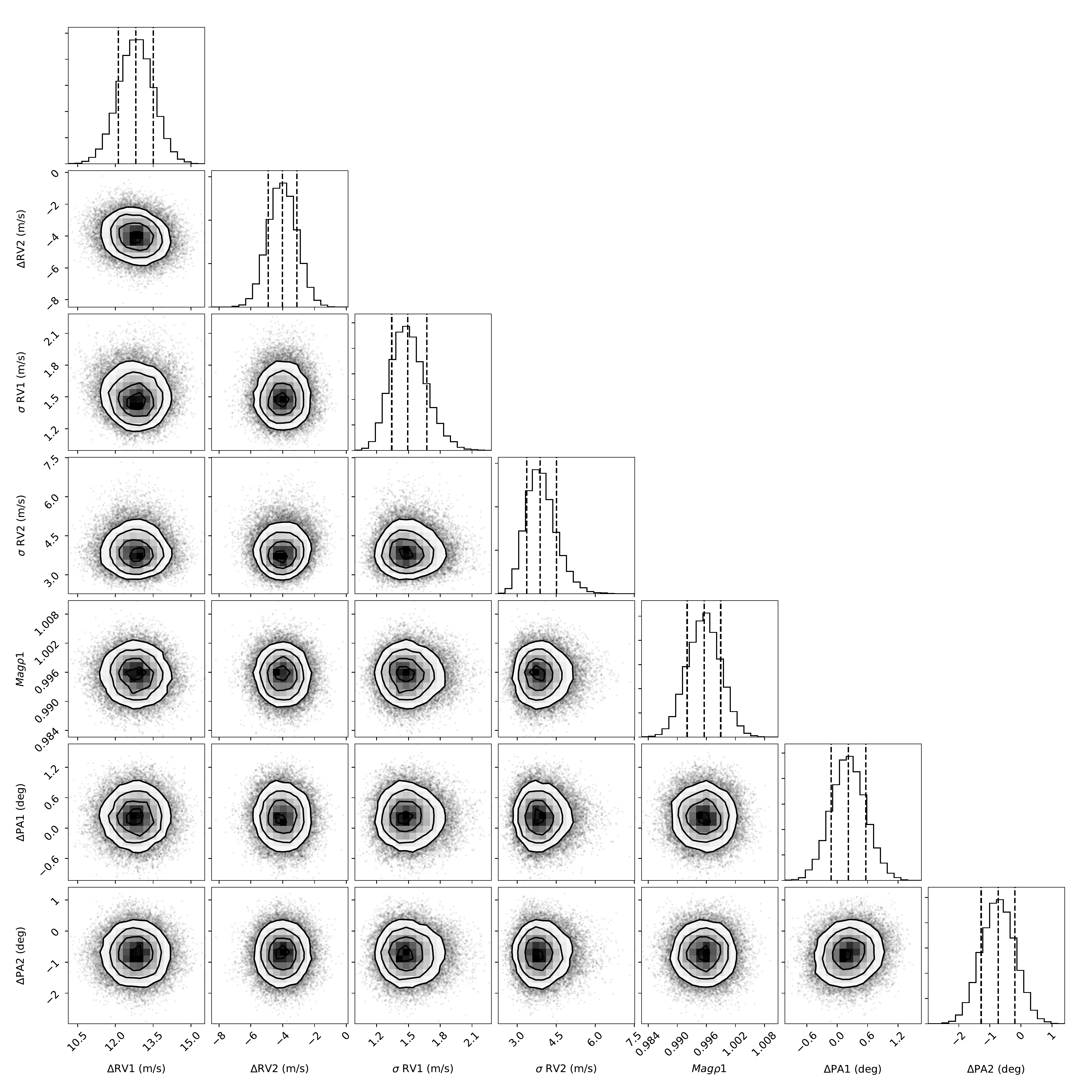}}}$
\caption{{Same as in Fig.~\ref{fig:cornerplot}, but for the masses of HD\,19467 A and B (\textit{left}) and for the RV offsets and jitters (1: HARPS, 2: HIRES) and the imaging offsets (1: Keck, 2: NaCo) (\textit{right}).}}
\label{fig:cornerplot_masses_rvoffjit}
\end{figure*}

\section{Construction of the color-magnitude diagrams using narrow-band photometry}
\label{sec:appcmd}

To build the diagrams shown in the top row of Fig.~\ref{fig:cmd}, we used spectra of M, L, and T dwarfs from the SpeX-Prism library \citep{2014ASInC..11....7B} and from \citet{Leggett2000} and \citet{ASchneider2015} to generate synthetic photometry in the SPHERE filter passbands. The zero points were computed using a flux-calibrated spectrum of Vega \citep{Hayes1985, Mountain1985}. We also considered the spectra of young and/or dusty free-floating objects from \citet{Liu2013}, \citet{Mace2013}, \citet{Gizis2015}, and of young companions \citep{2011ApJ...729..139W, Gauza2015, Stone2016, DeRosa2014, Lachapelle2015, Bailey2014, Rajan2017, Bonnefoy2014b, Patience2010, Lafreniere2010, Chauvin2017a, Delorme2017a, Cheetham2018b, Bonnefoy2018}. The colors and absolute fluxes of the benchmark companions and isolated T-type objects {were} generated from the distance and spectra of those objects in Appendix B in \citet{Bonnefoy2018}. To conclude, we used the spectra of Y dwarfs published in \citet{ASchneider2015}, \citet{Warren2007}, \citet{Delorme2008}, \citet{Burningham2008}, \citet{Lucas2010}, \citet{Kirkpatrick2012}, and \citet{Mace2013} to extend the diagrams in the late-T and early Y-dwarf domain. We used the distances of the field dwarfs reported in \citet{Kirkpatrick2000}, \citet{Faherty2012}, \citet{Dupuy2013}, \citet{Tinney2014}, \citet{Beichman2014}, and \citet{Luhman2016}. We considered those reported in \citet{Kirkpatrick2011}, \citet{Faherty2012}, \citet{ZapateroOsorio2014}, and \citet{Liu2016} for the dusty dwarfs. The companion distances {were} taken from \citet{vanLeeuwen2007} and \citet{Ducourant2014}.

\end{appendix}

\end{document}